\documentclass{aa}
\usepackage{natbib}
\usepackage{hyperref}
\usepackage[english]{babel}
\usepackage{graphicx}
\usepackage{amsmath}
\usepackage[dvipsnames]{xcolor}

\usepackage{slashbox}
\usepackage{color}
\begin{document}
% -----------------------------------------------------------------------------

\title{Uncertainties in gas kinematics arising from stellar continuum modelling in integral field spectroscopy data: the case of {\tt NGC2906} observed with {\tt MUSE/VLT}.}
\author{E. Bellocchi\inst{1,2}, Y. Ascasibar\inst{1,2,3}, L. Galbany\inst{4}, S. F. S\'anchez\inst{5}, H. Ibarra--Medel\inst{5,6}, M. Gavil\'an\inst{1},  \'A. D\'iaz\inst{1,2,3}}
\institute{
$^1$ Universidad Aut\'onoma de Madrid, Departamento de F\'isica Te\'orica, 28049, Cantoblanco, Madrid (Spain)\\
$^2$ Astro--UAM, UAM, Unidad Asociada CSIC (Spain)\\
$^3$ CIAFF--UAM, Centro de Investigaci\'on Avanzada en F\'isica Fundamental (Spain)\\
$^4$ PITT PACC, Department of Physics and Astronomy of University of Pittsburgh, PA 15260, Pittsburgh (USA)\\
$^5$ Instituto de Astronom\'ia, Universidad Nacional Aut\'onoma de M\'exico, A. P. 70-264, C.P. 04510, M\'exico, D.F. Mexico \\
$^6$ Department of Astronomy, University of Illinois Champaign--Urbana, 61801, Illinois (USA)\\
\email{enrica.bellocchi@gmail.com}\\  
}

\date{Received 22 October 2018 / Accepted 12 March 2019}
 
% -----------------------------------------------------------------------------
\abstract
% -----------------------------------------------------------------------------
{
Integral field spectroscopy (IFS) provides detailed information about galaxy kinematics at high spatial and spectral resolution, and the disentanglement of the gaseous and stellar components is a key step in the analysis of the data.
}
{
We study how the use of several stellar subtraction methods and line fitting approaches can affect the derivation of the main kinematic parameters (velocity and velocity dispersion fields) of the ionized gas component.
}
{
The target of this work is the nearby galaxy NGC 2906, observed with the {\tt MUSE} instrument at Very Large Telescope ({\tt VLT}). 
A sample of twelve spectra is selected from the inner (nucleus) and outer (spiral arms) regions, characterized by different ionization mechanisms.
We compare three different methods to subtract the stellar continuum ({\tt FIT3D, STARLIGHT} and {\tt pPXF}), combined with one of the following stellar libraries: {\tt MILES, STELIB} and {\tt GRANADA+MILES}.
} 
{
{The choice of the stellar subtraction method is the most important ingredient affecting the derivation of the gas kinematics, followed by the choice of the stellar library and by the line fitting approach.}
In our data, typical uncertainties in the observed wavelength and width of the H$\alpha$ and [NII] lines are of the order of $\langle\delta\lambda\rangle_{\rm rms} \sim 0.1$~\AA\ and $\langle\delta\sigma\rangle_{\rm rms} \sim 0.2$~\AA\ (i.e. $\sim 5$ and $10$~km~s$^{-1}$, respectively).
The results obtained from the [NII] line seem to be slightly more robust, as it is less affected by stellar absorption than H$\alpha$.
All methods considered yield statistically consistent measurements once a mean systemic contribution $\Delta \bar\lambda = \Delta \bar\sigma = 0.2\,\Delta_{\tt MUSE}$ is added in quadrature to the line fitting errors, where $\Delta_{\tt MUSE} = 1.1$~\AA\ $\sim 50$~km~s$^{-1}$ denotes the instrumental resolution of the {\tt MUSE} spectra.
}
{
Although the subtraction of the stellar continuum is critical in order to recover line fluxes, any method (including none) can be used in order to measure the gas kinematics, as long as an additional component $\Delta \bar\lambda = \Delta \bar\sigma = 0.2\,\Delta_{\tt MUSE}$ is added to the error budget.
}


\titlerunning{Stellar subtraction methods and gas kinematics}
\authorrunning{Bellocchi et al.}

\maketitle

% -----------------------------------------------------------------------------
\section{Introduction}
% -----------------------------------------------------------------------------

The kinematic characterization of different galaxy populations is a key observational input to distinguish between different evolutionary paradigms, such as the `major merger' \citep{T06, T08, kartal12} and `steady cold gas accretion' \citep{keres05, Ocv08, dekel09, ceverino10} scenarios, since it allows one to determine the fraction of rotating disks to mergers at different cosmic epochs \citep[see e.g.,][]{genzel01, T08, dekel09_a, FS09, lemoine09, lemoine10, Bou11, Epi11, Glazebrook13, Bellocchi16}.
Kinematics is also important in order to study the physical processes that govern the formation and evolution of the galaxies, providing a powerful diagnostic to infer the main source of dynamic support \citep{puech07, Epi09, Cappellari13, Zhu18}, to distinguish between relaxed virialized systems and merger events \citep{Flores06, S08, Bellocchi12, BBallesteros15, Bellocchi16}, to infer fundamental galaxy quantities like the dynamical mass \citep{Co05, Bellocchi13, AOrtiz18} and also to detect and characterize radial motions associated with feedback mechanisms, like massive gas outflows \citep{S09, rupke13, LCoba17_2, LCoba17, Maiolino17}.

With the aid of integral field spectroscopy (IFS), we can resolve the kinematical status and internal processes at work within galaxies and more clearly understand the role of the dominant mechanisms involved at early epochs of galaxy formation.
This technique provides much more information of the observed galaxies than was previously available through classical long--slit spectroscopy: on the one hand, the full two--dimensional spatial information (i.e., maps, rather than radial profiles) can be reconstructed at high resolution in regions of bright emission; on the other hand, one of the advantages of IFS is that large areas may be combined in order to properly characterize weak signals, such as in the galaxy outskirts.
The observed spectrum at each spaxel (or area) includes gas in different phases, traced by distinctive features at different wavelengths \citep[e.g., H$\alpha$ emission allows one to study the warm ionized emission, NaD absorption traces the neutral gas, the CO emission helps to trace the cold molecular clouds, etc.; see][]{Bekeraite16, Leung18, Levy18} as well as the underlying stellar population (a smooth continuum with absorption features.
To properly characterize their physical properties, we need to apply methods that are able to separate the contribution of gas and stars to the total emission.
To this aim, in the last ten years, many routines have been developed to subtract the stellar continuum.
Some of them like {\tt STARLIGHT} \citep{CidFer05}, {\tt FIT3D} \citep{Sanchez16}, {\tt pPXF} \citep{Cappellari04}, {\tt STECKMAP} \citep{Ocvirk06}, {\tt sedfit} \citep{Walcher06} and {\tt PyPARADISE} (Husemann et al. in prep.) are widely used nowadays.
These routines allow to study in detail the underlying stellar population(s) of the source and derive interesting parameter/properties such as the total stellar mass, the star formation rate \citep[e.g.,][]{Perez13, GD17}, the age and metallicities, as well as the kinematics of the stars.

In this paper we want to investigate how the choice of different stellar subtraction methods affects the derivation of the gas kinematics, traced by the ionized component through the H$\alpha$ and [NII] lines.
We also analyze how the choice of different line fitting approaches influence the derivation of the kinematic parameters (mean velocity and velocity dispersion).
The target of this study is a nearby galaxy observed with {\tt MUSE/VLT}, from which we obtain high resolution spectra, very suitable to our aims (a related discussion has been recently carried out by \citet{Belfiore19} in the context of the MaNGA Data Analysis Pipeline). We describe the observations in Section~\ref{sec_observations}. In Section 3 we present the selection of the spectra, a brief description of the methods involved for the stellar continuum subtraction, and the line fitting approaches considered.
Section 4 is devoted to the discussion of the results obtained by different methods, and our main conclusions are summarized in Section 5.
Appendix~\ref{sec_appendix} presents some details on the line fitting analysis along with complementary kinematic results and the quality of the fits is discussed in Appendix B. Throughout the paper we consider H$_0$ = 70 km s$^{-1}$ Mpc$^{-1}$, $\Omega_M$ = 0.3 and $\Omega_\Lambda$ = 0.7.

% -----------------------------------------------------------------------------
\section{Observations}
\label{sec_observations}
% -----------------------------------------------------------------------------

\begin{figure*}
   \centering
  \includegraphics[width=0.83\textwidth]{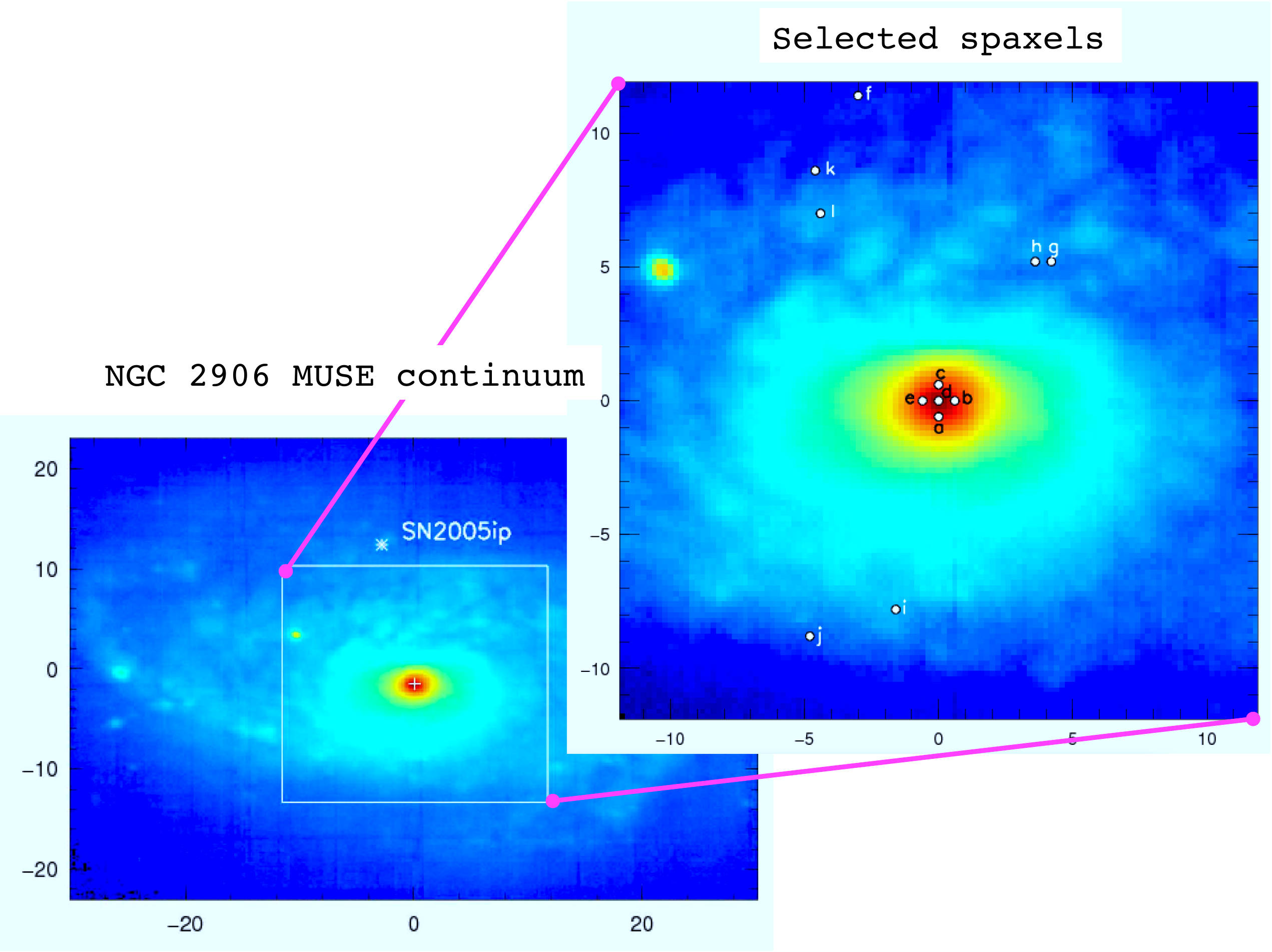}
\vskip-3mm    
   \includegraphics[width=0.83\textwidth]{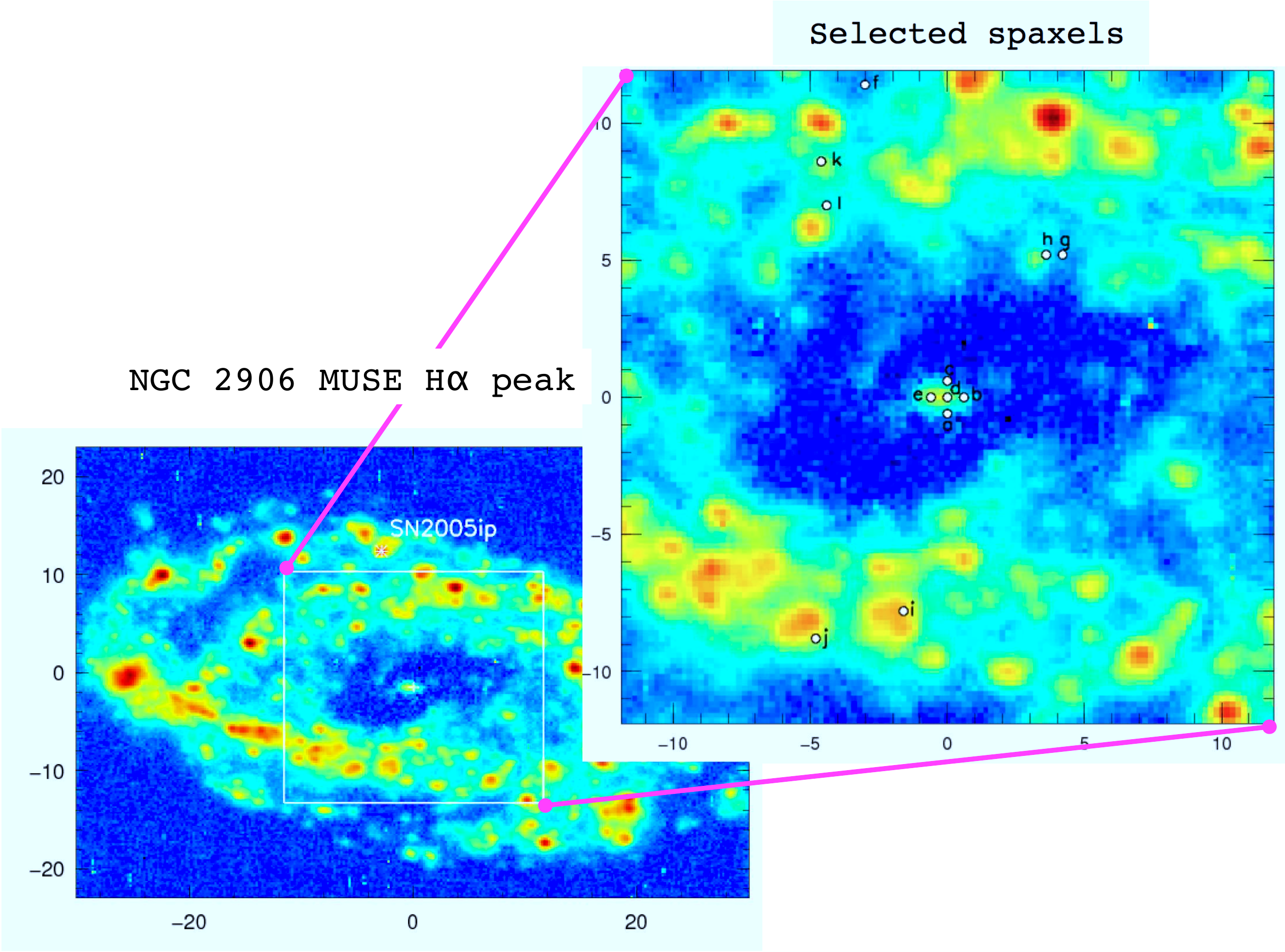}
\caption{\small {\it Top:} Continuum {\tt MUSE} image of NGC 2906. The continuum image has been obtained as the mean value of the median values in two wavelength ranges: 5620 -- 5750 \AA\ and 6890 -- 7000 \AA. {\it Left:} The nuclear peak is shown by a cross symbol and the position of the supernova SN2005ip is represented by the star symbol. The region selected by the white box is zoomed in on the right. {\it Right:} Region showing the distribution of the spectra (white dots) which have been chosen for the analysis. Each spectrum has been identified by a letter ({\tt a, b, ..., k, l}) in order to allow the reader to easily find its position. {\it Bottom:} H$\alpha$ peak emission of the same regions as described in the top figure.} 
 \label{figura_zoom}
\end{figure*}

NGC 2906\footnote{This galaxy was previously observed with the integral field spectrograph PMAS/PPak mounted on the 3.5m telescope at the Calar Alto Observatory. It is part of the Third Public Data Release (DR3) of the Calar Alto Legacy Integral Field Area (CALIFA) survey \citep{Sanchez16c}.} is a nearby (z = 0.007138, d = 30 Mpc) `normal' star--forming galaxy which was observed during the science verification of the {\tt MUSE} instrument \citep{Bacon04}, and included in the pilot study of the All--weather MUse Supernova Integral--field of Nearby Galaxies \citep[AMUSING;][]{Galbany16} given that it hosts the type IIn supernova (SN) 2005ip \citep{Boles05, Modjaz05}. 
According to its morphology, this object is classified as {\it Scd} galaxy. In Fig.~\ref{figura_zoom} the continuum and the H$\alpha$\footnote{The H$\alpha$ flux emission is actually the flux peak of the line having removed the continuum emission computed in a region close to the H$\alpha$--[NII] complex.} emissions of this object are shown.

The {\tt MUSE} instrument provides a FoV of 1 arcmin$^2$ with spaxels of 0.2$^{\prime\prime} \times 0.2^{\prime\prime}$ in Wide Field Mode (WFM). At this redshift, the arcsec to kpc conversion gives 0.147 kpc arcsec$^{-1}$. In this configuration the data cube is composed by 320 $\times$ 322 spaxels with 3680 elements of spectral sampling.
The wavelength coverage ranges between 4750--9350 \AA\, with a mean resolution of R$\sim$3000\footnote{See the {\tt MUSE} spectral resolution (R) as a function of the wavelength at:\\ \href{https://www.eso.org/sci/facilities/paranal/instruments/muse/inst.html}{\tt https://www.eso.org/sci/facilities/paranal/instruments/\\muse/inst.html}. At the H$\alpha$ wavelength, it corresponds to a spectral resolution R $\sim$ 2500, deriving a FWHM of 2.6~\AA.} and sampling of 1.25 \AA. This object has been observed with a seeing (FWHM) of 0.88$^{\prime\prime}$ ($\sim$0.13 kpc). With the {\tt MUSE} instrument it is possible to combine good seeing conditions with large FoV and a spatial sampling sufficient to properly sample the point spread function (PSF).
Thus, it is the suitable instrument to study in detail the kinematics of our object.

% -----------------------------------------------------------------------------
\section{Analysis}
% -----------------------------------------------------------------------------

\begin{figure*}
   \centering
   \vskip-3mm
   \includegraphics[width=0.175\textwidth, height=0.115\textwidth]{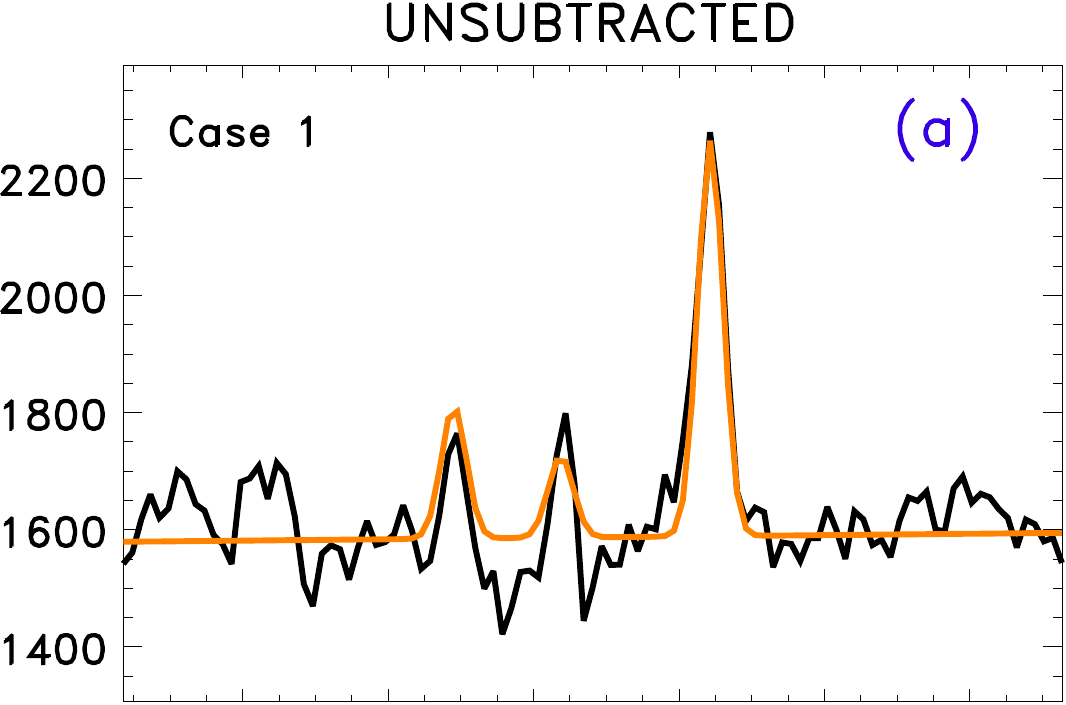}
    \includegraphics[width=0.175\textwidth, height=0.115\textwidth]{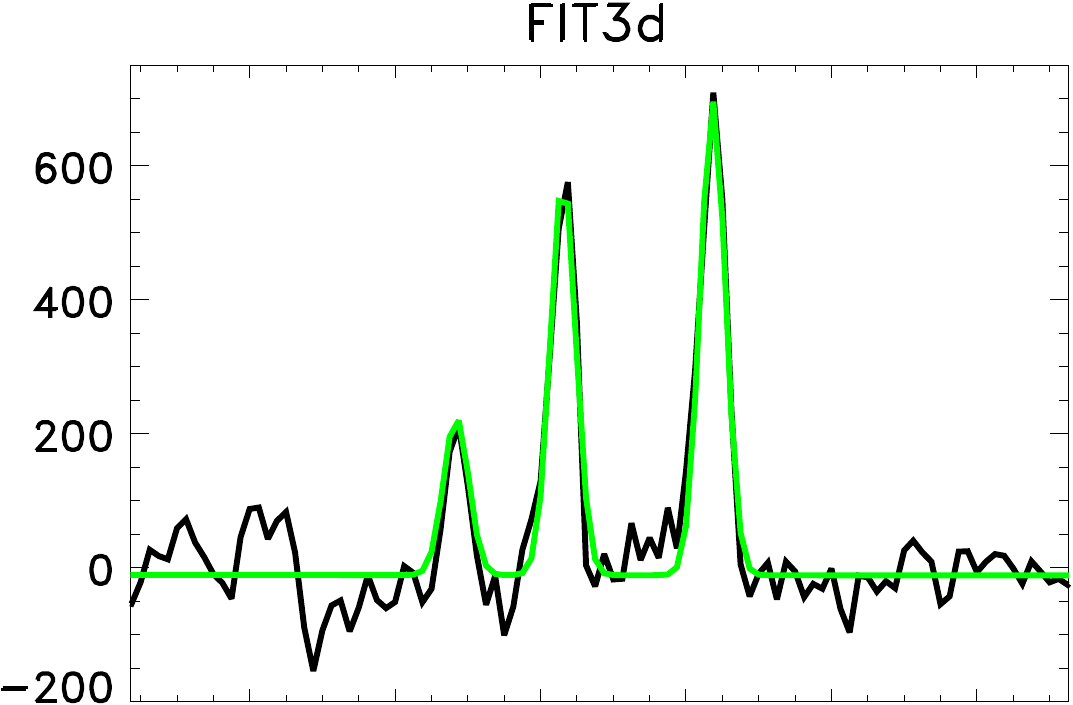}
  \includegraphics[width=0.175\textwidth, height=0.115\textwidth]{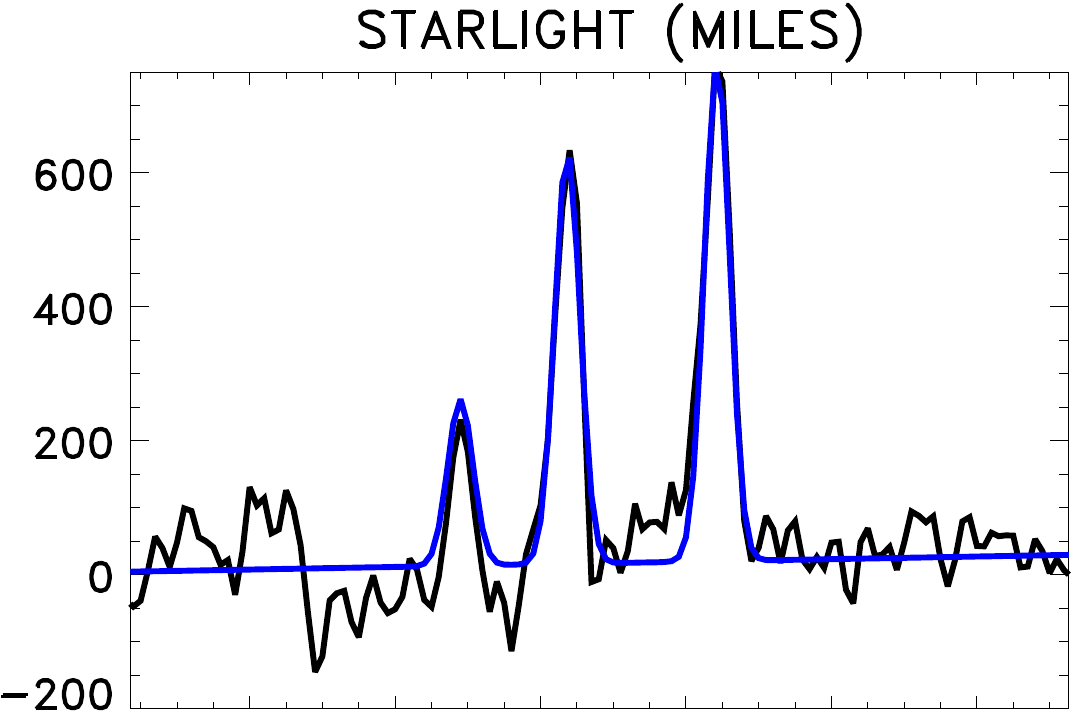}
   \includegraphics[width=0.175\textwidth, height=0.115\textwidth]{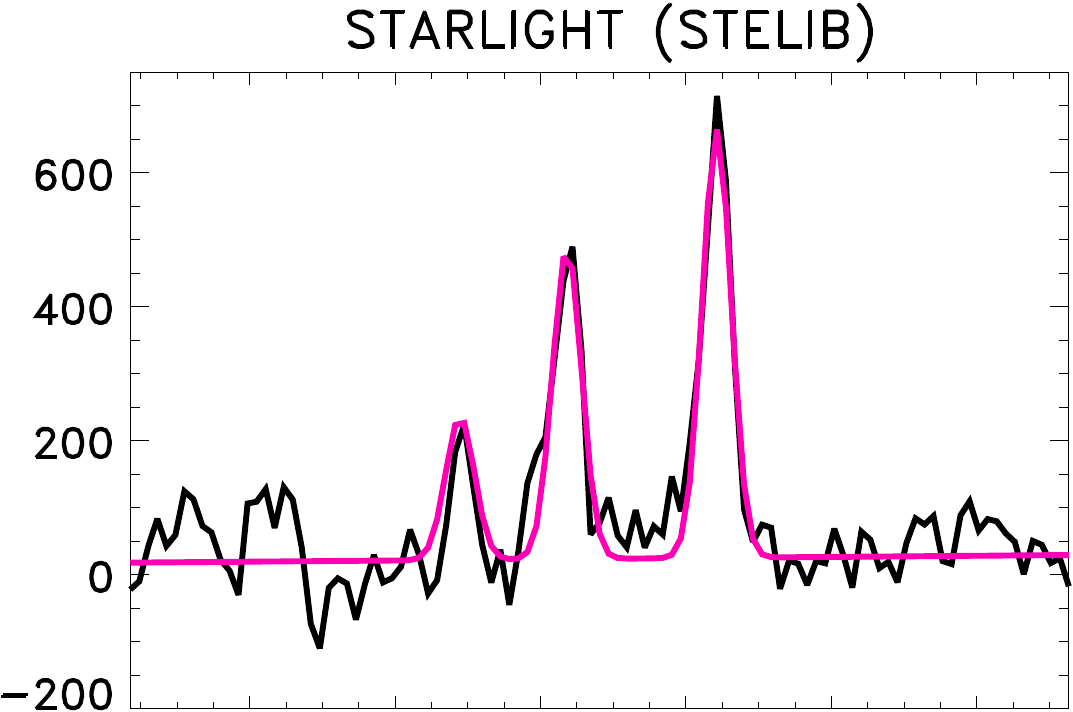}
   \includegraphics[width=0.175\textwidth, height=0.115\textwidth]{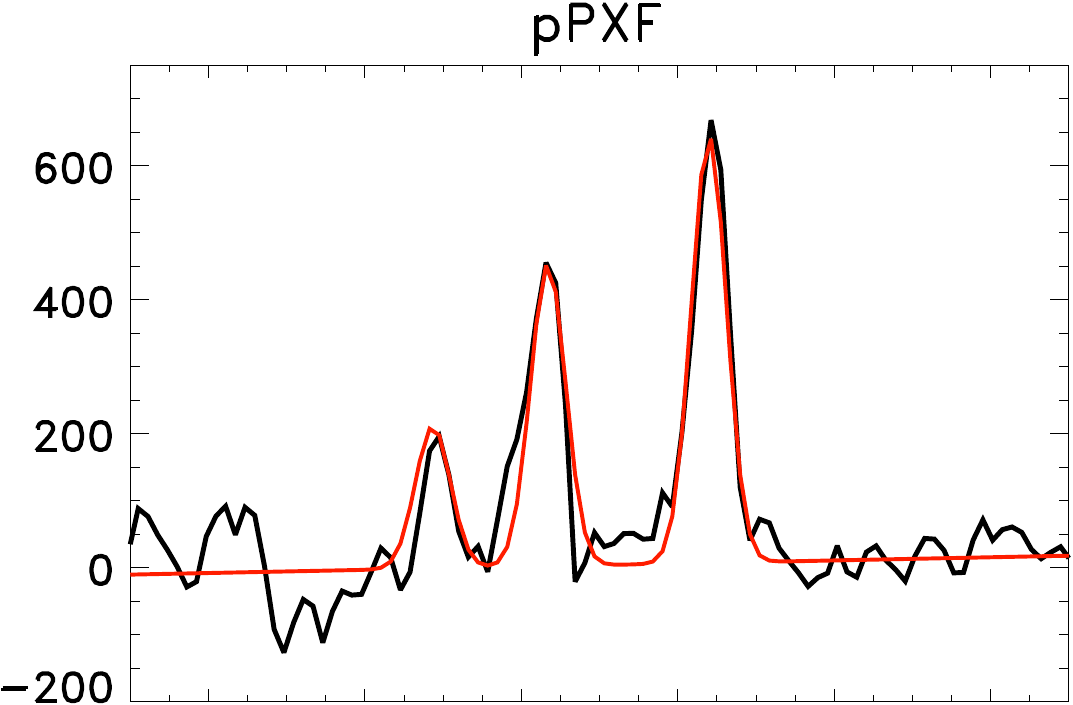}
 \vskip-0.3mm  
  \includegraphics[width=0.175\textwidth, height=0.115\textwidth]{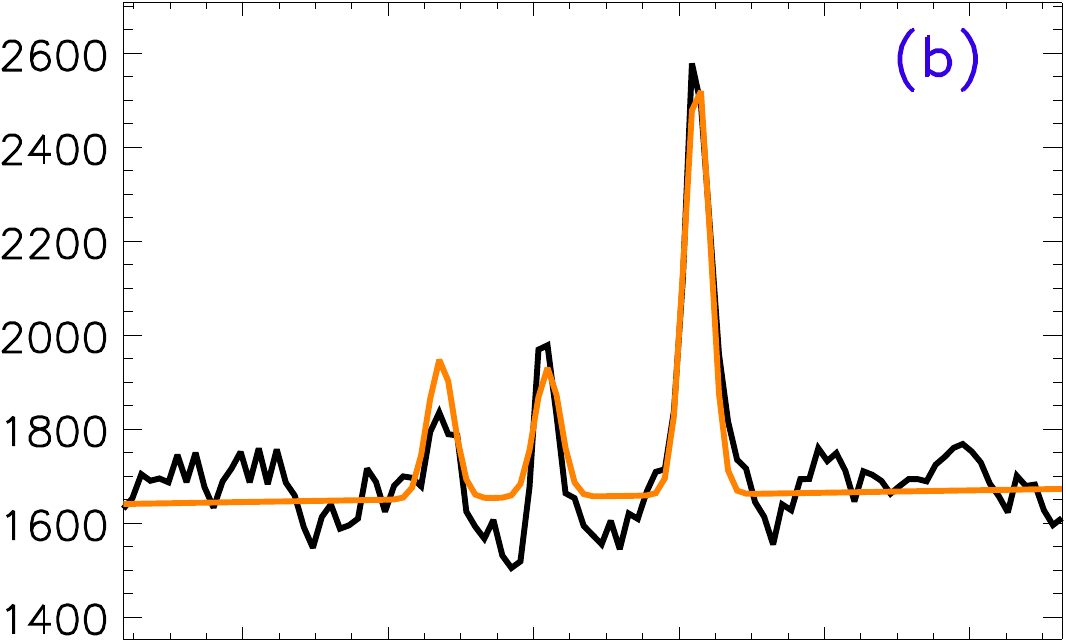}
   \includegraphics[width=0.175\textwidth, height=0.115\textwidth]{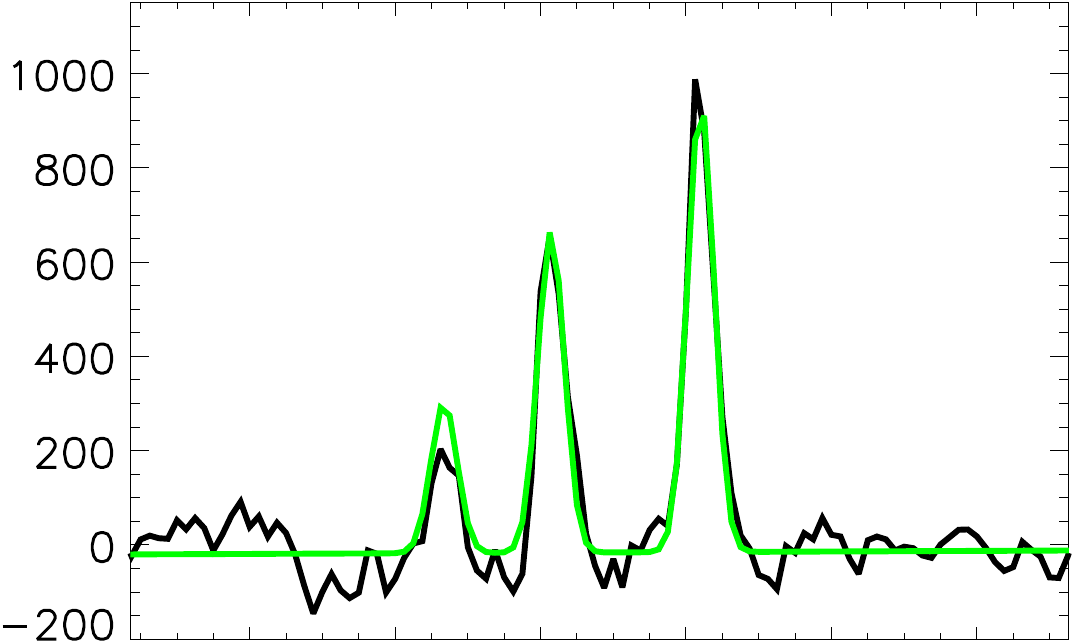}
   \includegraphics[width=0.175\textwidth, height=0.115\textwidth]{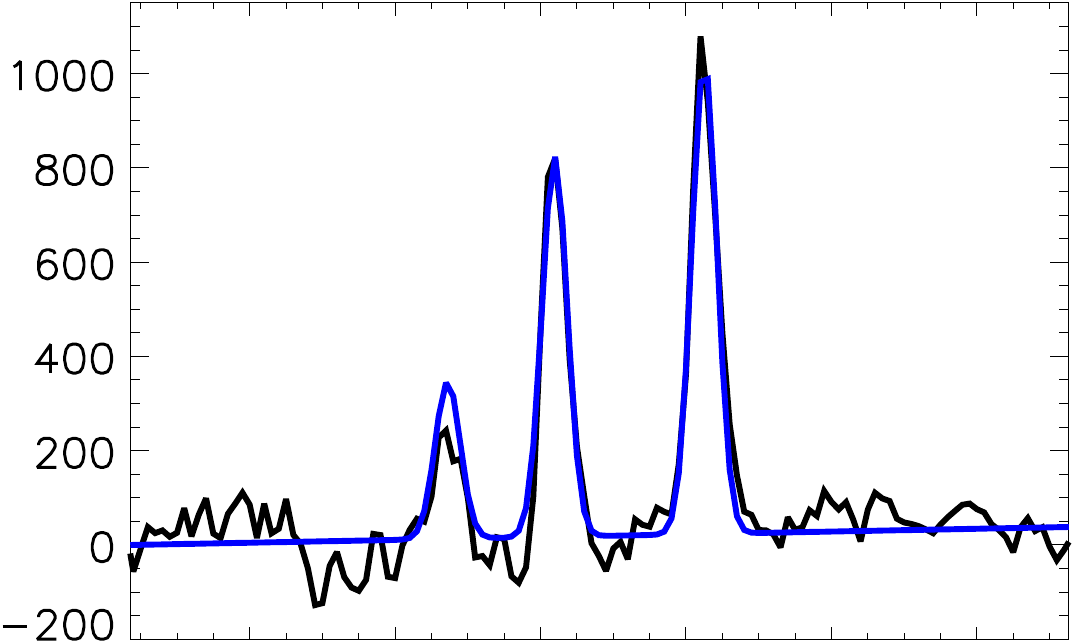}  
   \includegraphics[width=0.175\textwidth, height=0.115\textwidth]{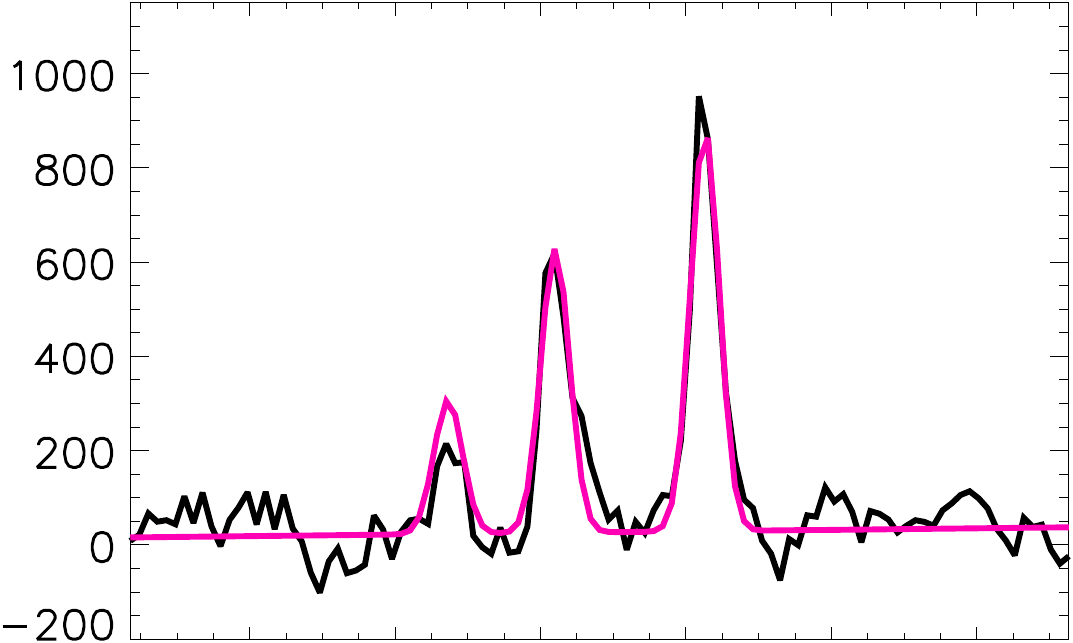}
  \includegraphics[width=0.1755\textwidth, height=0.115\textwidth]{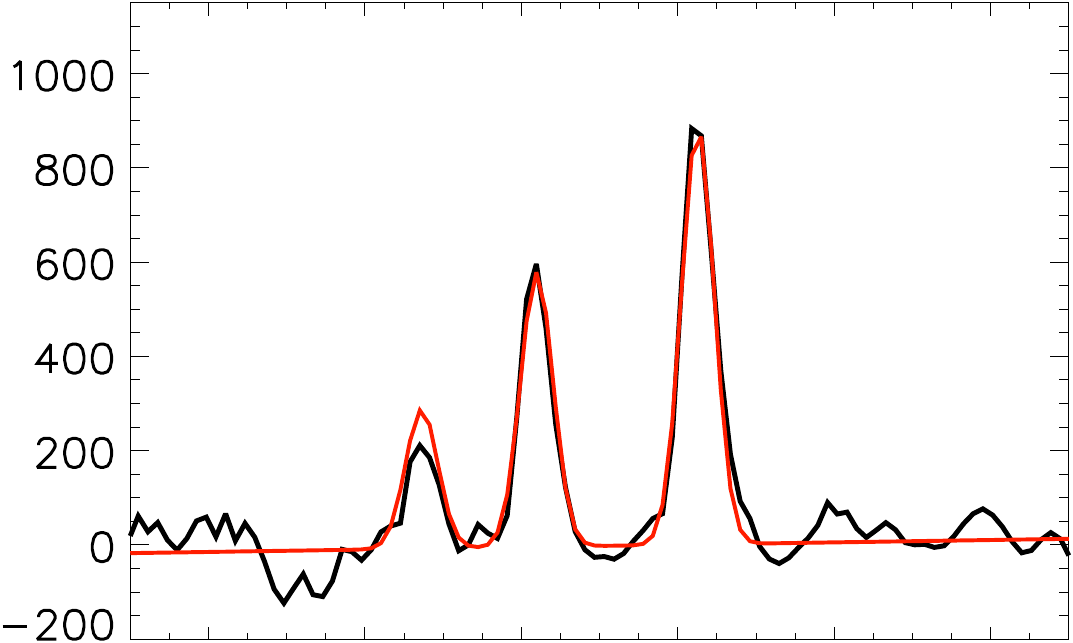}
 \vskip-0.3mm  
  \includegraphics[width=0.175\textwidth, height=0.115\textwidth]{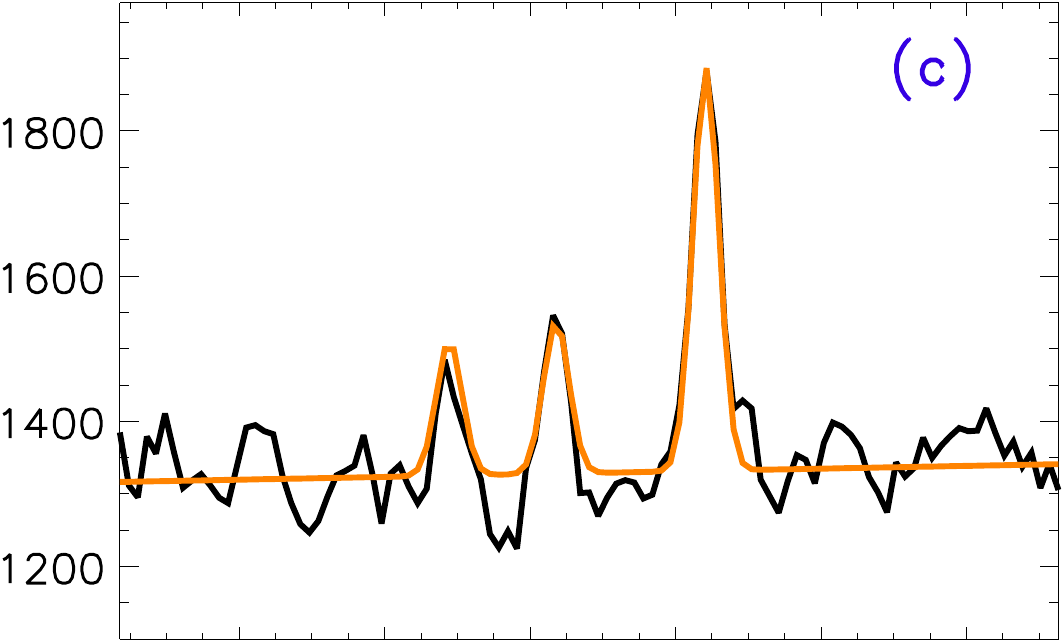}
   \includegraphics[width=0.175\textwidth, height=0.115\textwidth]{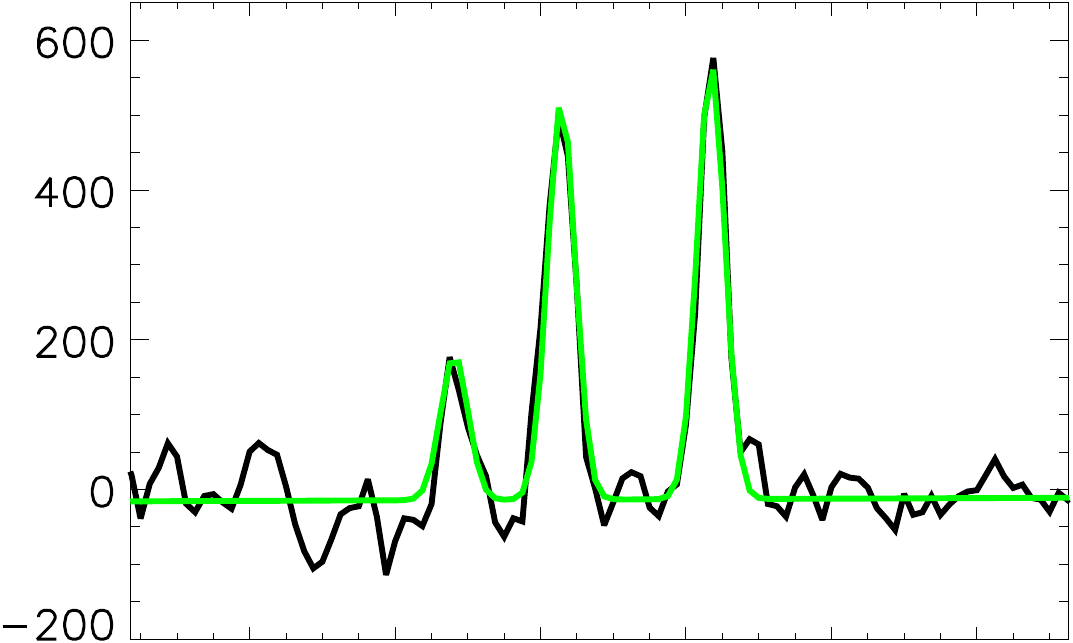}  
   \includegraphics[width=0.175\textwidth, height=0.115\textwidth]{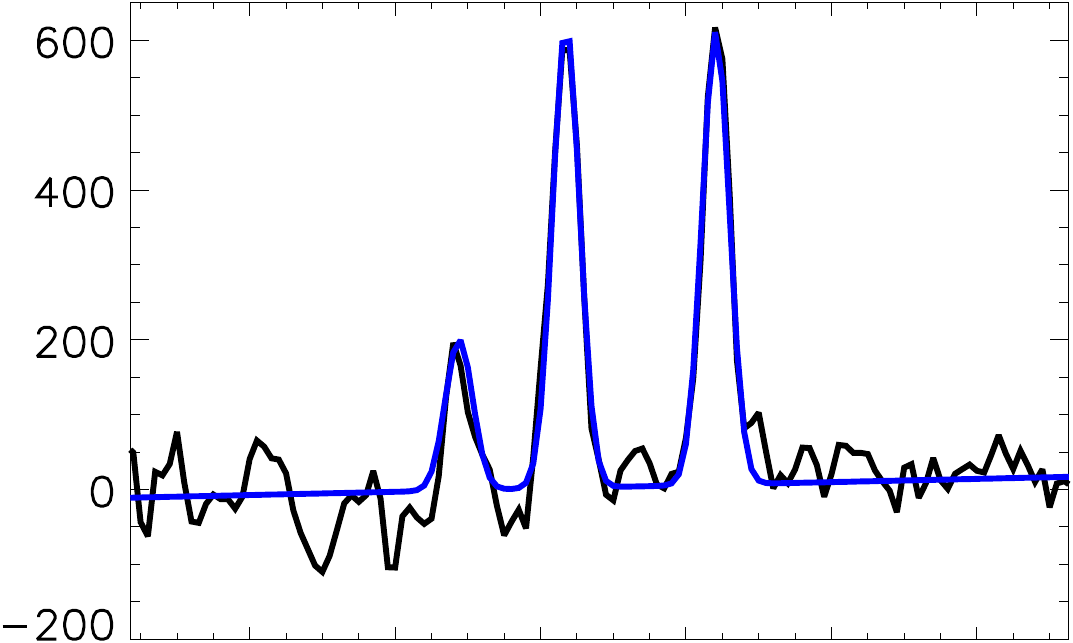}  
   \includegraphics[width=0.175\textwidth, height=0.115\textwidth]{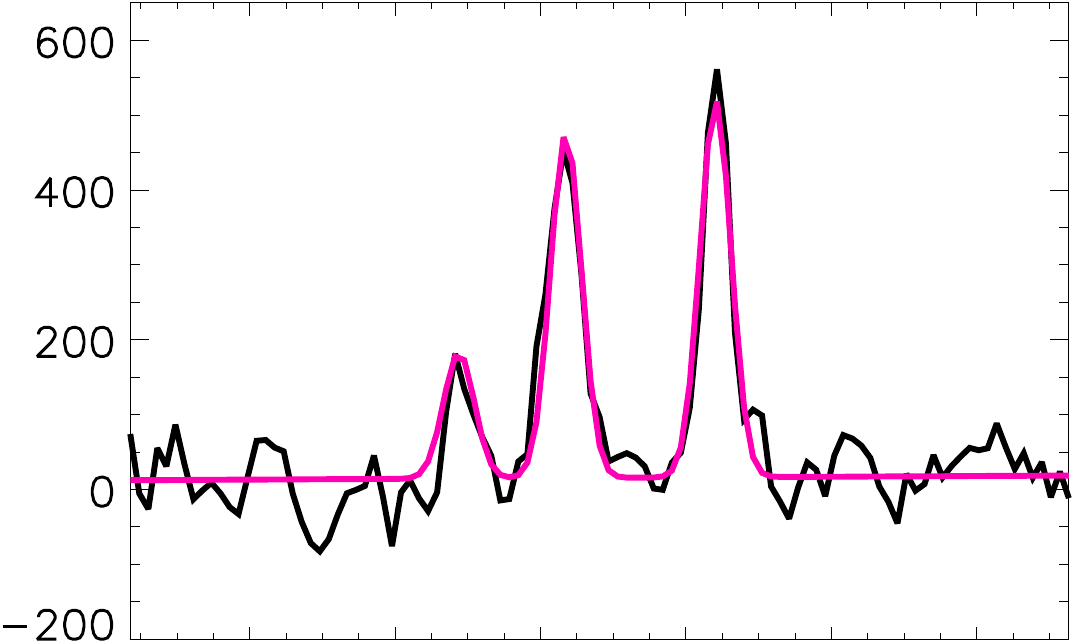}  
   \includegraphics[width=0.175\textwidth, height=0.115\textwidth]{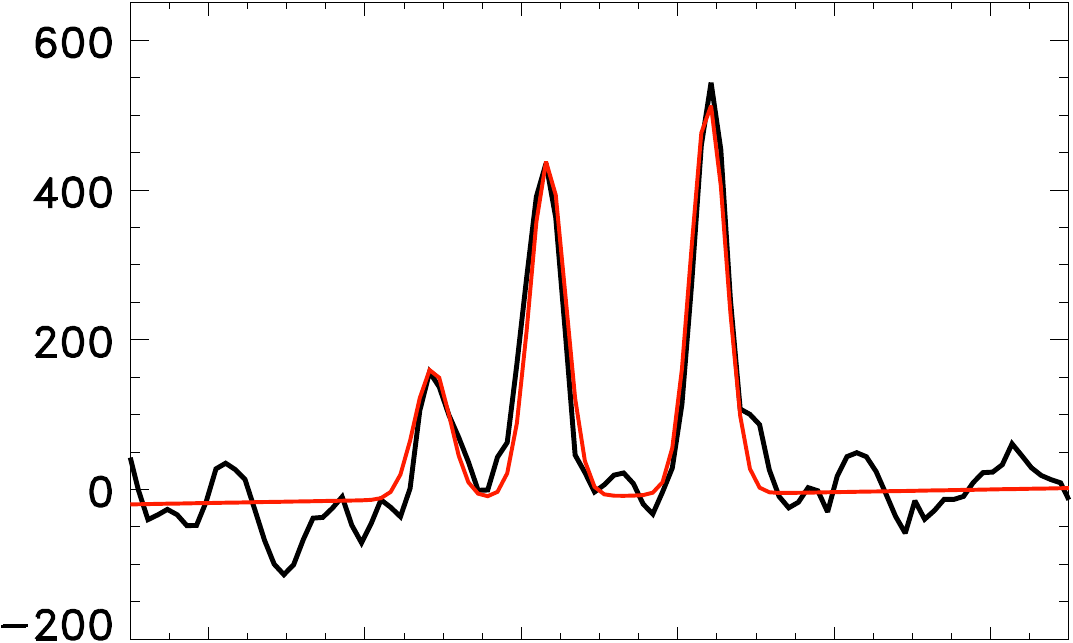}   
 \vskip-0.3mm  
  \includegraphics[width=0.175\textwidth, height=0.115\textwidth]{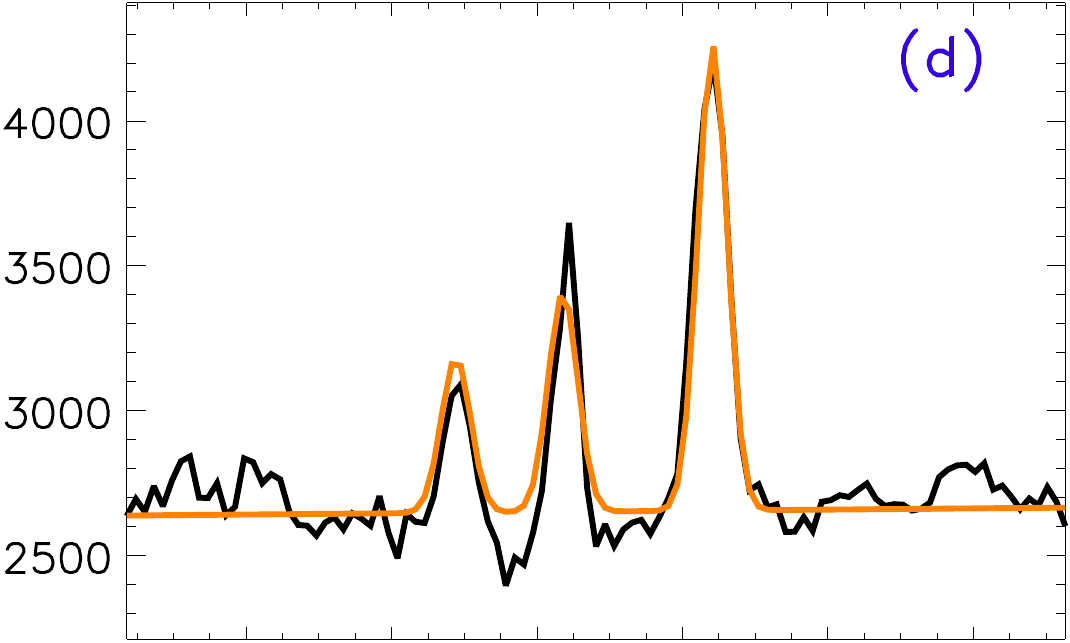}
   \includegraphics[width=0.175\textwidth, height=0.115\textwidth]{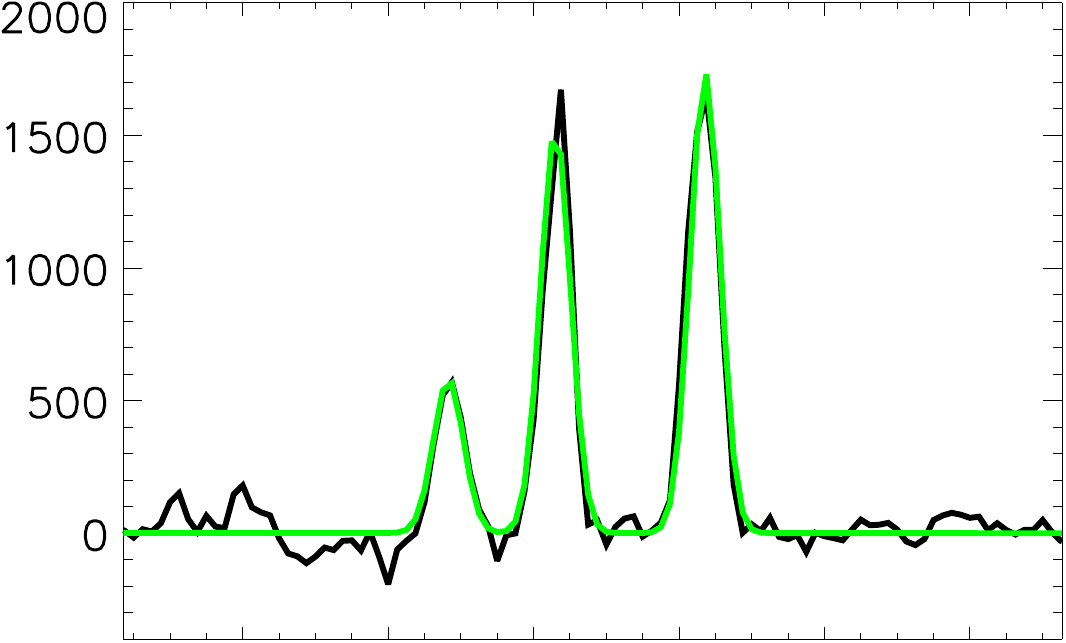}
   \includegraphics[width=0.175\textwidth, height=0.115\textwidth]{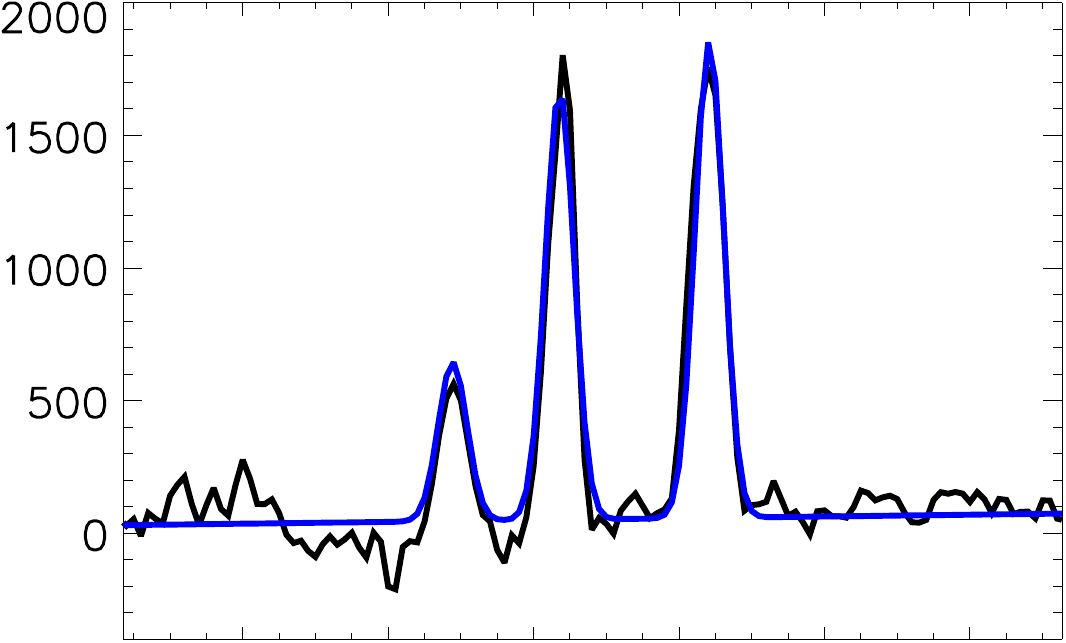}
   \includegraphics[width=0.175\textwidth, height=0.115\textwidth]{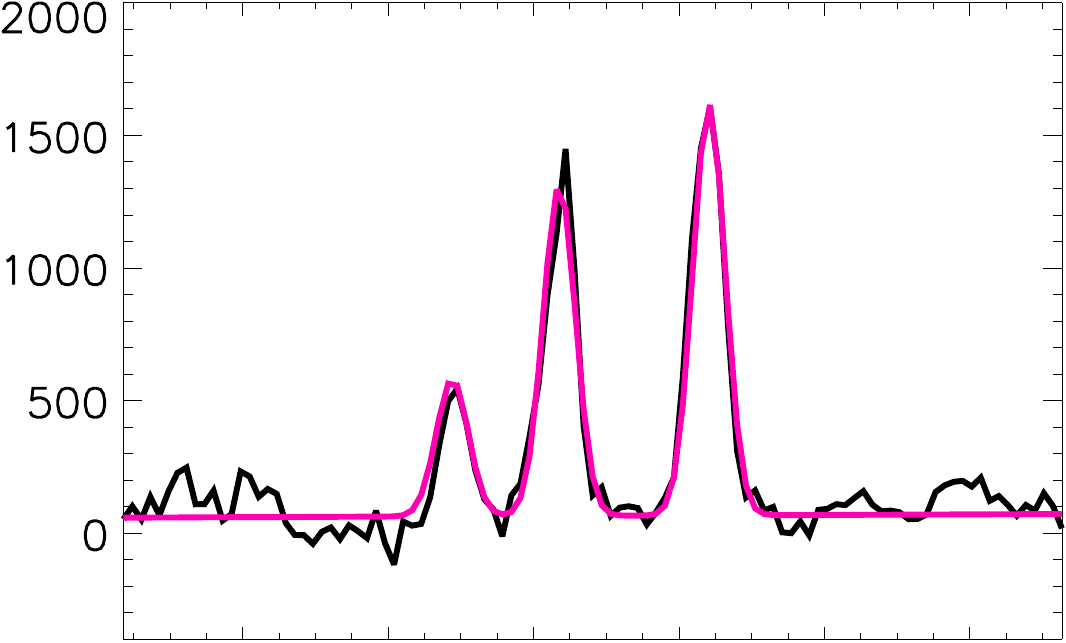}
    \includegraphics[width=0.175\textwidth, height=0.115\textwidth]{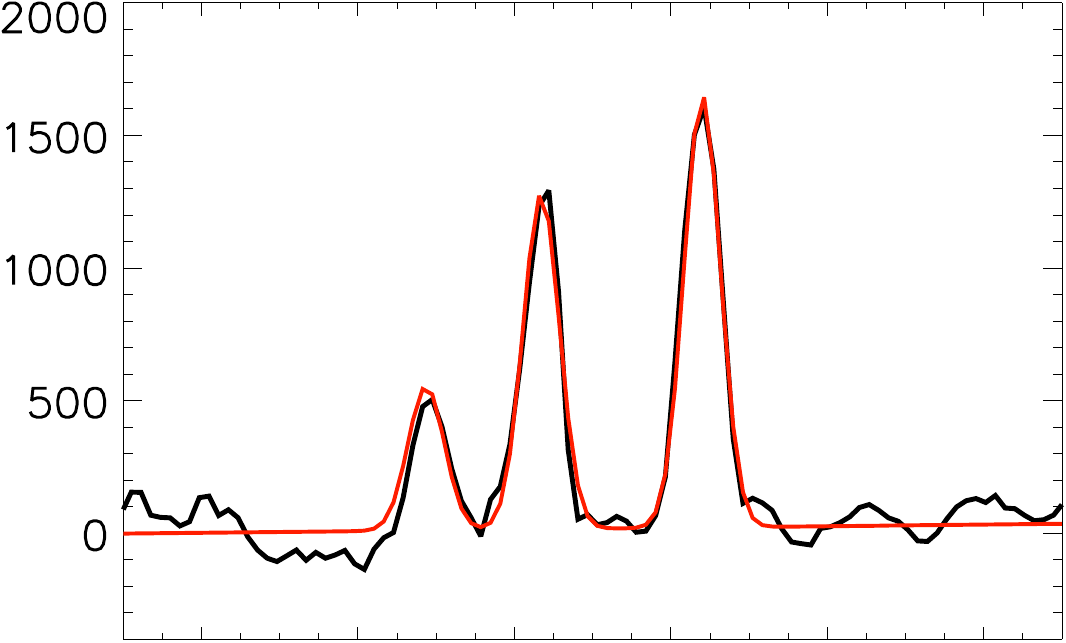}    
 \vskip-0.3mm  
 \includegraphics[width=0.175\textwidth, height=0.115\textwidth]{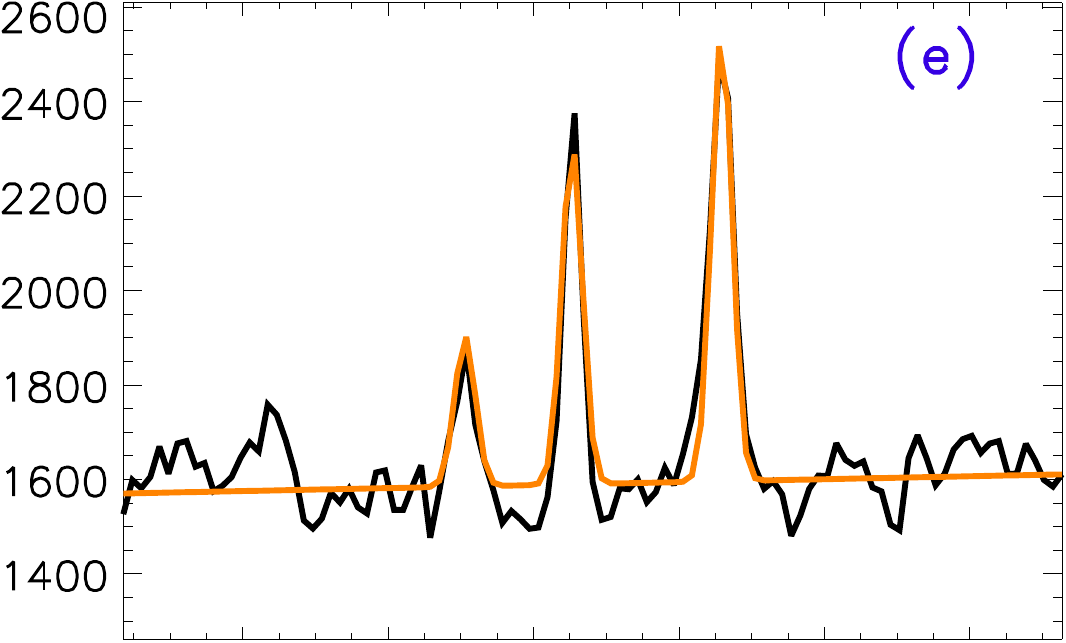}
  \includegraphics[width=0.175\textwidth, height=0.115\textwidth]{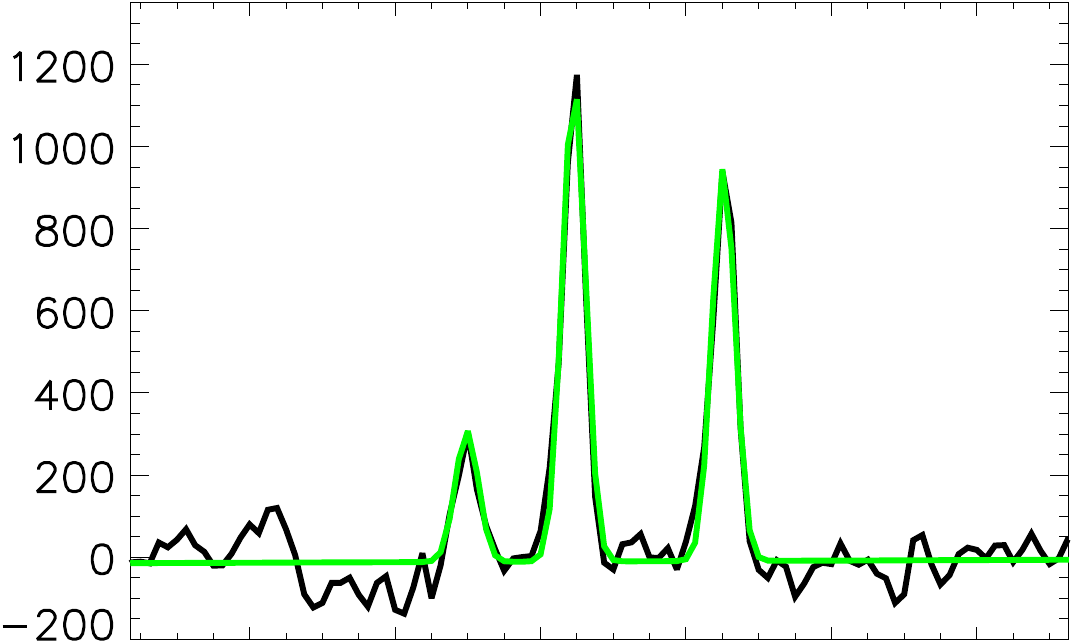}
   \includegraphics[width=0.175\textwidth, height=0.115\textwidth]{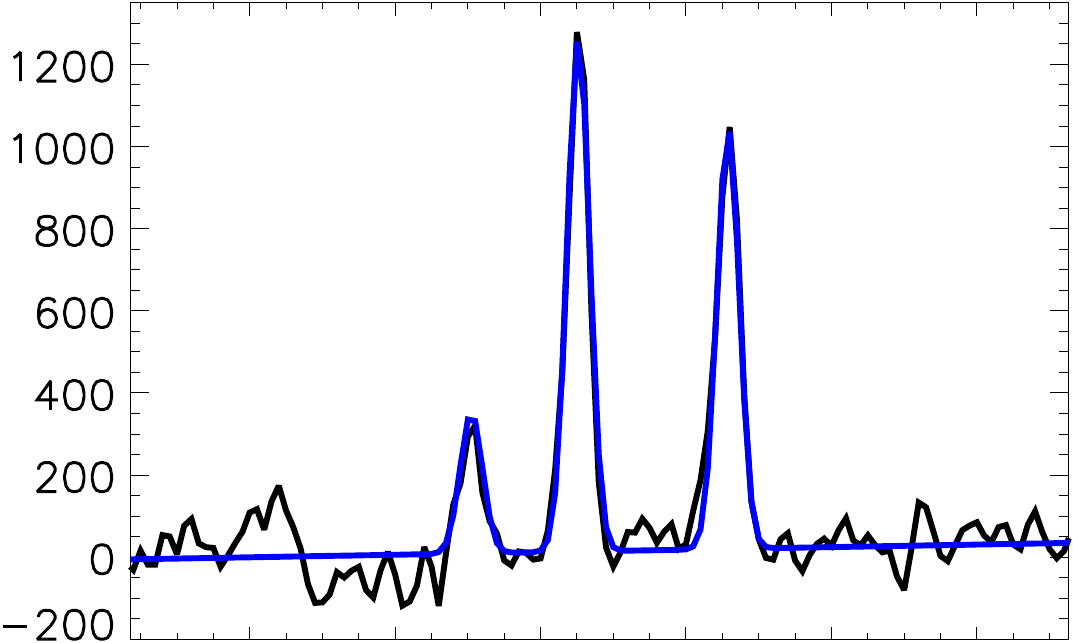}
   \includegraphics[width=0.175\textwidth, height=0.115\textwidth]{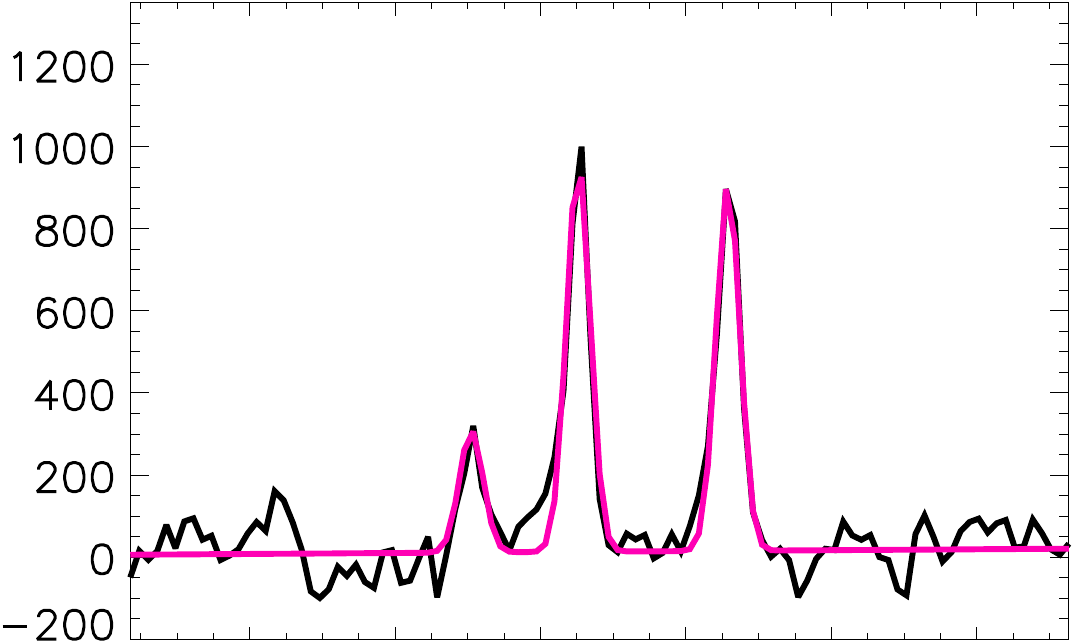}  
   \includegraphics[width=0.175\textwidth, height=0.115\textwidth]{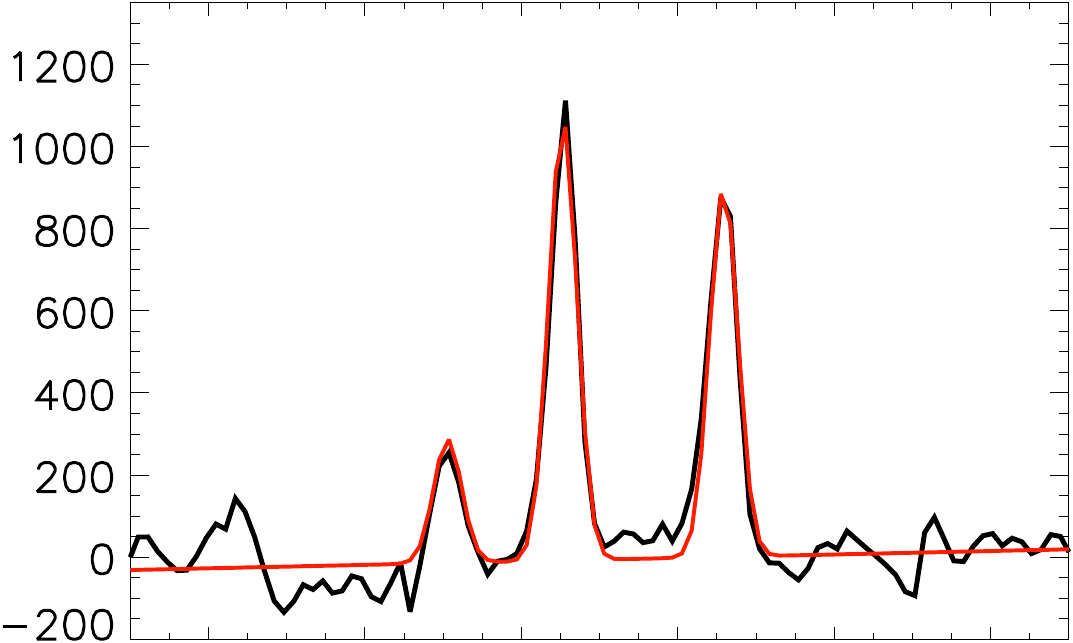}   
 \vskip-0.3mm  
 \includegraphics[width=0.175\textwidth, height=0.115\textwidth]{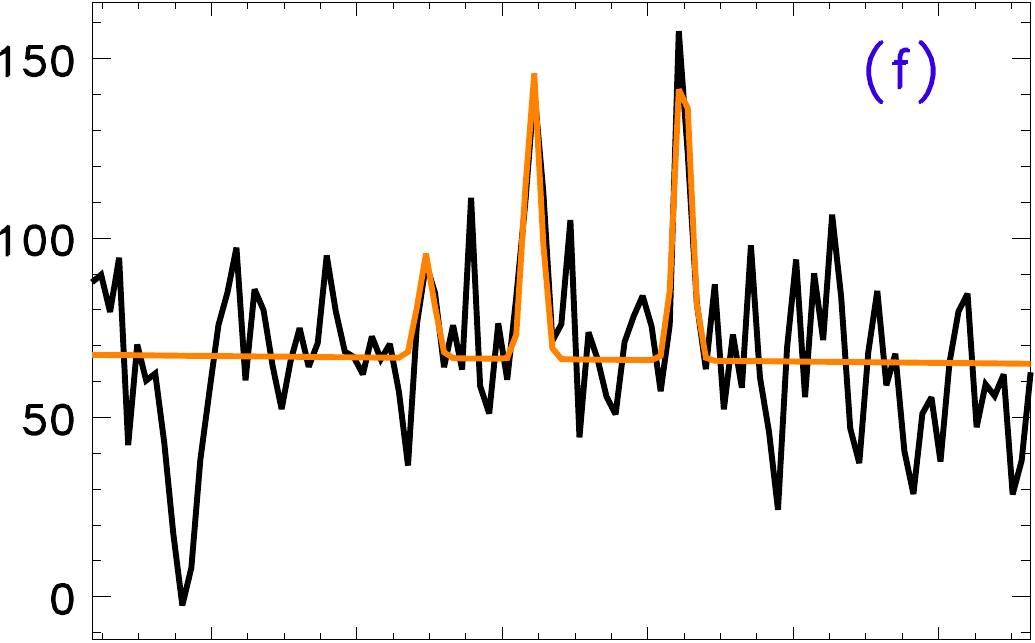}   
   \includegraphics[width=0.175\textwidth, height=0.115\textwidth]{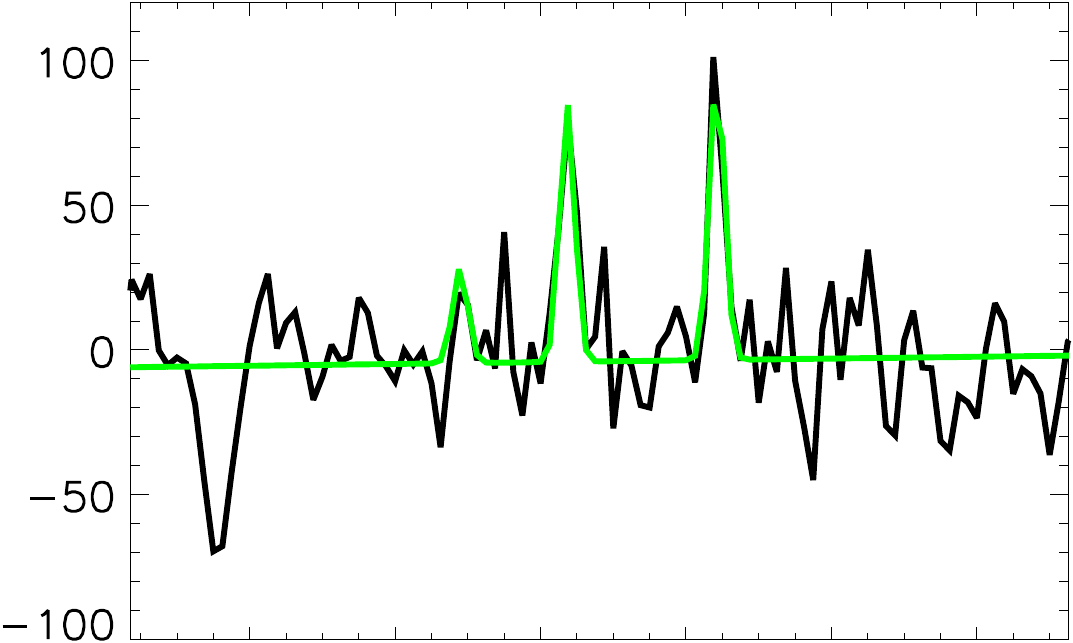}   
   \includegraphics[width=0.175\textwidth, height=0.115\textwidth]{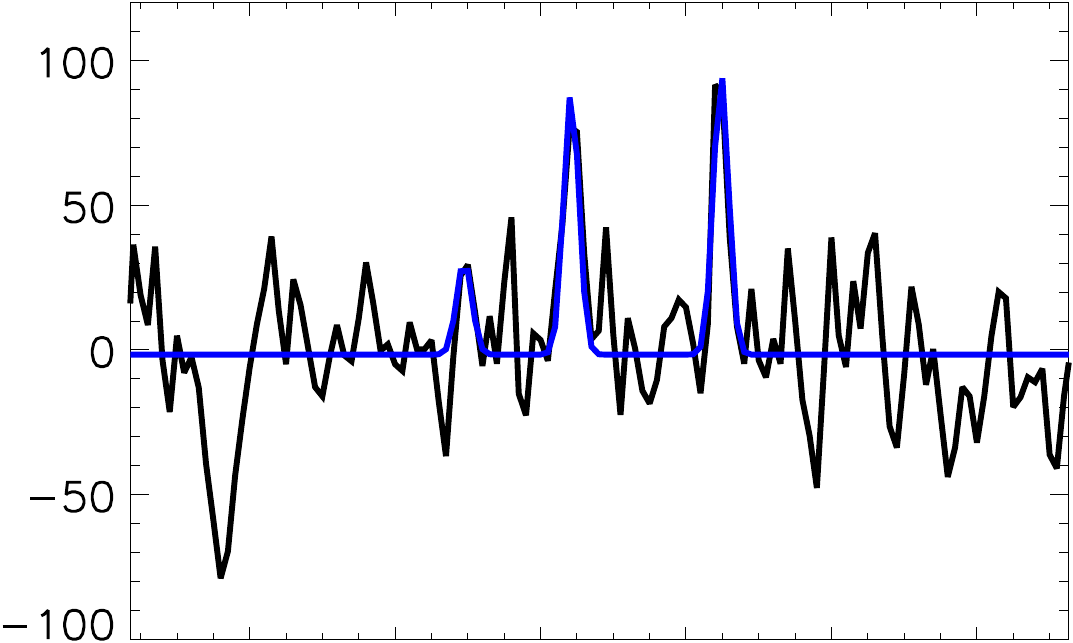}   
   \includegraphics[width=0.175\textwidth, height=0.115\textwidth]{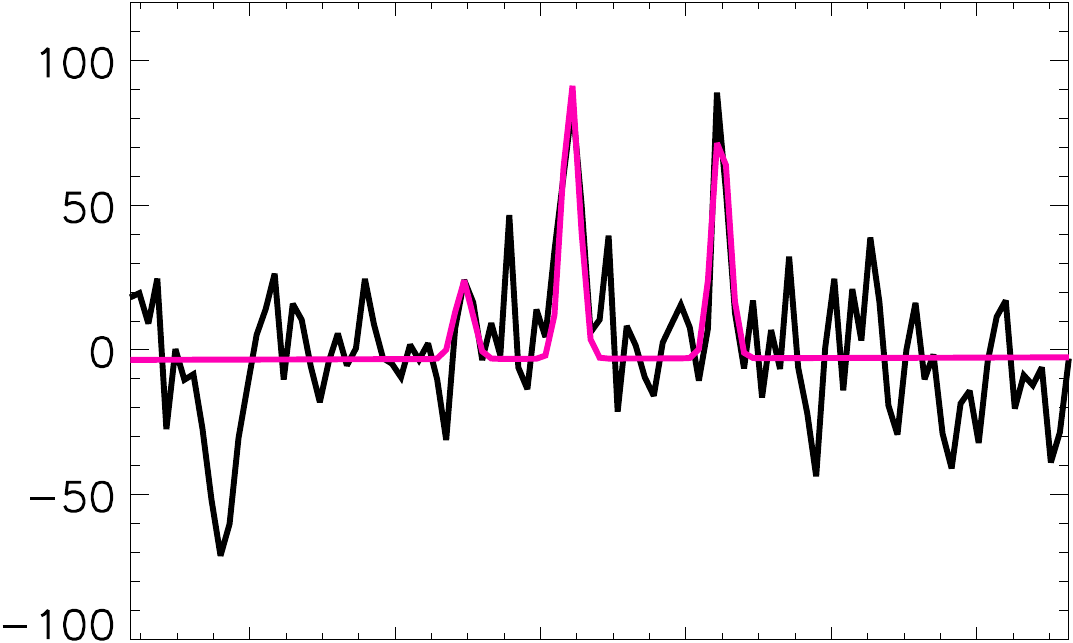}   
   \includegraphics[width=0.175\textwidth, height=0.115\textwidth]{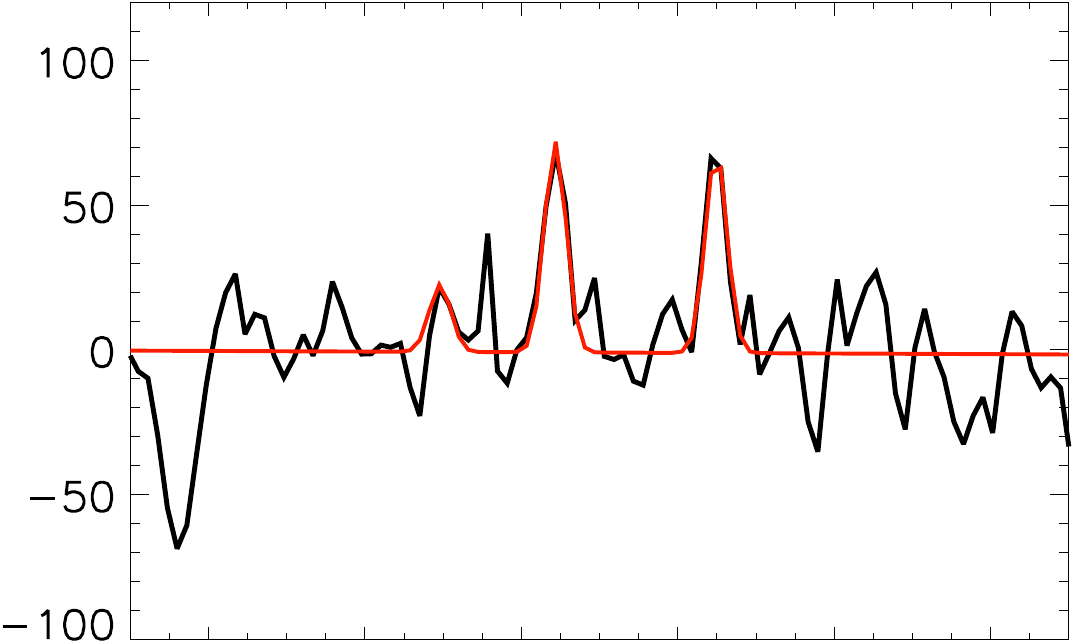}      
   \begin{minipage}[h]{17.5cm}
\vskip1.5mm\caption{ {\it From left to right:} {\tt Unsubtracted} {\tt MUSE} spectra at the different positions (identified by the letter) along with the pure gas spectra obtained from applying the three different stellar subtraction methods, respectively: {\tt FIT3D}, {\tt STARLIGHT (MILES and STELIB)} and {\tt pPXF} methods. The H$\alpha\lambda$6563 and the two forbidden [NII]$\lambda\lambda$6548, 6583 lines are shown. The line fitting results are shown using the colored solid line: the {\tt unsubtracted} data are in orange, {\tt FIT3D} data in green, {\tt STARLIGHT (MILES)} data in dark blue, {\tt STARLIGHT (STELIB)} ones in magenta and {\tt pPXF} ones in red. 
The line fitting {\tt case[1]} (all parameters {\tt fixed}) is considered. The flux is in units of 10$^{-20}$ erg~s$^{-1}$~cm$^{-2}$~\AA$^{-1}$ and the wavelength is in \AA. }
\label{fig_spectra_all_fixed}
\end{minipage}
%\vspace{-1pt}
\end{figure*}

\begin{figure*}
   \centering
 \vskip-0.3mm  
 \includegraphics[width=0.175\textwidth, height=0.115\textwidth]{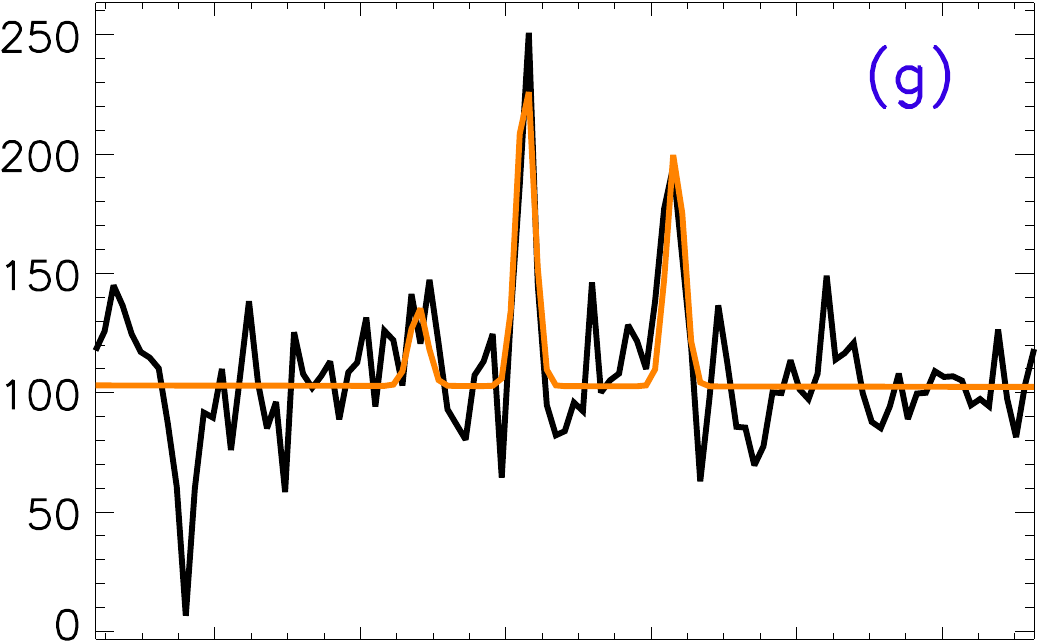}   
   \includegraphics[width=0.175\textwidth, height=0.115\textwidth]{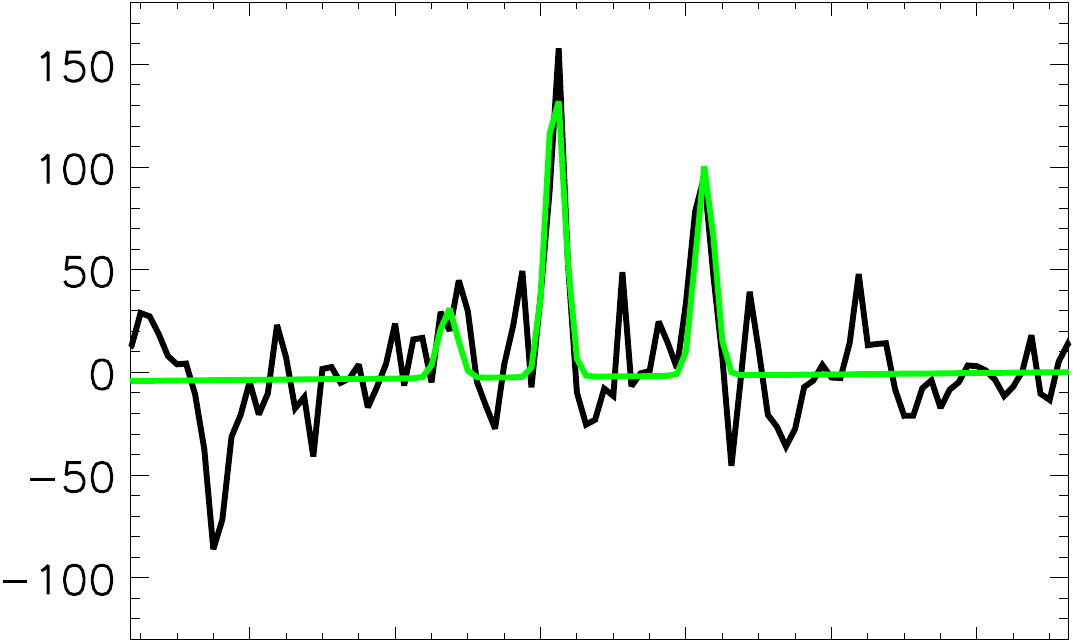}
   \includegraphics[width=0.175\textwidth, height=0.115\textwidth]{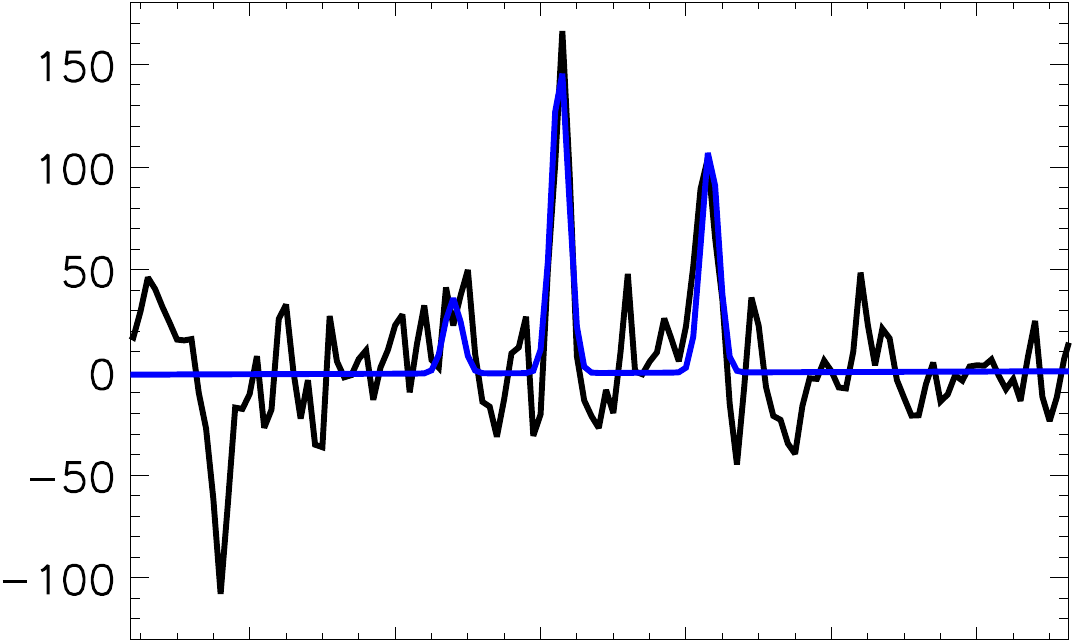}
   \includegraphics[width=0.175\textwidth, height=0.115\textwidth]{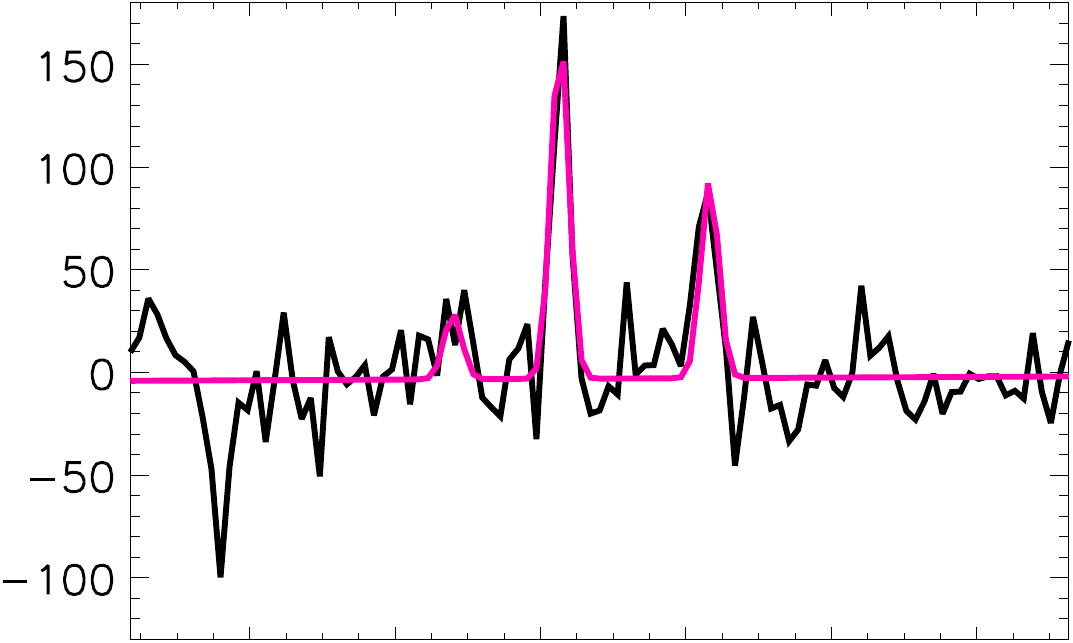}
   \includegraphics[width=0.175\textwidth, height=0.115\textwidth]{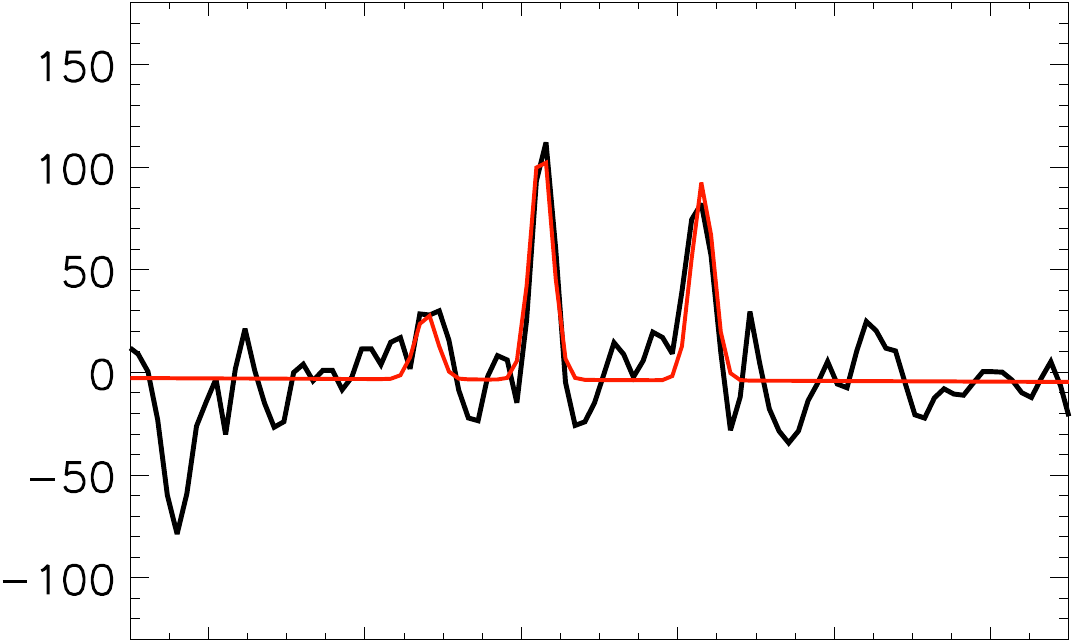}   
 \vskip-0.3mm  
 \includegraphics[width=0.175\textwidth, height=0.115\textwidth]{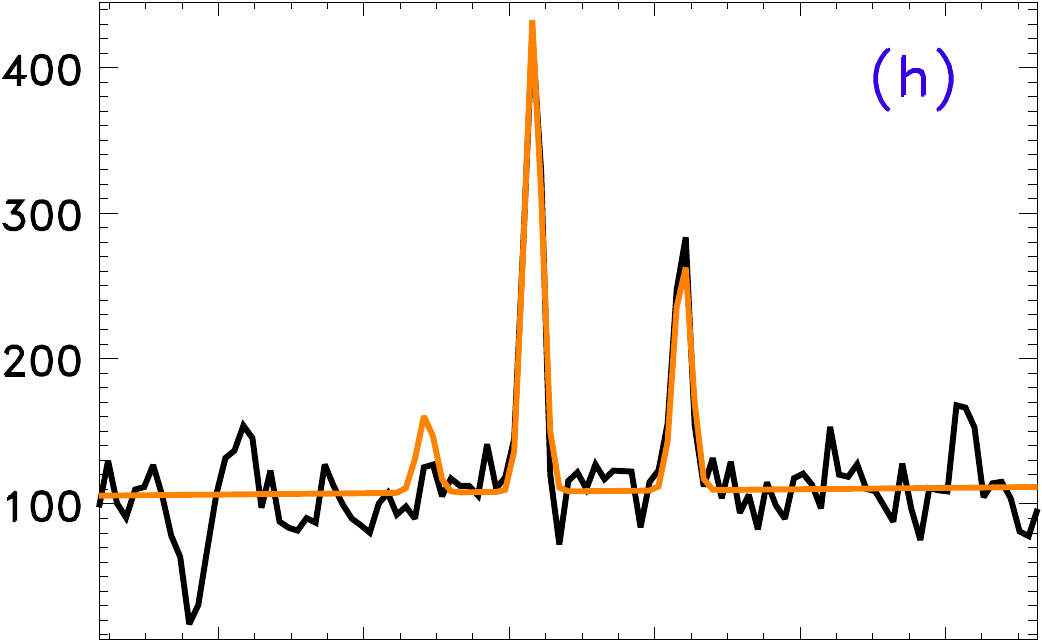}   
   \includegraphics[width=0.175\textwidth, height=0.115\textwidth]{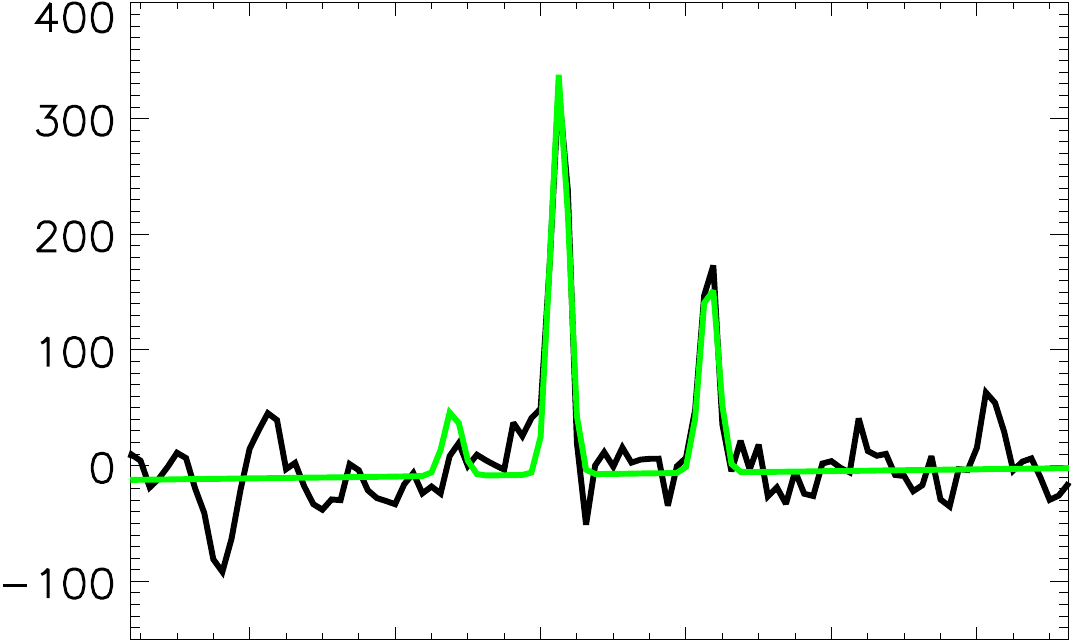}
   \includegraphics[width=0.175\textwidth, height=0.115\textwidth]{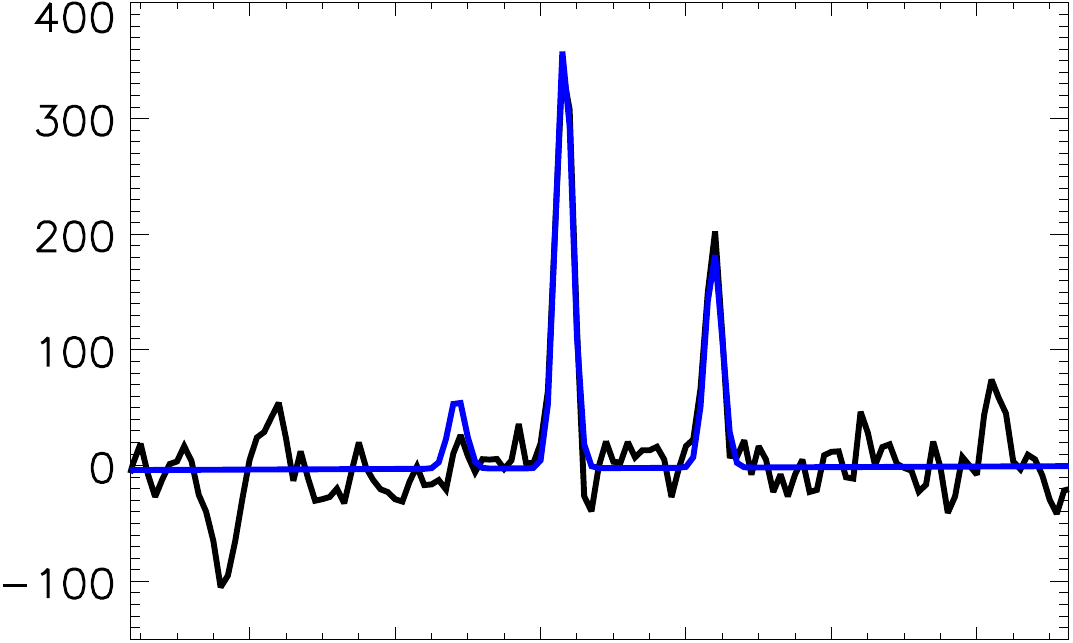}
   \includegraphics[width=0.175\textwidth, height=0.115\textwidth]{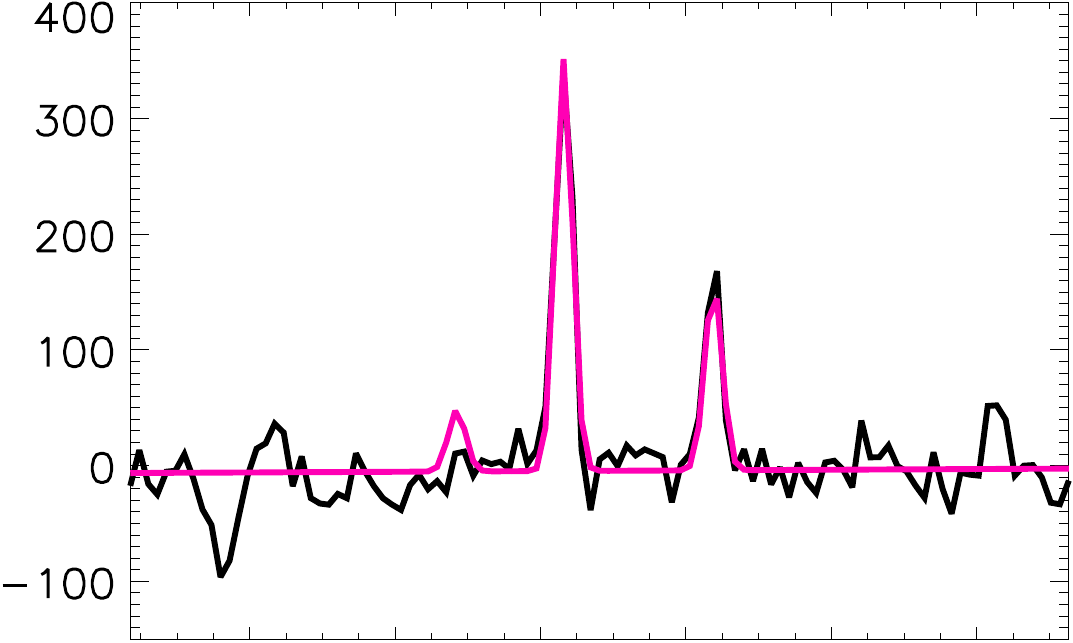}
   \includegraphics[width=0.175\textwidth, height=0.115\textwidth]{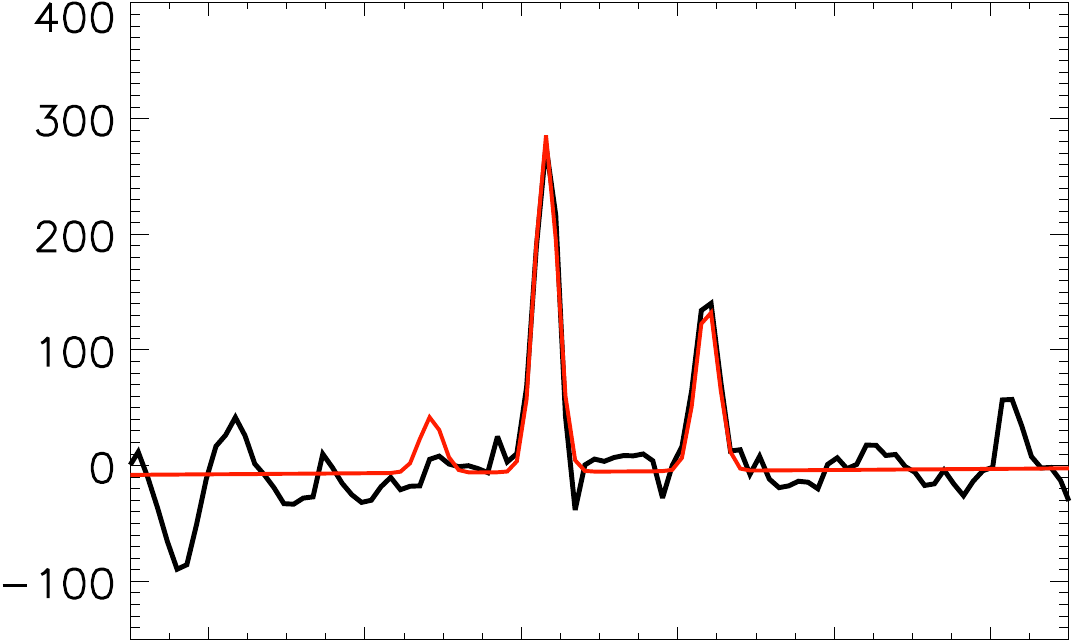}   
 \vskip-0.3mm  
 \includegraphics[width=0.175\textwidth, height=0.115\textwidth]{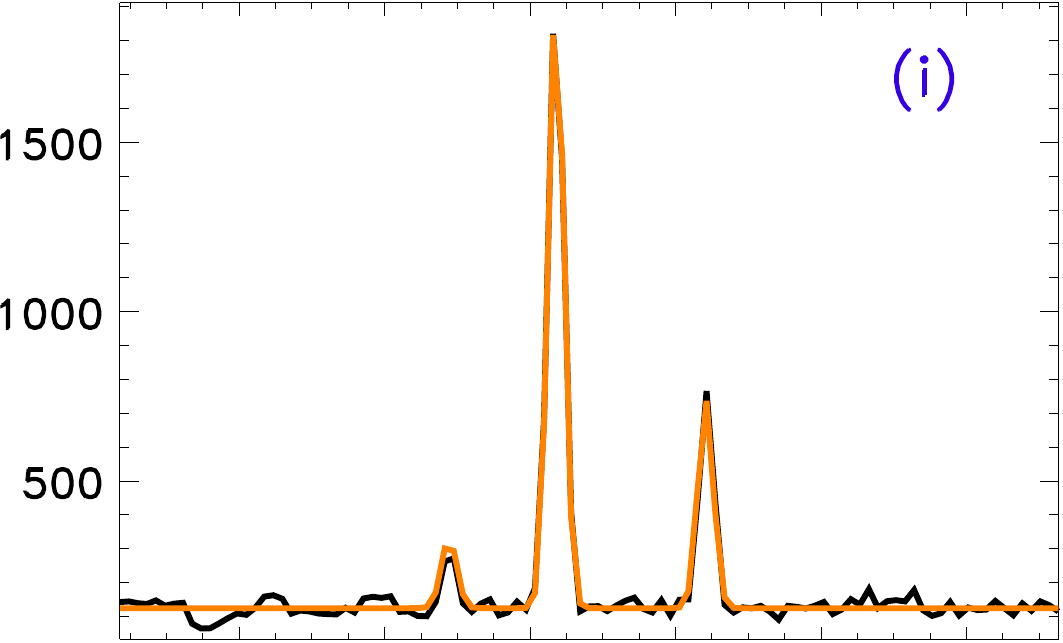}   
   \includegraphics[width=0.175\textwidth, height=0.115\textwidth]{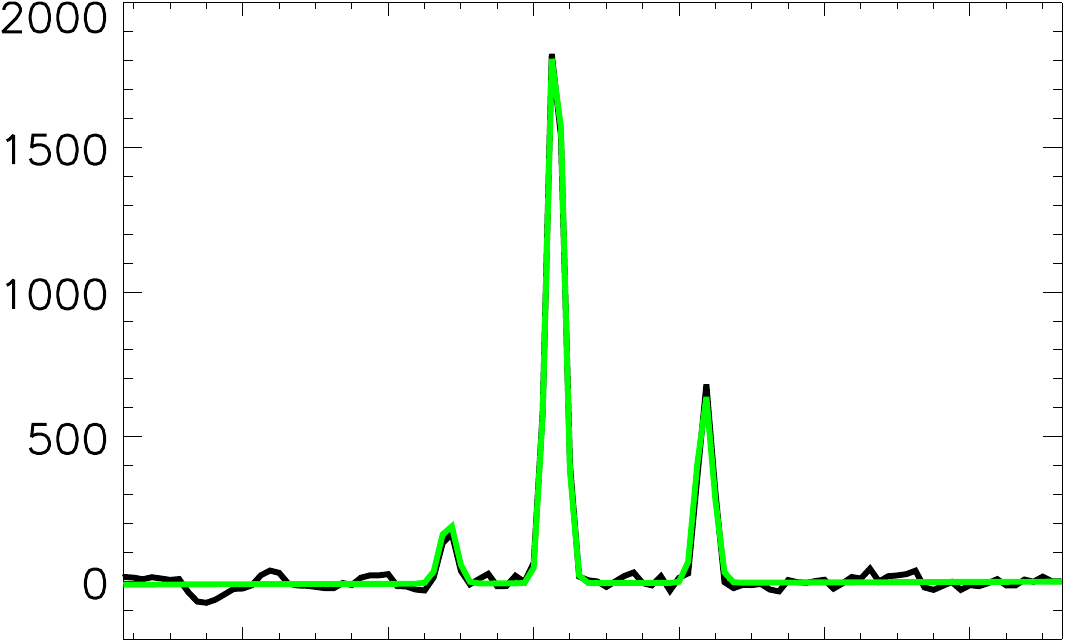}
   \includegraphics[width=0.175\textwidth, height=0.115\textwidth]{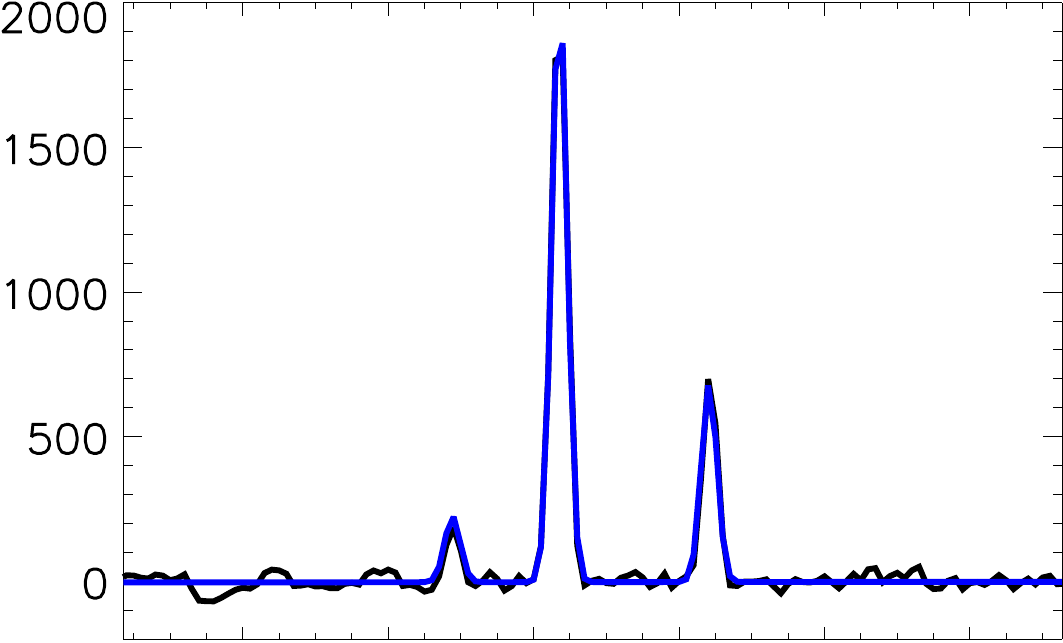}
   \includegraphics[width=0.175\textwidth, height=0.115\textwidth]{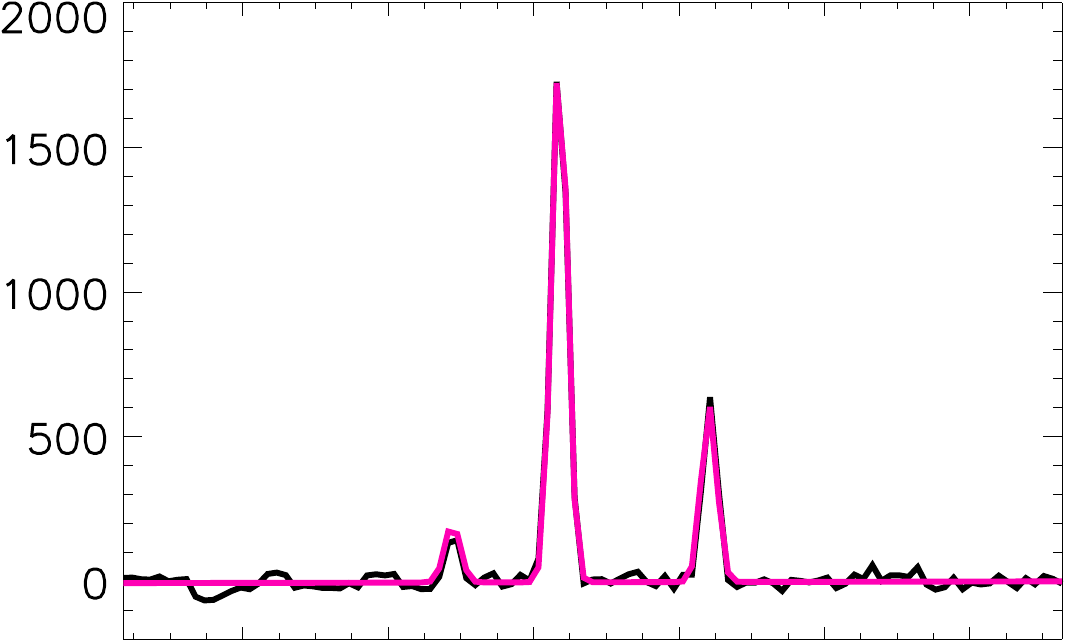}
   \includegraphics[width=0.175\textwidth, height=0.115\textwidth]{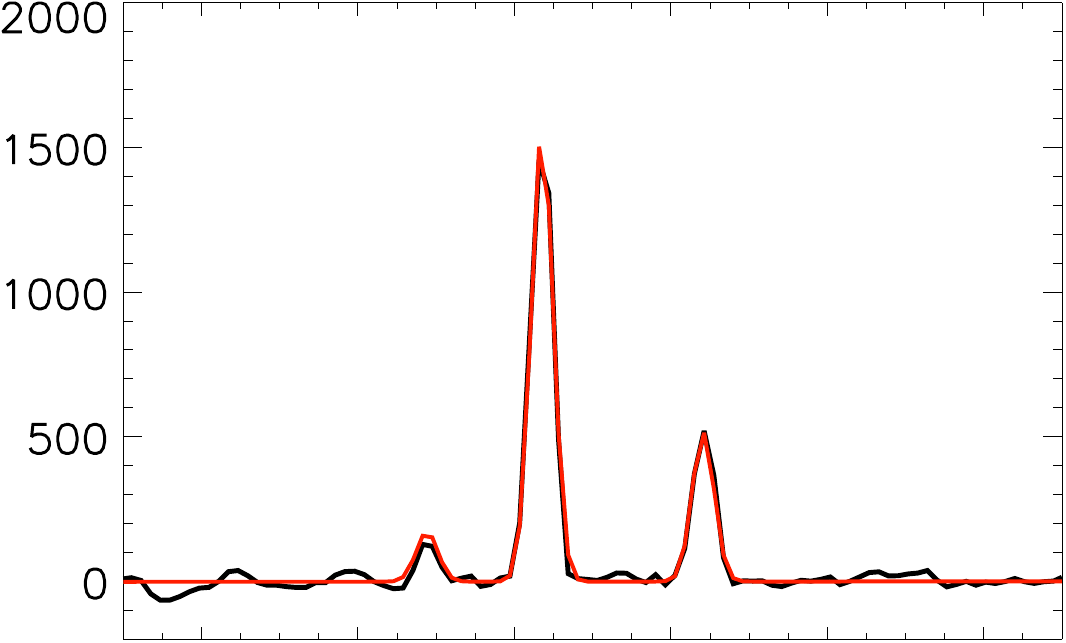}  
 \vskip-0.3mm  
 \includegraphics[width=0.175\textwidth, height=0.115\textwidth]{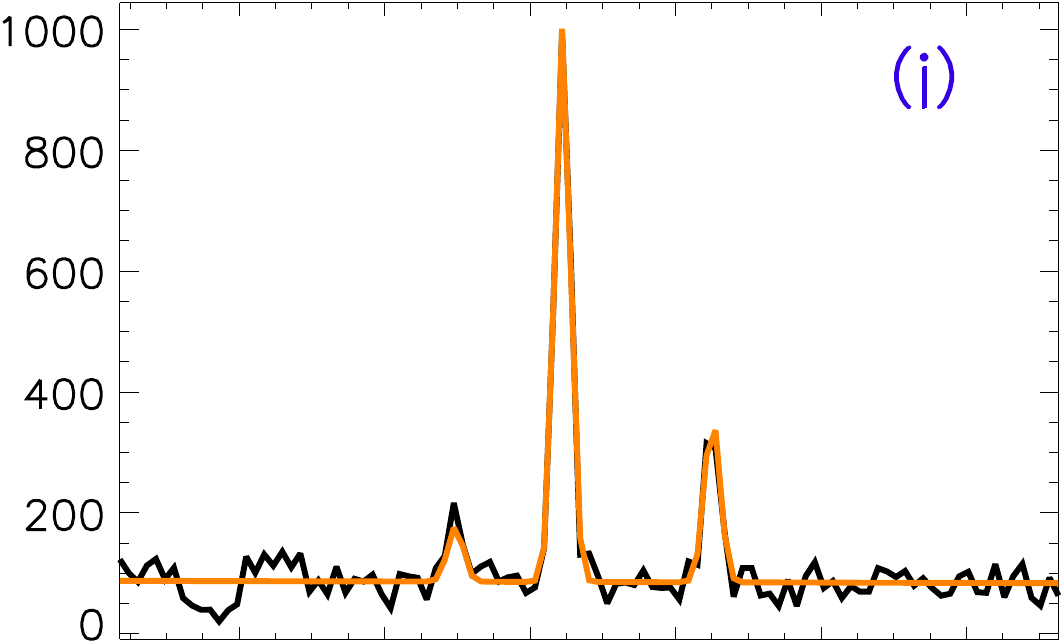}   
  \includegraphics[width=0.175\textwidth, height=0.115\textwidth]{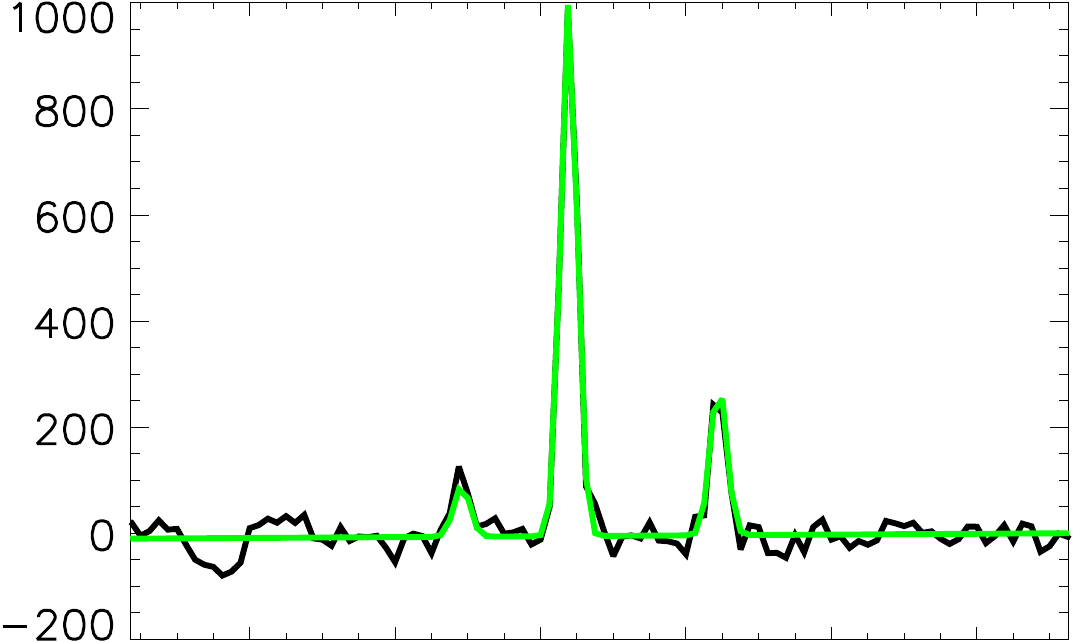}
   \includegraphics[width=0.175\textwidth, height=0.115\textwidth]{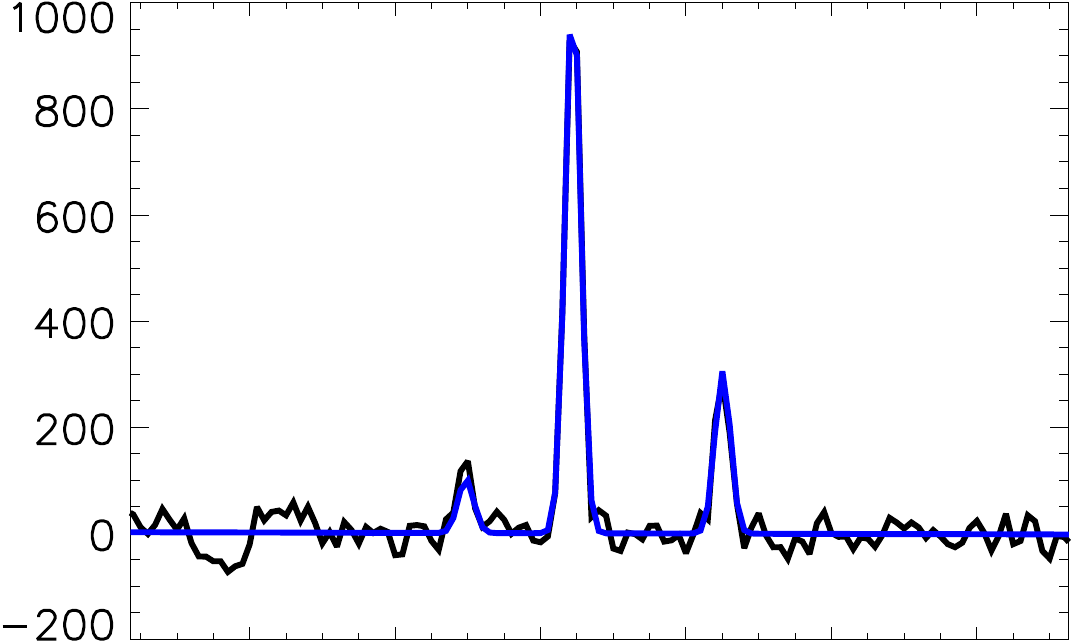}
   \includegraphics[width=0.175\textwidth, height=0.115\textwidth]{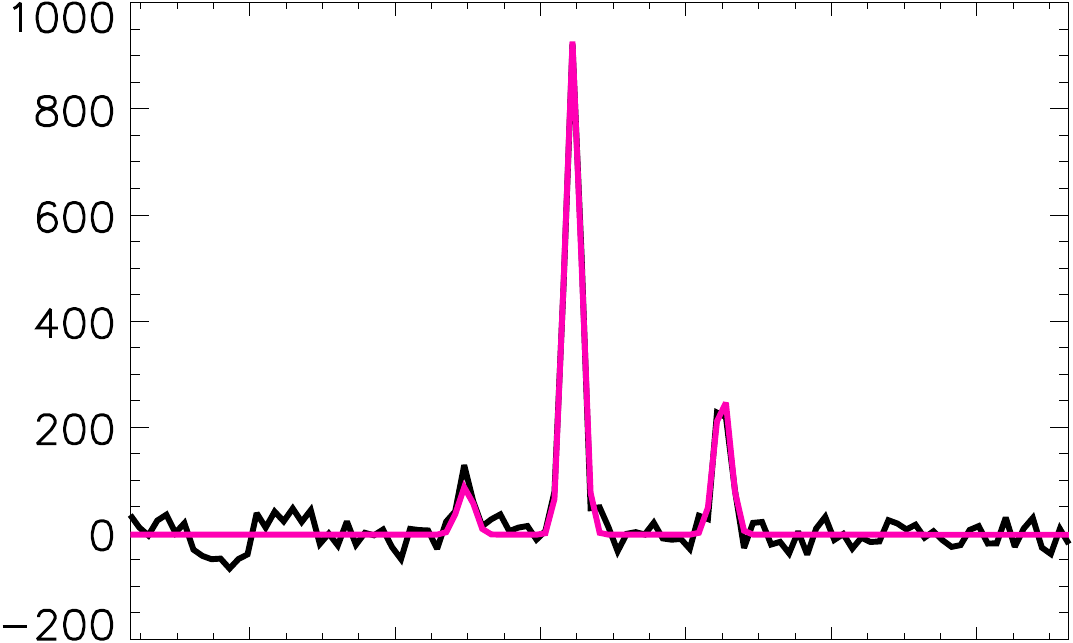}
   \includegraphics[width=0.175\textwidth, height=0.115\textwidth]{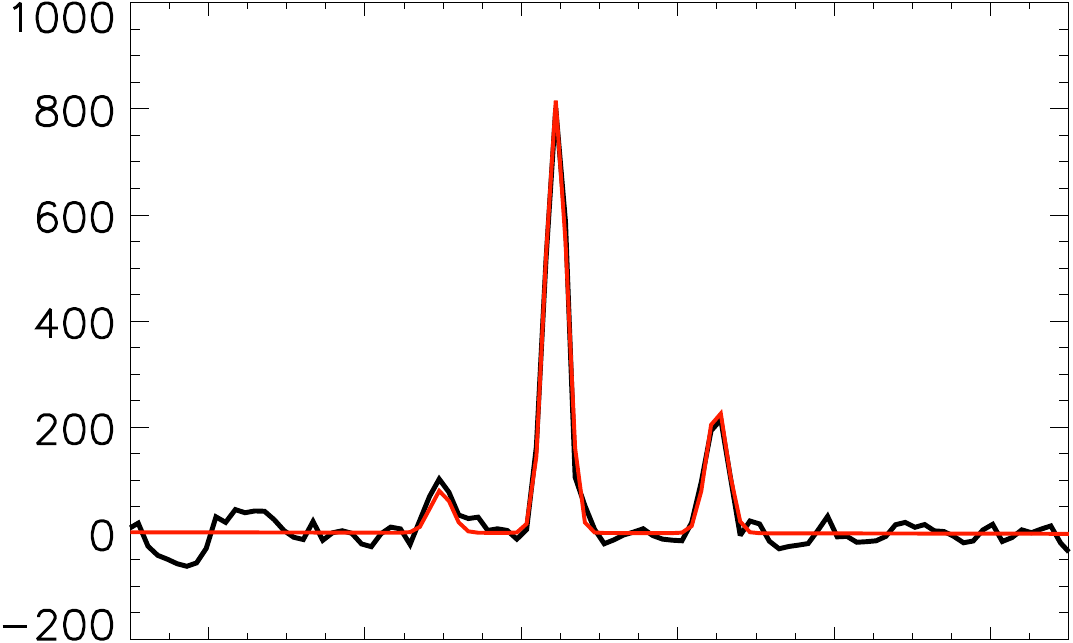}
 \vskip-0.3mm  
 \includegraphics[width=0.175\textwidth, height=0.115\textwidth]{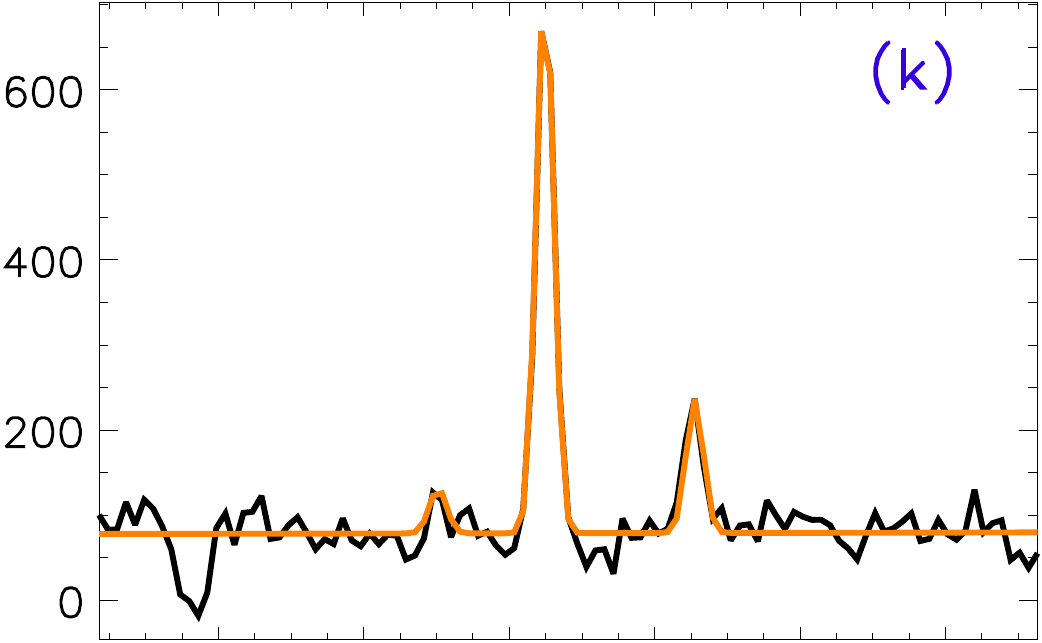}   
   \includegraphics[width=0.175\textwidth, height=0.115\textwidth]{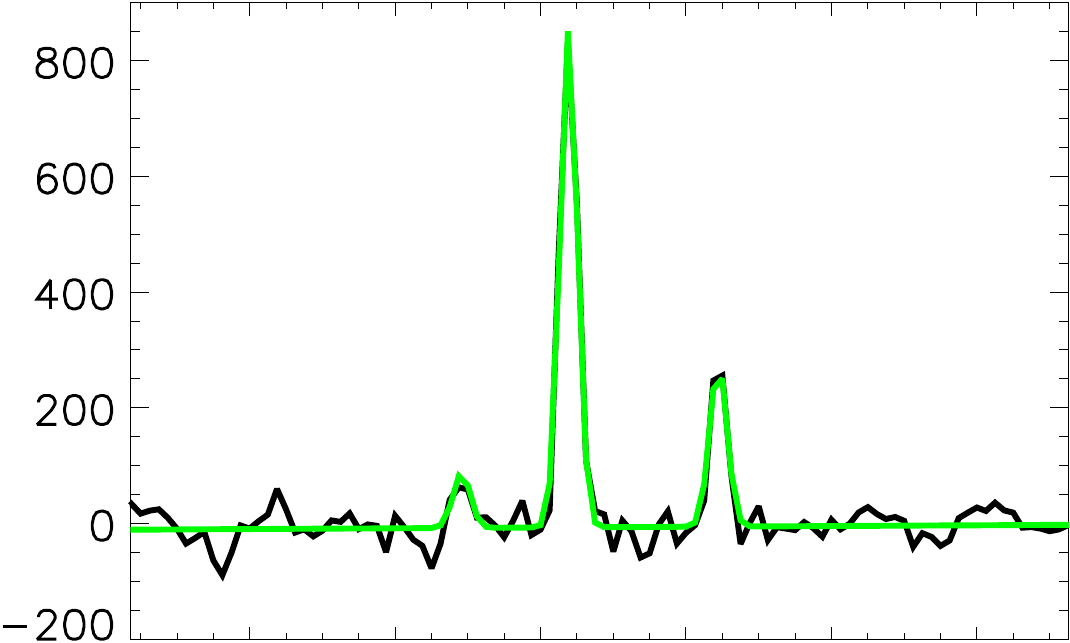}
   \includegraphics[width=0.175\textwidth, height=0.115\textwidth]{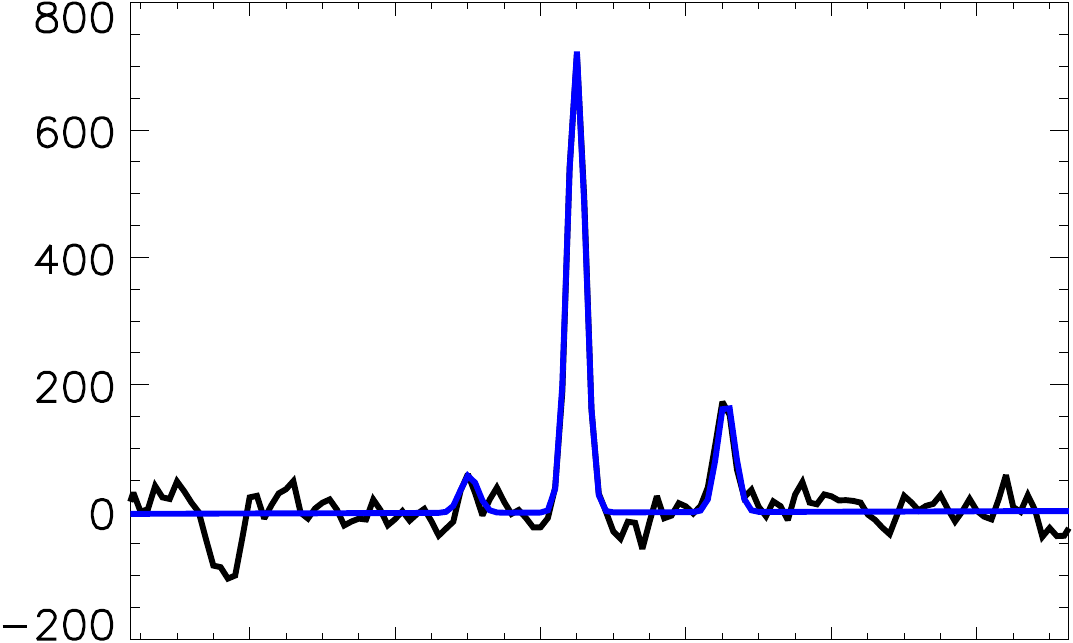}
   \includegraphics[width=0.175\textwidth, height=0.115\textwidth]{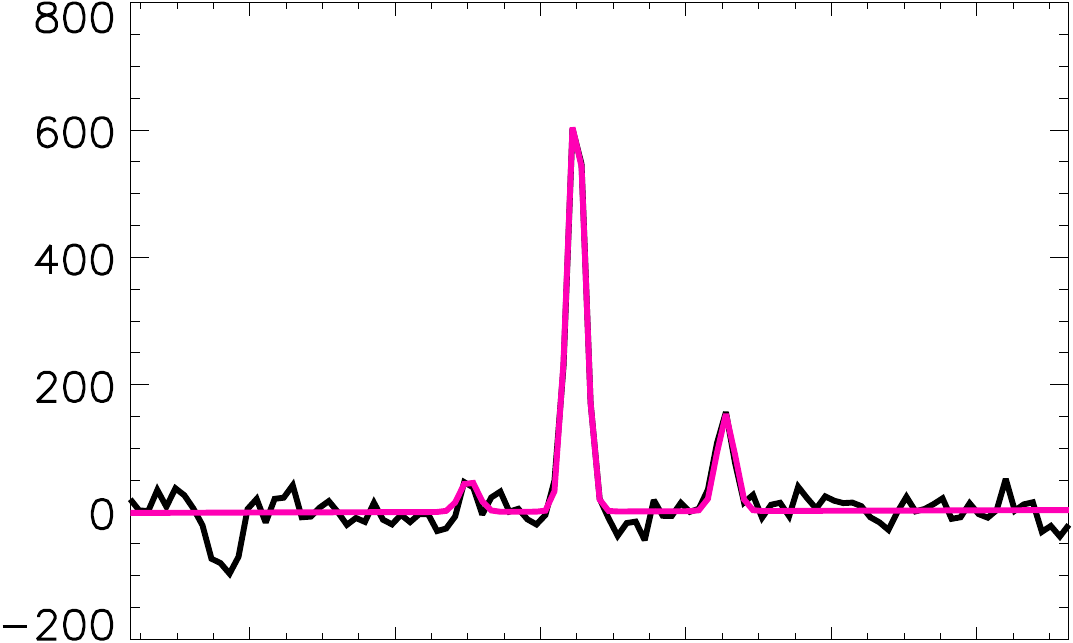}
   \includegraphics[width=0.175\textwidth, height=0.115\textwidth]{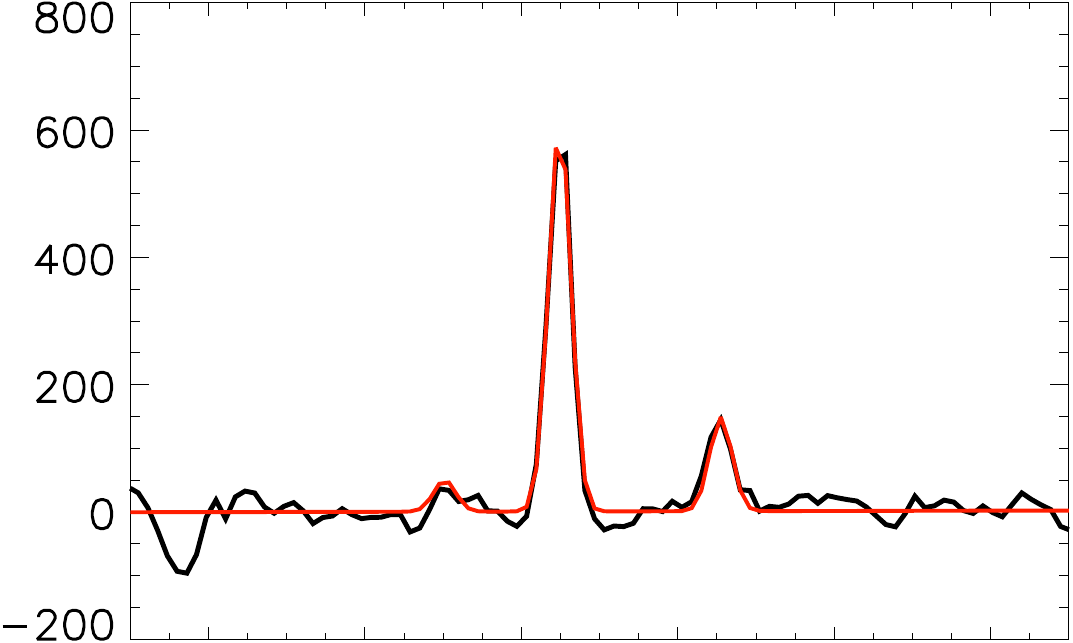}
 \vskip-0.2mm  
 \includegraphics[width=0.175\textwidth, height=0.115\textwidth]{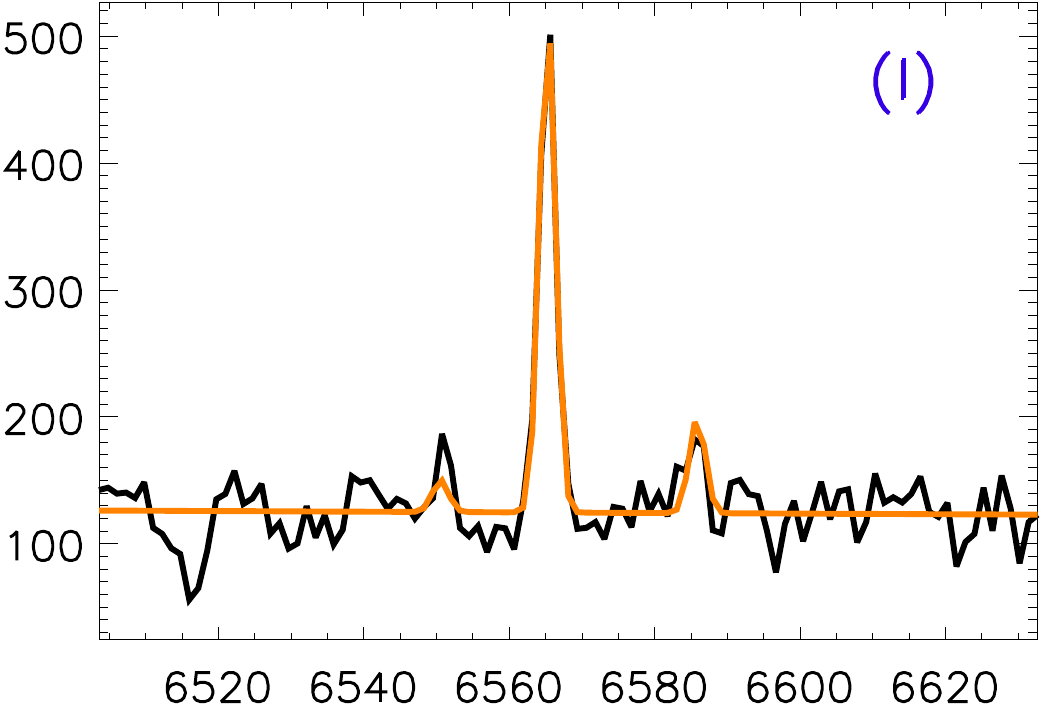}   
   \includegraphics[width=0.175\textwidth, height=0.115\textwidth]{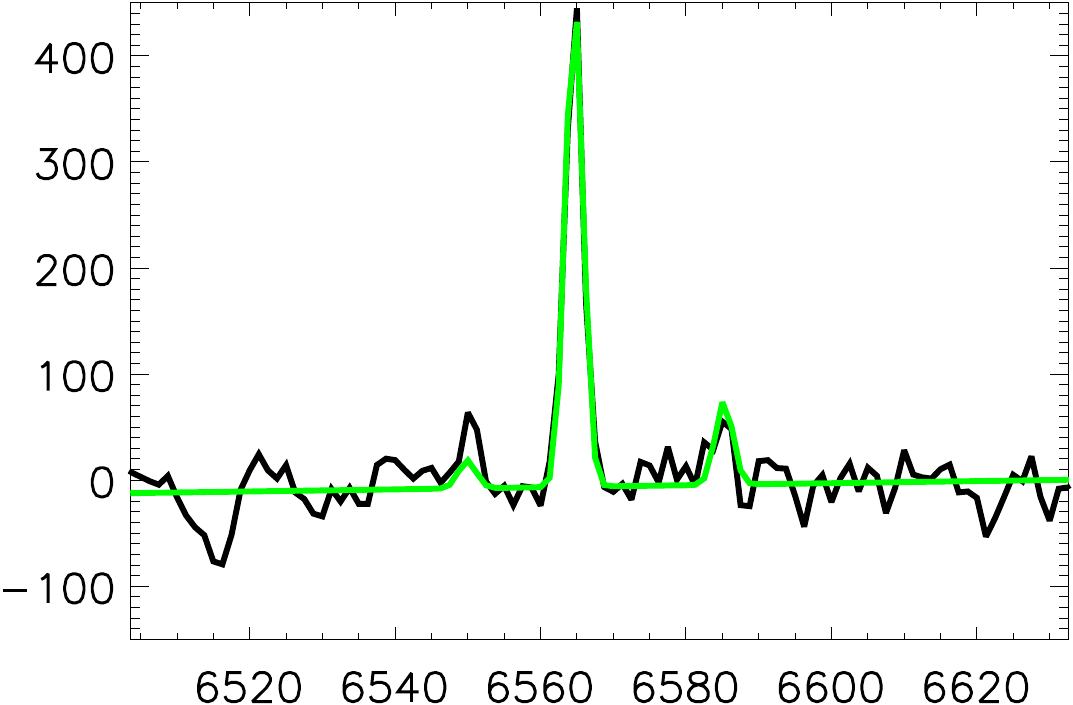}
   \includegraphics[width=0.175\textwidth, height=0.115\textwidth]{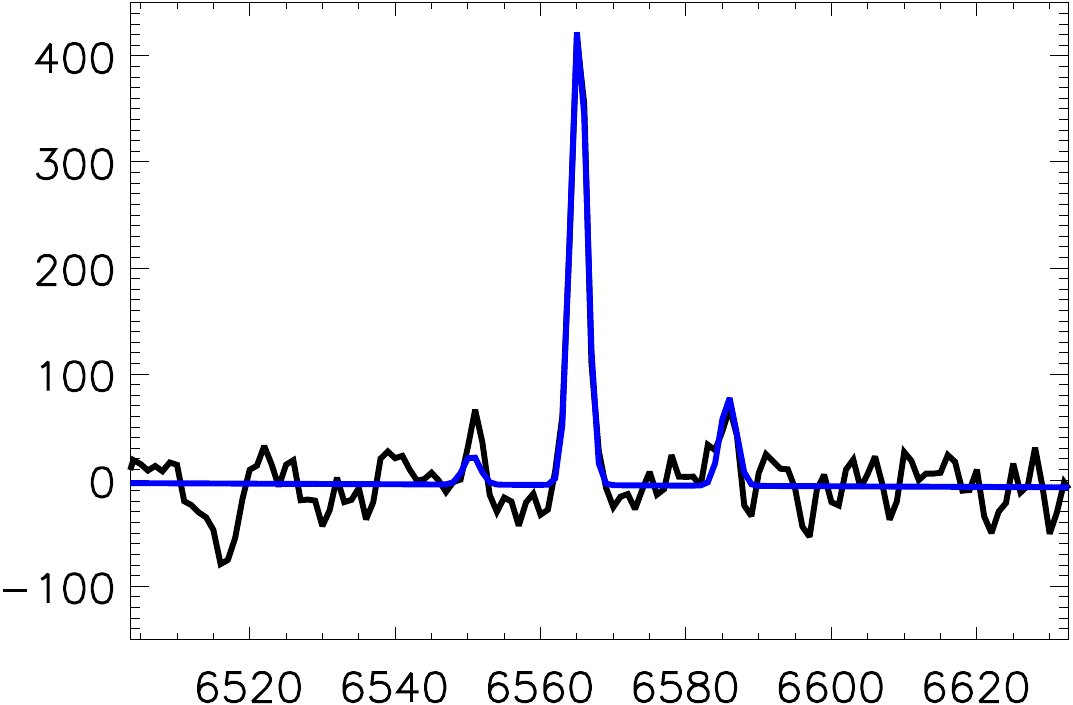}
   \includegraphics[width=0.175\textwidth, height=0.115\textwidth]{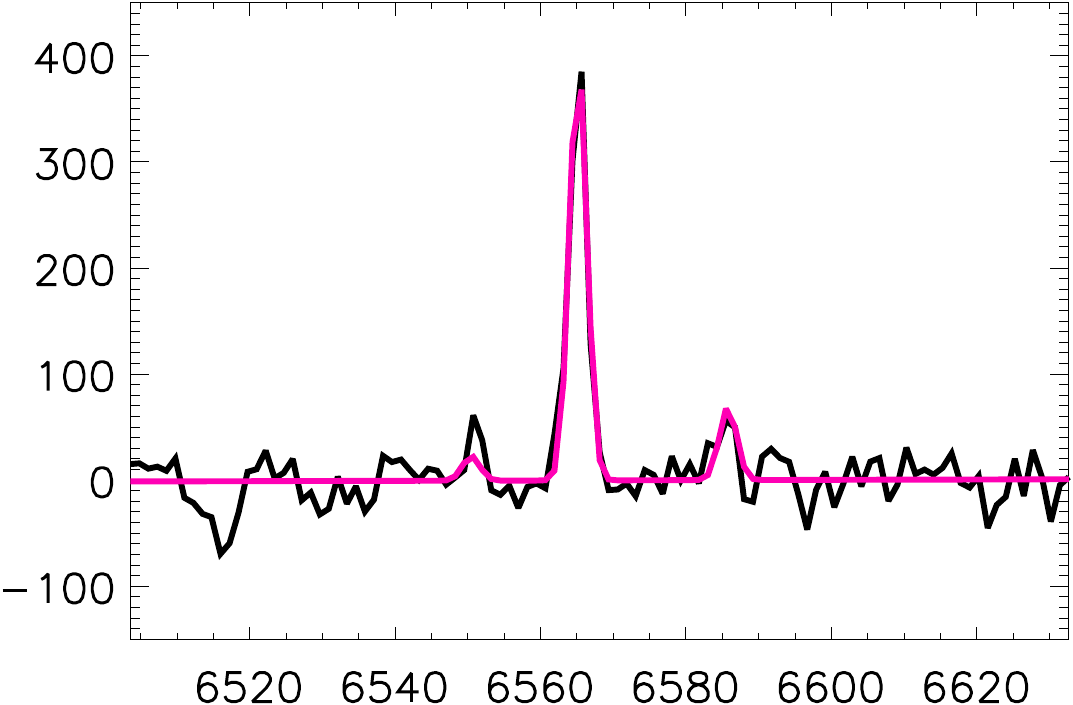}
   \includegraphics[width=0.1755\textwidth, height=0.115\textwidth]{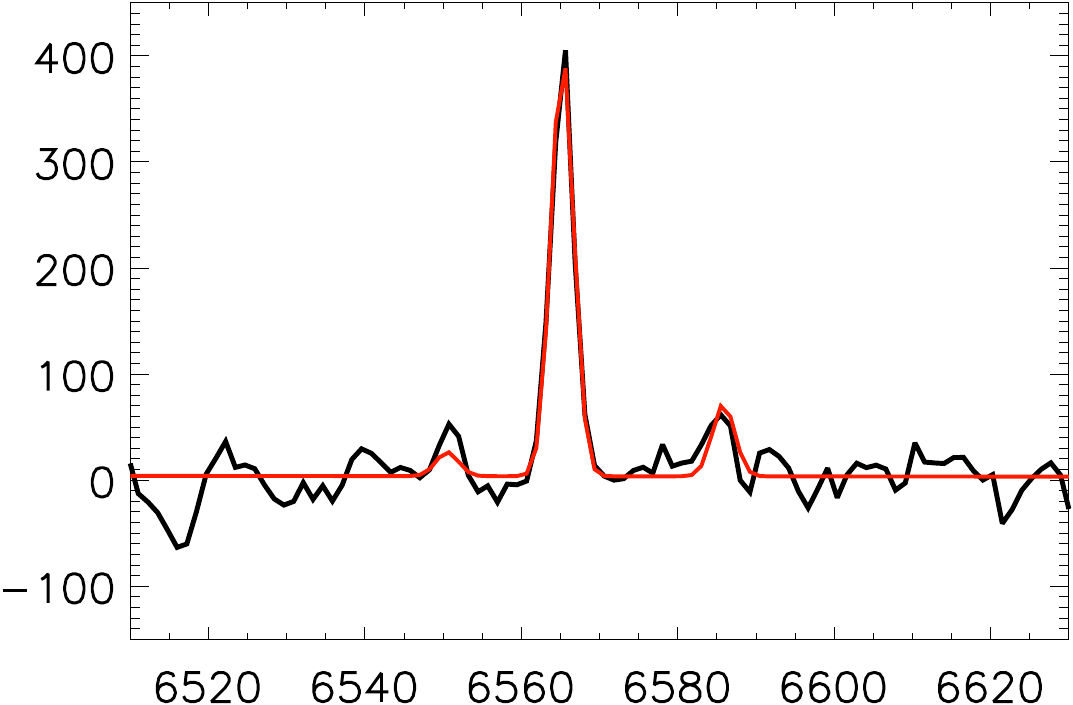}   
   \begin{minipage}[h]{17.5cm}
\vskip1.5mm\caption{Same figure caption as in Fig.~\ref{fig_spectra_all_fixed}.}
\label{fig_spectra_all_fixed_cont}
\end{minipage}
%\vspace{-1pt}
\end{figure*}

In this section we first present how the spectra have been selected, then describe the codes used to subtract the stellar continuum emission from the observed unsubtracted data as well as the stellar libraries involved, and finally discuss the different approaches that have been applied to fit the H$\alpha$--[NII] line complex.

% -----------------------------------------------------------------------------
\subsection{Selected spectra}

This work focuses on the analysis of twelve spectra.
Half of them are located in the nuclear region, while the remaining ones are placed in different parts of the galaxy.
For practical reasons, we refer to the different spaxels using letters ({\tt a, b, ..., k, l}), as shown in Figs.~\ref{figura_zoom}, ~\ref{fig_spectra_all_fixed} and ~\ref{fig_Ha_NII_results}, ordered according to their (decreasing) log([NII]$\lambda$6583/H$\alpha$) flux ratios\footnote{The flux ratios have been computed applying the most general line fitting approach ({\tt case~[4]}; see Sect.~\ref{LFIT}).}.

The spatial distribution of the selected spaxels is shown in Fig.~\ref{figura_zoom}. The intensity peak in the continuum image (spectrum {\tt d}) has been considered the origin of our system of reference, and the offsets of all spectra with respect to that origin are listed in Tab~\ref{single_values_sum}. The minimum separation between the spectra is 0.6$^{\prime\prime}$, reached in the region close to the nucleus, up to a maximum separation of 11.8$^{\prime\prime}$ from the centre.

We have tried to ensure that these spectra are representative of a wide range of physical conditions and signal--to--noise (S/N) ratios.
The original spectra, extracted from from the {\tt MUSE} datacube, are shown on the left column of Fig.~\ref{fig_spectra_all_fixed}, whereas the other columns display the results obtained after subtracting the stellar continuum with each of the four methods described below.
Although the precise line fluxes and equivalent widths are rather sensitive to the adopted method, it is evident from Fig.~\ref{fig_spectra_all_fixed} that the values of the $\log([{\rm NII}]\lambda6583/{\rm H}\alpha)$ ratio illustrate the whole range of values typically found in a \citet{BPT81} diagram (see also Tab.~\ref{single_values_sum}).

% -----------------------------------------------------------------------------
\subsection{Stellar continuum subtraction methods}

In order to characterize nebular emission, the stellar continuum emission needs to be subtracted. The H$\alpha$ and H$\beta$ emission line profiles may be significantly affected from Balmer absorption especially in evolved stellar populations \citep[e.g.,][]{Sarzi07, Cresci15, Gomes16}.

This is clearly visible in NGC 2906, where the H$\alpha$ emission in the central spectra is quite absorbed (from {\tt a} to {\tt d} spectra in Fig.~\ref{fig_spectra_all_fixed}), while for the outer spectra this effect does not seem to be so relevant (from {\tt f} to {\tt l} spectra in Figs.~\ref{fig_spectra_all_fixed} and \ref{fig_spectra_all_fixed_cont}).

Here we remove the stellar contribution using three different software programs that are publicly available and have been widely used in the literature:

\begin{enumerate}
 \item {\tt FIT3D}\footnote{Specifically, the pipeline {\tt Pipe3D} was used which is based on the {\tt FIT3D} code.\\ \url{http://www.astroscu.unam.mx/~sfsanchez/FIT3D/}} \citep{Sanchez16b, Sanchez16}.
 \item {\tt STARLIGHT}\footnote{\url{http://www.starlight.ufsc.br}} \citep{CidFer04, CidFer05, CidFer13}.
 \item Penalized Pixel--Fitting\footnote{\url{http://www-astro.physics.ox.ac.uk/~mxc/software/}} \citep[{\tt pPXF};][]{Cappellari04, Cappellari17}.
\end{enumerate}

These spectral synthesis codes combine the spectra from a base of simple stellar populations of various ages and metallicities in order to match an observed spectrum.
The search for an optimal model and the coefficients of the population vector associated with the base elements also allows to derive interesting outputs, such as the star formation and chemical enrichment histories of a galaxy or region, as well as to recover non--linear parameters, such as the amount of dust extinction and the main kinematic parameters (velocity shift and dispersion) of both stars and gas.
The reader is referred to the corresponding articles for an in--depth discussion of each method.

\begin{table*}
\centering %\begin{minipage}[h]{10.5cm}
\caption{Set of SSP models and stellar libraries used for the spectral fitting.}
\label{stelibra}
\begin{tiny}
 \begin{tabular}{l cccccccc}
\hline\hline\noalign{\smallskip}
Method &	 Model 	& Library 	& Isochrone  & IMF  & Metallicities  & N$_t$   & Ages [Gyr]  & FWHM [\AA]\\
{\smallskip} 
 (1) & (2) &(3) & (4) & (5) & (6) & (7) & (8) & (9) \\
\hline\noalign{\smallskip}
{\tt FIT3D}  & SED@ & {\tt GRANADA(+MILES)} & Pad00  &Salp 	& 0.002, 0.008, 0.02, 0.03   	&  156 &0.001--13 & $<$ 2\\
{\tt STARLIGHT}  & CB07   	& {\tt MILES }	& Pad94 	& Chab 	& 0.004, 0.008, 0.02, 0.05 	& 66 	& 0.001--18 &2.5 \\
{\tt STARLIGHT} & BC03  	& {\tt STELIB} 	& Pad94	& Chab   		& 0.0001, 0.0004, 0.004, 0.008, 0.02, 0.05 &  150  &  0.001--18 & 3.0\\
{\tt pPXF} & V10 	& {\tt MILES} 	& Pad00 	& Salp 	& 0.0004, 0.001, 0.004, 0.008, 0.019, 0.03  &  156 & 0.1--18 & 2.5  \\
\hline\hline\noalign{\smallskip}
\end{tabular}
\vskip0.cm
\end{tiny}
\begin{minipage}[h]{18cm}
\footnotesize
{\bf Notes:} Col (1): Stellar subtraction method. Col (2): Model. Col (3): Name of the stellar library. Col (4): Isochrone. Padova 1994 and Padova 2000 are considered (see text). Col (5): Initial mass function. {\tt Salpeter} and {\tt Chabrier} are used. Col (6): Range of metallicity. Col (7): Number of templates used in the fit. Col (8): Age of the stellar population in Gyr. Col (9): Resolution of the stellar library (full width at half maximum) in~\AA. 
\end{minipage}
\end{table*}

In any case, the characteristics of the stellar libraries adopted as a basis are as important as the fitting algorithm.
The choice of the stellar library generally depends on the resolution of the data.
In our case, {\tt MUSE} has a nominal spectral resolution FWHM $\sim$ 2.6\footnote{The intrinsic spectral resolution is slightly lower than the theoretical one (FWHM $\sim$ 2.2 \AA).} \AA, and we combined each code with one of the following stellar libraries in order to achieve a similar (or lower) resolution:

\begin{enumerate}
\item {\tt STELIB:} empirical library from \citet[; hereafter BC03]{BC03};
\item {\tt MILES:} `Medium-resolution Isaac Newton Telescope Library of Empirical Spectra' from \citet{Vazdekis10}; 
\item {\tt GRANADA \& MILES:} a combination of the theoretical stellar libraries from \citet{GD05} and \citet{GD10} with \citet{Vazdekis10}. 
\end{enumerate}

The {\tt STELIB}\footnote{The {\tt STELIB} stellar library can be found at \href{http://webast.ast.obs-mip.fr/stelib}{\tt http://webast.ast.obs-mip.fr/stelib}.} library \citep{LeB03} consists of an homogeneous library of 249 stellar spectra in the visible range (3200 to 9500 \AA), with an intermediate spectral resolution ($\leq$3 \AA) and sampling (1 \AA). Only 187 over 249 stars have measured metallicity and can be used to compute the predicted spectra. This library includes stars of various spectral types and luminosity classes, spanning a relatively wide range in metallicity (Z = 0.0001 -- 0.05). 

The {\tt MILES}\footnote{The {\tt MILES} stellar library can be found at \href{http://www.iac.es/proyecto/miles/pages/webtools/tune-ssp-models.php}{\tt http://www.iac.es/proyecto/miles/pages/webtools/tune-\\ssp-models.php}.} stellar library \citep{Vazdekis10} consists of stars spanning a large range in atmospheric parameters. The 985 spectra were obtained at the 2.5m Isaac Newton Telescope (INT) and cover the range 3525--7500~\AA\ \citep{SBlazquez06} at 2.51 \AA\ (FWHM) spectral resolution \citep{FBarroso11}. These models cover ages between $\sim$0.1 and $\sim$18 Gyr and metallicities between 0.0004 -- 0.03. Vazdekis/Miles models are based on the previous models from \citet{Vazdekis99} and \citet{Vazdekis03} and they use Padova2000 isochrones. The Padova isochrones (1994 and 2000) are presented in \citet{Girardi02}\footnote{Available at {\tt http://pleiadi.pd.astro.it/}} based on the (solar scaled mixture) tracks from \citet{Bertelli94} and \citet{Girardi00}, respectively.

{\tt GRANADA\footnote{The {\tt GRANADA} library is a high resolution library.} \& MILES} stellar library ({\tt gsd156}) comprises 156 templates that cover 39 stellar ages (1 Myr to 13 Gyr), and 4 metallicities (Z = 0.002, 0.008, 0.02 and 0.03). These templates have been extracted from a combination of the synthetic stellar spectra from the {\tt GRANADA} (\citep{Martins05}) and the SSP libraries provided by the {\tt MILES} project. This SSP--library uses the Salpeter (1955) Initial Mass Function (IMF) and \citet{Girardi00} and Geneva\footnote{The Geneva isochrones have been computed with the isochrone program presented in \citet{Mey95} and they follow the prescriptions quoted in \citet{Cervinho01} from the evolutionary tracks from \citet{Schaller92}, \citet{Charbonnel93}, \citet{Schaerer93a} and \citet{Schaerer93b}.} tracks. This library is described in detail in \citet{CidFer13}. 

The `code--stellar library' combinations we used are: 

\begin{enumerate}

\item the {\tt FIT3D} code with the {\tt GRANADA and MILES} libraries;
\item the {\tt STARLIGHT} code with the {\tt MILES} library \citep[see][]{Galbany16} and references therein for more information\footnote{This model is considered a `modified' version of the BC03 models (i.e., CB07; see \citet{Bruzual07} for further details).});
\item the {\tt STARLIGHT} code with the {\tt STELIB} library;
\item the {\tt pPXF} code with the {\tt MILES} library.
\end{enumerate}

In Tab.~\ref{stelibra} we summarize all combinations `code--stellar library' used in this analysis, including details on the isochrones and the age of the stellar populations considered.

% -----------------------------------------------------------------------------
\subsection{Line fitting cases}
\label{LFIT}

In this work we focus on the H$\alpha$$\lambda$6563\AA\ and the forbidden [NII]$\lambda$$\lambda$6548, 6583\AA\ emission lines\footnote{We actually do not discuss the results of [NII]$\lambda$6548\AA\ since this line is three times fainter than [NII]$\lambda$6583\AA\ (according to atomic physics) and more subjected to higher uncertainties in the fit. Furthermore, in most cases, it would share the same kinematics as the [NII]$\lambda$6583\AA\ line.}.
The routine {\tt MPFITEXPR} \citep[implemented in {\tt IDL} code by][]{Markwardt09} is quite commonly used to derive the kinematic parameters \citep[e.g.,][]{Arribas08, Bellocchi12, Bellocchi13, Cazzoli14, Cairos15, Bellocchi16, Poggianti17}.
This algorithm allows to fit the observed lines of the individual spectra to Gaussian profiles and the continuum emission as a line with a certain slope. It derives the best set of lines that match the available data. In case of adjusting multiple lines, the line flux ratios can be fixed according to the atomic physics (as in the case of the [NII] doublet). This routine gives in output the wavelength (i.e., centroid $\lambda$), the width ($\sigma$) and flux of the line. It also allows one to combine in different ways the $\lambda$ and $\sigma$ parameters keeping them fixed or let free to vary. Thus, when fitting the H$\alpha$--[NII] complex, there are four cases that can be taken into account:

\begin{enumerate}
 \item {\underline{\tt $\lambda$ \& $\sigma$ {\tt FIXED}}}: the wavelengths of the different lines are fixed according to the atomic physics and their widths are constrained to be equal for all lines \citep[e.g.,][]{Bellocchi13, Bellocchi16};
\item {\underline{\tt $\lambda$ {\tt FREE}, $\sigma$ {\tt FIXED}}}: the wavelengths of the different lines are free to vary but their widths are constrained to be equal for all lines;
 \item {\underline{\tt $\lambda$ {\tt FIXED}, $\sigma$ {\tt FREE}}}: the wavelengths of the different lines are fixed according to the atomic physics but their widths are left free to vary;
\item {\underline{\tt $\lambda$ \& $\sigma$ {\tt FREE}}}: both wavelengths and widths are free to vary for all lines.
\end{enumerate}

All these cases are summarized in Tab.~\ref{cases_tab}. 
In the next section the results of our analysis are shown and discussed. It is worth mentioning that in this work the {\tt MUSE} instrumental profile ($\sigma_{INS}$) has not been subtracted\footnote{To have a rough estimate of the instrumental profile we apply the line fitting {\tt case 1} to the [OI]$\lambda$6300 sky line in a relatively large region of the galaxy, where the [OI] emission is present. We derive a typical (median) value for the wavelength of $\lambda_{INS}~\sim$~6299.99 \AA\ and $\sigma_{INS}\sim$ 0.88~\AA, with the respective mean values of 6299.99 ($\pm$ 0.91) \AA\ and 0.96 ($\pm$ 0.48) \AA.}.

 \begin{table}
    \caption{Different cases considered for the line fitting analysis for the wavelength and for the width.}
    \label{cases_tab}
   \centering
    \begin{tabular}{c|cc} % Column formatting, @{} suppresses leading/trailing space
       \hline\hline
 { Cases }   & { Wavelength} ($\lambda$) & { Width }($\sigma$)\\
       \hline
       {\tt [1]}    & {\tt fixed}& {\tt fixed} \\
       {\tt [2]}    & {\tt free} & {\tt fixed} \\
       {\tt [3]}    & {\tt fixed}& {\tt free} \\
       {\tt [4]}    & {\tt free} & {\tt free} \\
       \hline\hline
    \end{tabular}
 \end{table}

% -----------------------------------------------------------------------------
\section{Results}
% -----------------------------------------------------------------------------

\begin{figure*}
   \centering
 \hskip-15mm
   \includegraphics[width=0.9\textwidth]{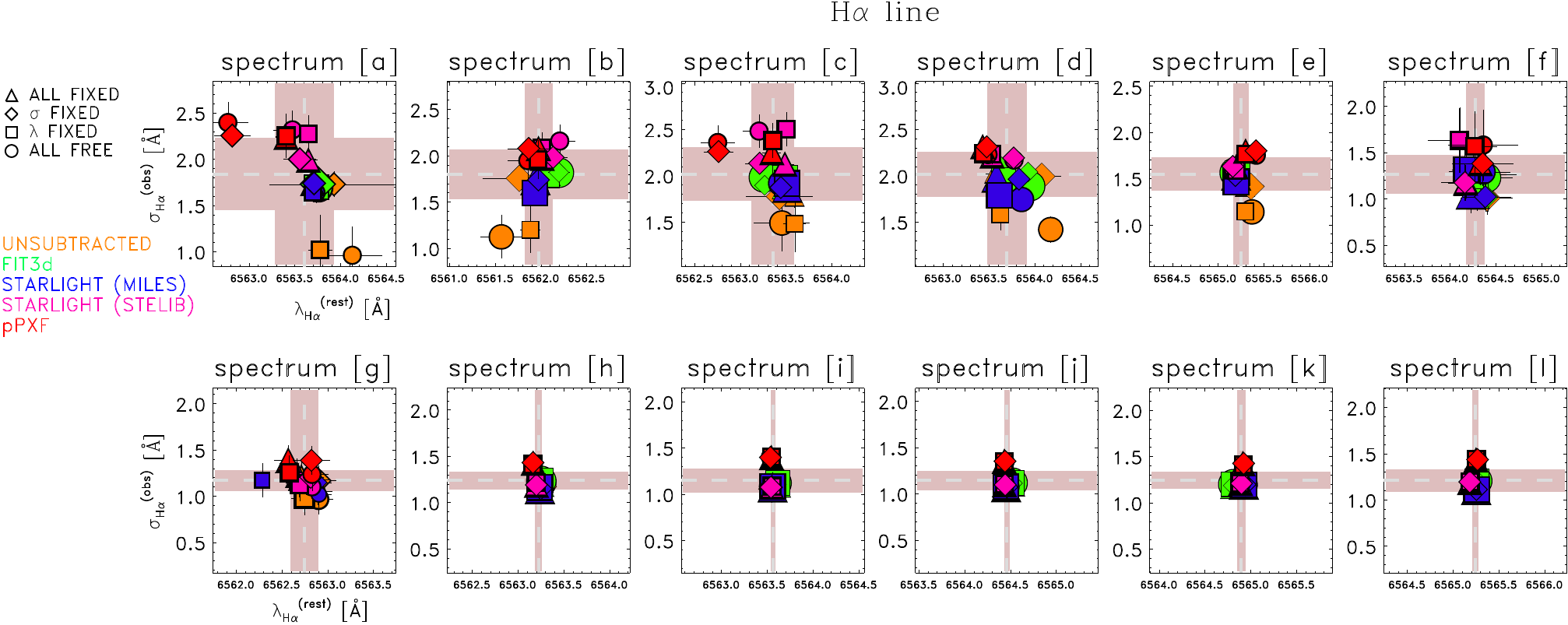}
 \vskip3mm
    \includegraphics[width=0.83\textwidth]{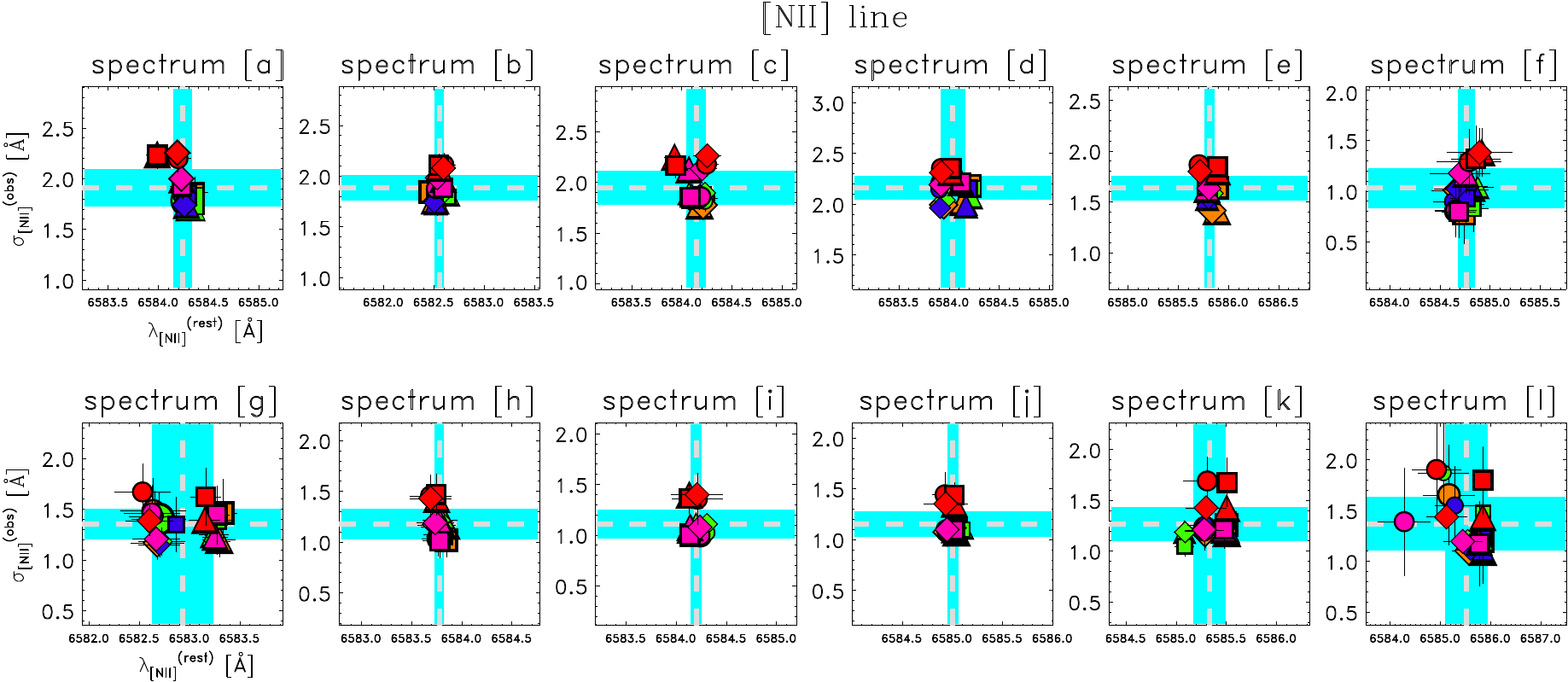}
\caption{\small Kinematic results for the individual spectra when using different stellar subtraction methods and line fitting approaches for the H$\alpha$ ({\it top panels}) and [NII] ({\it bottom panels}) emission lines. The four line fitting cases and the five stellar subtraction methods used are highlighted according to the different symbols and colors (as in Fig.~\ref{fig_spectra_all_fixed}), respectively (twenty values for each spectrum). The mean wavelength and dispersion ($\overline\lambda_i$, $\overline\sigma_i$) as well as their standard deviation ($\Delta\overline\lambda_i$, $\Delta\overline\sigma_i$) are represented using dashed light grey lines and the shaded area (light red and cyan), respectively.}
\label{fig_Ha_NII_results}
\end{figure*}

In this section we present and discuss the kinematic results obtained when different stellar subtraction methods and line fitting approaches are considered. The raw observed data (hereafter, unsubtracted) are also included in the analysis in order to study the case in which the stellar emission is not removed at all and thus understand the importance of removing (or not) the stellar component from the data for the derivation of the kinematics of the gas component. 

In Sect. \ref{Indiv_part} we focus on the results derived for each individual spectrum to better study the dispersion related to the choice of a specific stellar subtraction method as well as the one associated to the use of a particular line fitting approach. 
In Sect. \ref{General_results} we then analyze the mean trends obtained when considering all spectra averaged according to the number of methods or the number of cases considered.

% -----------------------------------------------------------------------------
\subsection{Mean and dispersion for individual spectra}
\label{Indiv_part}

\begin{table*}
\caption{Mean values and standard deviations for the wavelength and line width of the H$\alpha$ and [NII] emission lines obtained when combining all line fitting cases and stellar continuum subtraction methods for each individual spaxel. }
\label{single_values_sum}
\begin{tabular}{ccc|cccc} % Column formatting, @{} suppresses leading/trailing space
\hline\hline  
{\tt  Spectrum} &  {\tt Offset} & {\tt log(NII/H$\alpha$)}
& $\overline\lambda_{\rm H\alpha} \pm \Delta\overline\lambda_{\rm H\alpha}$
& $\overline\sigma_{\rm H\alpha} \pm \Delta\overline\sigma_{\rm H\alpha}$
& $\overline\lambda_{\rm [NII]} \pm \Delta\overline\lambda_{\rm [NII]}$
& $\overline\sigma_{\rm [NII]} \pm \Delta\overline\sigma{\rm [NII]}$ \\
  {\tt name} &	($^{\prime\prime}$, $^{\prime\prime}$)	& & {\tiny [\AA]}  & {\tiny [\AA]}  &{\tiny [\AA]}  & {\tiny [\AA]} \\ 
(1) & (2) & (3) & (4) & (5) & (6) & (7)\\ 
\hline
{\tt a} & $(0.0, -0.6)$ 	&0.74 		&	  6563.60 $\pm$ 0.32		&	 1.84 $\pm$ 0.39		&	   6584.240 $\pm$ 0.097	&	 1.91 $\pm$ 0.19  \\
{\tt b} & $(+0.6, 0.0)$ &0.59 		&	  6561.97 $\pm$ 0.15		&	 1.80 $\pm$ 0.27        &        6582.554 $\pm$ 0.048	&	 1.89 $\pm$ 0.13  \\
{\tt c} & $(0.0, +0.6)$ 	&0.46 		&	  6563.35 $\pm$ 0.23		&	 2.02 $\pm$ 0.29         &        6584.141 $\pm$ 0.094	&	 1.95 $\pm$ 0.17  \\
{\tt d} & $(0.0, 0.0)$  	&0.39 		&	  6563.69 $\pm$ 0.21		&	 2.02 $\pm$ 0.24         &        6584.04 $\pm$ 0.12	&	 2.16 $\pm$ 0.12  \\
{\tt e} & $(-0.6, 0.0)$ 	&0.20 		&	  6565.249 $\pm$ 0.085	&	 1.55 $\pm$ 0.18        &        6585.810 $\pm$ 0.051	&	 1.64 $\pm$ 0.12  \\
{\tt f} & $(-3.0,+11.4)$ 	&--0.06 	&	  6564.27 $\pm$ 0.11		&	 1.27 $\pm$ 0.21          &        6584.761 $\pm$ 0.086	&	 1.03 $\pm$ 0.19  \\
{\tt g} & $(+4.2, +5.2)$	&--0.12 	&	  6562.75 $\pm$ 0.15		&	 1.17 $\pm$ 0.11          &        6582.93 $\pm$ 0.31	&	 1.36 $\pm$ 0.15  \\
{\tt h} & $(+3.6, +5.2)$	&--0.33 	&	  6563.224 $\pm$ 0.039	&	 1.239 $\pm$ 0.097   &        6583.776 $\pm$ 0.047	&	 1.17 $\pm$ 0.15  \\
{\tt i} & $(-1.6, -7.8)$ 		&--0.49 	&	  6563.559 $\pm$ 0.026	&	 1.15 $\pm$ 0.13        &        6584.197 $\pm$ 0.056	&	 1.11 $\pm$ 0.14  \\
{\tt j} & $(-4.8, -8.8)$ 		&--0.53 	&	  6564.460 $\pm$ 0.031	&	 1.15 $\pm$ 0.11          &        6585.004 $\pm$ 0.053	&	 1.15 $\pm$ 0.13  \\
{\tt k} & $(-4.5, +8.6)$ 	&--0.61 	&	  6564.893 $\pm$ 0.046	&	 1.242 $\pm$ 0.093   &        6585.33 $\pm$ 0.16	&	 1.27 $\pm$ 0.17  \\
{\tt l} & $(-4.4, +7.0)$ 		&--0.62 	&	  6565.241 $\pm$ 0.031	&	 1.21 $\pm$ 0.12          &        6585.51 $\pm$ 0.43	&	 1.37 $\pm$ 0.27  \\
\hline\hline
\end{tabular}
\begin{minipage}[h]{18cm}\vskip3mm
{\bf Notes.} Col (1): Name of the spectrum as explained in the text. 
Col (2): Offset in arcsec ($^{\prime\prime}$) with respect to the continuum intensity peak  $(0.0, 0.0)$. 
Col (3): Logarithmic ratio between the [NII] and the H$\alpha$ fluxes. 
Cols (4, 5, 6, 7): Mean kinematic parameters ($\overline\lambda$, $\overline\sigma$) and their uncertainties ($\Delta\overline\lambda$, $\Delta\overline\sigma$) for the H$\alpha$ (Cols. 4 and 5) and [NII] (Cols. 6 and 7) lines, respectively.
The uncertainties for both kinematic parameters, and for both lines, have been derived as the square root of the variance computed using the twenty (five methods $\times$ four cases) values for each spectrum.
\end{minipage}
\end{table*}

\begin{figure*}
   \centering
  \includegraphics[width=0.6\textwidth, height=0.25\textwidth]{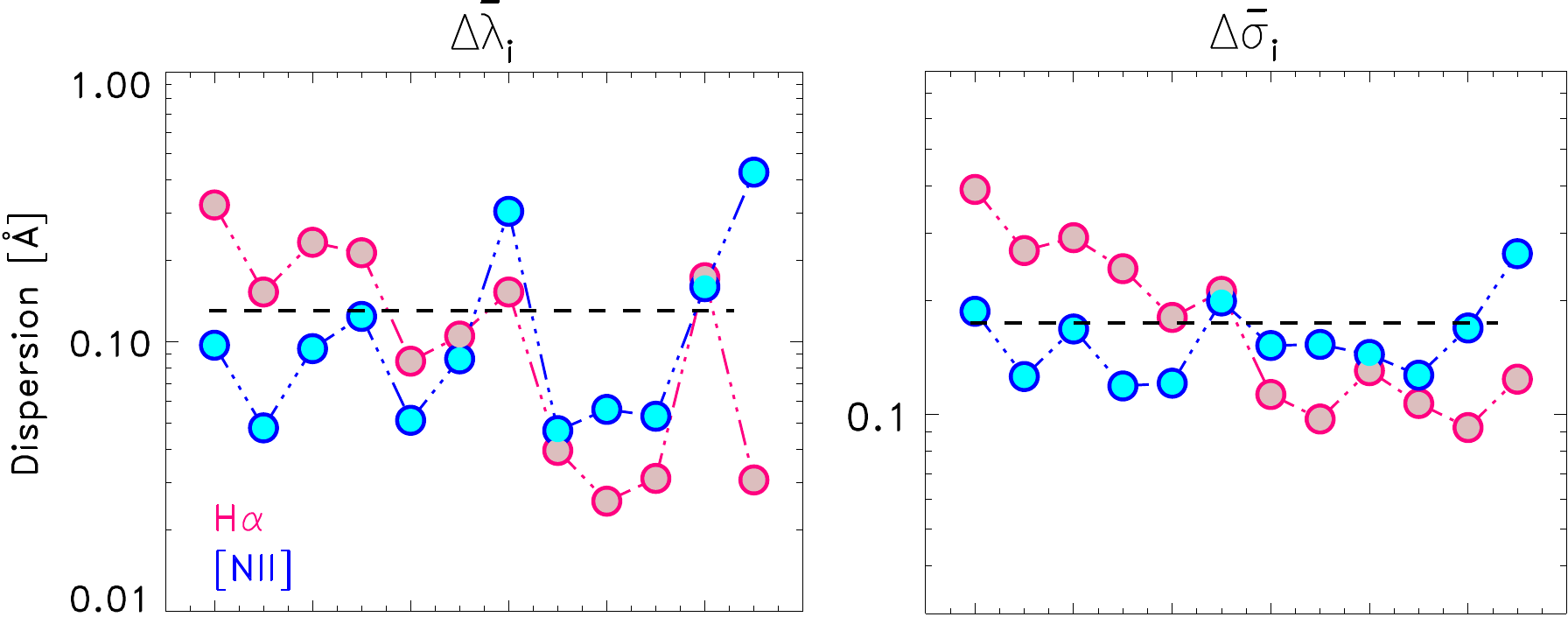}
\vskip4mm
\includegraphics[width=0.6\textwidth, height=0.3\textwidth]{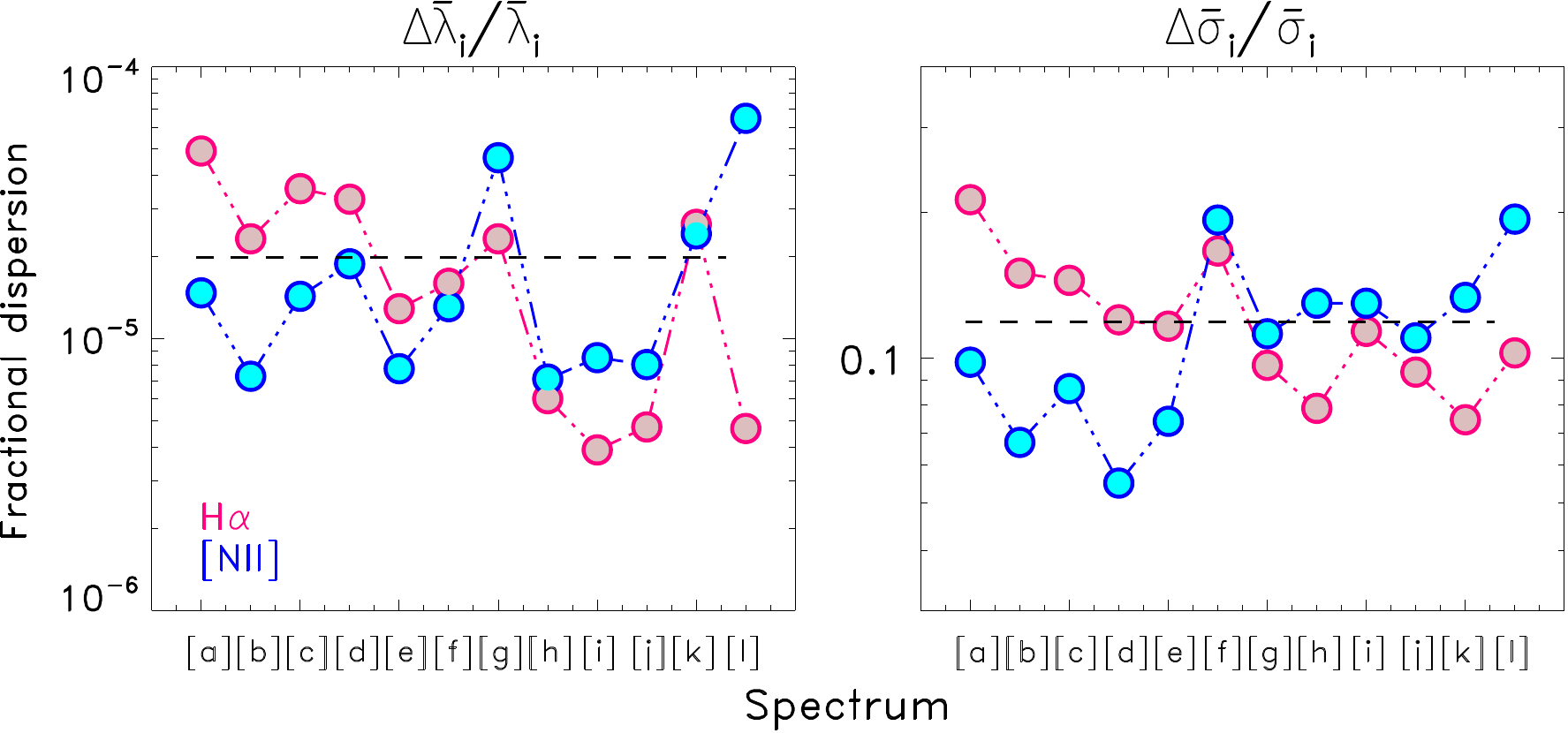}
    \caption{\small Trend of the linear (top) and fractional (bottom) dispersion values of the $\lambda$ ($\Delta\overline\lambda_i$ and $\Delta\overline\lambda_i$/$\overline\lambda_i$; left) and $\sigma$ ($\Delta\overline\sigma_i$/$\overline\sigma_i$; right) parameters. They have been obtained for each spectrum for both H$\alpha$ (light red) and [NII] (light blue) dots as due to the use of different stellar subtraction methods and line fitting approaches.}
   \label{fig_Dispersion_lam_sig}
\end{figure*}

In this section the derivation of the kinematic results of each individual spectrum ($\lambda_{i, mc}$,  $\sigma_{i, mc}$) is presented, where the index $i$ runs over our 12 individual spectra, and a combination of twenty values per spectrum is analyzed (i.e., 5~methods $\times$ 4 cases; $N_{mc} = N_m\,N_c = 20$).
We remind that, according to our selection criterion, the [NII] contribution is very significant in the first six spectra (from spectrum {\tt [a]} to {\tt [f]}), while the H$\alpha$ flux emission dominates in the remaining six spectra (from spectrum {\tt [g]} to {\tt [l]}).

We first determined the wavelength $\lambda_{i, mc}$ and the velocity dispersion $\sigma_{i, mc}$ for both lines in each spectrum (Fig.~\ref{fig_Ha_NII_results}) as derived when applying different line fitting approaches (`cases') and continuum--subtraction methods.
The more asymmetric shape of the H$\alpha$ profile in the first spectra, characterized by lower S/N as a consequence of the H$\alpha$ absorption, reflects in larger dispersions when deriving the kinematic values when using different values. 
Conversely, the quality of the fit of the [NII] line is also affected by the corresponding S/N ratio, although the differences are not as marked as for the Balmer line.

Considering that each measurement provides a wavelength $\lambda_{i, mc}$ and a velocity dispersion $\sigma_{i, mc}$, we defined the mean values for each spectrum among all twenty combinations as
\begin{equation}
\overline\lambda_i = \frac{1}{N_m} \frac{1}{N_c} \sum_{m=1}^{N_m} \sum_{c=1}^{N_c} \lambda_{i, mc}
~;~
\overline\sigma_i = \frac{1}{N_m} \frac{1}{N_c} \sum_{m=1}^{N_m} \sum_{c=1}^{N_c} \sigma_{i, mc}
\end{equation}
and we characterize the behavior of the different methods and cases in terms of the offset from the appropriate mean value:
\begin{equation}
\hskip3mm\delta\lambda_{i, mc} =  \lambda_{i, mc} - \overline\lambda_i
~;~~
\delta\sigma_{i, mc} = \sigma_{i, mc} - \overline\sigma_i.
\end{equation}

As the true kinematic parameters of every individual spectrum are unknown, we cannot determine the optimal combination of subtraction method and fitting approach (in each particular case), but we may quantify the uncertainties associated to such choice in a statistical sense, computing the standard deviations
\begin{equation}
\hskip3mm\Delta\overline\lambda_i = \sqrt{ \frac{\sum_{mc} \delta\lambda_{i, mc}^2}{N_m\,N_c} }
~;~~
\Delta\overline\sigma_i = \sqrt{ \frac{\sum_{mc} \delta\sigma_{i, mc}^2}{N_m\,N_c} }
\end{equation}
arising from our $N_{mc}=20$ combinations.

The mean values of the wavelength $\overline{\lambda}_i$ and the velocity dispersion $\overline{\sigma}_i$ obtained for each spectrum, along with the measurement dispersions (uncertainty ranges) $\Delta\overline\lambda_i$ and $\Delta\overline\sigma_i$, are listed in Tab.~\ref{single_values_sum} and shown in Fig.~\ref{fig_Ha_NII_results}. 
The latter, as well as the fractional uncertainties ($\Delta\overline\lambda_i$/$\overline{\lambda}_i$ and $\Delta\overline\sigma_i$/$\overline{\sigma}_i$) are plotted in Fig.~\ref{fig_Dispersion_lam_sig}.
We find that, for H$\alpha$, $\Delta\overline\lambda_i$ decreases from 0.3 \AA\  down to 0.03 \AA\ when moving from spectrum {\tt [a]} to {\tt [l]}, and a (much milder) decreasing trend is also found for the velocity dispersion uncertainty $\Delta\overline\sigma_i$, which varies from $\sim$0.4 down to $\sim$0.1. The respective fractional uncertainties for the wavelength parameter, $\Delta\overline\lambda_i$/$\overline\lambda_i$, thus cover the range of values between 4.9$\times$10$^{-5}$~\AA\ down to 3.9$\times$10$^{-6}$ \AA\ from spectrum {\tt [a]} to {\tt [l]}, while those obtained for the velocity dispersion, $\Delta\overline\sigma_i$/$\overline\sigma_i$, varies from $\sim$ 0.21 down to $\sim$0.08.
The opposite trend is derived for the [NII] line, reflecting the line intensity ratios, although more stable and constant values are obtained for both line wavelength and width for different spaxels.

It is important to bear in mind that, on the one hand, we are quantifying the dispersion of the measured parameters with respect to the average value, which may be different from the true solution. On the other hand, assuming that the estimated central wavelengths and line widths inferred from the various codes and combinations of stellar population synthesis (SPS) models are statistically independent is conservative, in the sense that any off--diagonal terms in the covariance matrix would likely reduce the quoted systematic uncertainties.
For these reasons, our estimate of the dispersion between codes may be actually regarded as a upper limit to the true systematic uncertainty associated to the choice of a specific method.

\begin{figure*}
\centering
\parbox{0.3\textwidth}
{
\includegraphics[width=0.27\textwidth]{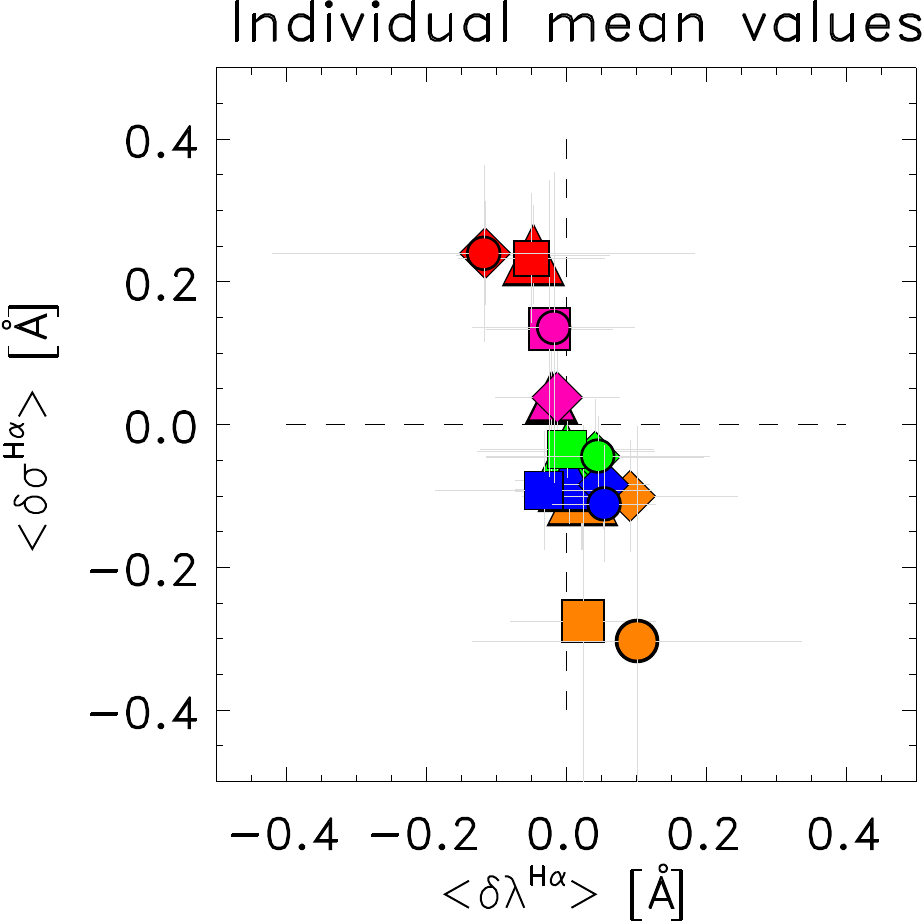}
}
\parbox{0.65\textwidth}
{
\centering
\includegraphics[width=0.65\textwidth]{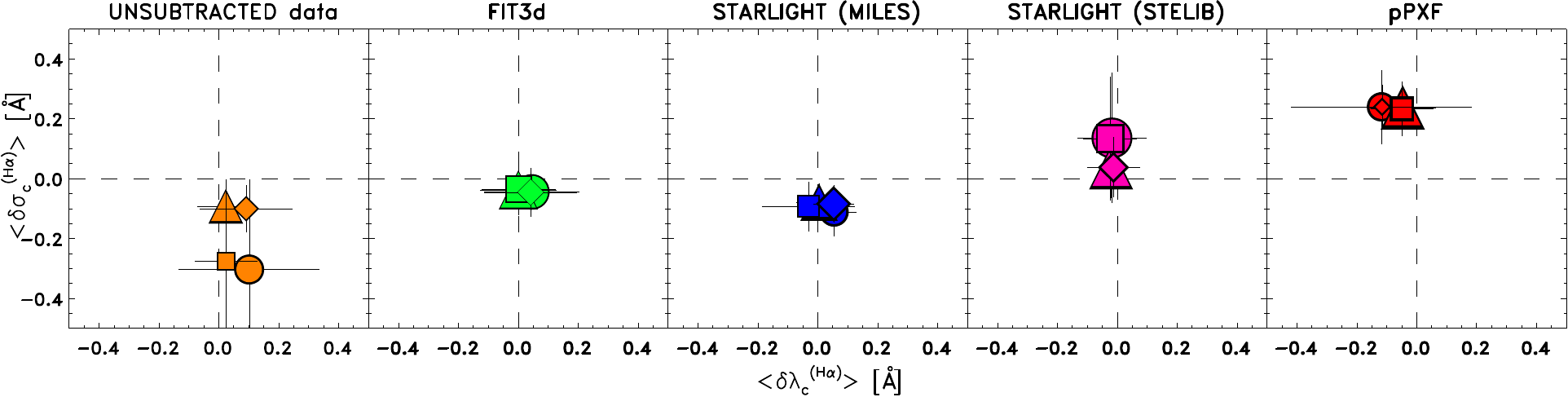}\\
\includegraphics[width=0.53\textwidth]{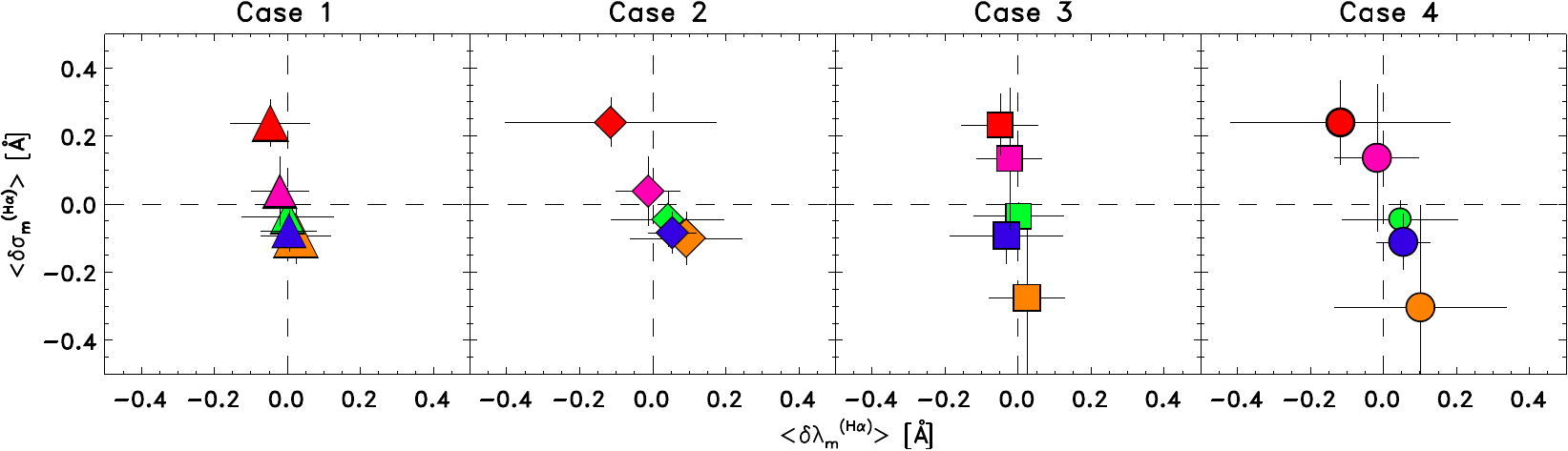}
}

\vspace*{5mm}
\parbox{0.3\textwidth}
{
\includegraphics[width=0.27\textwidth]{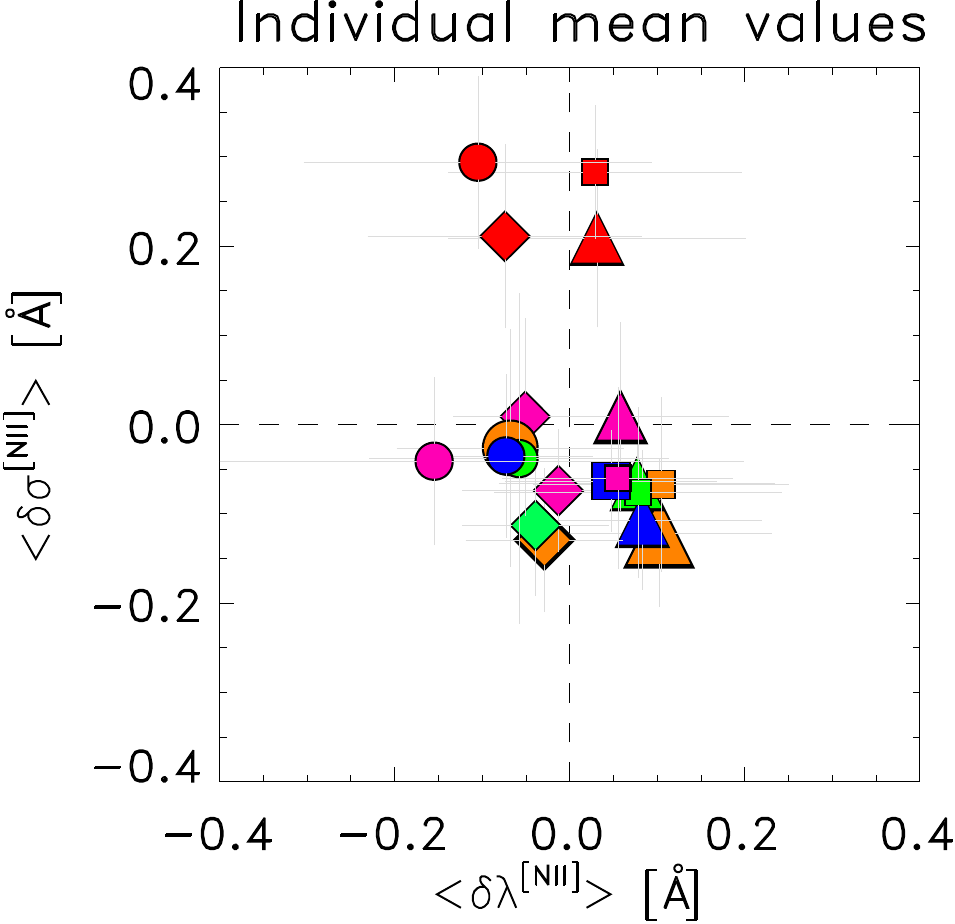}
}
\parbox{0.65\textwidth}
{
\centering
\includegraphics[width=0.65\textwidth]{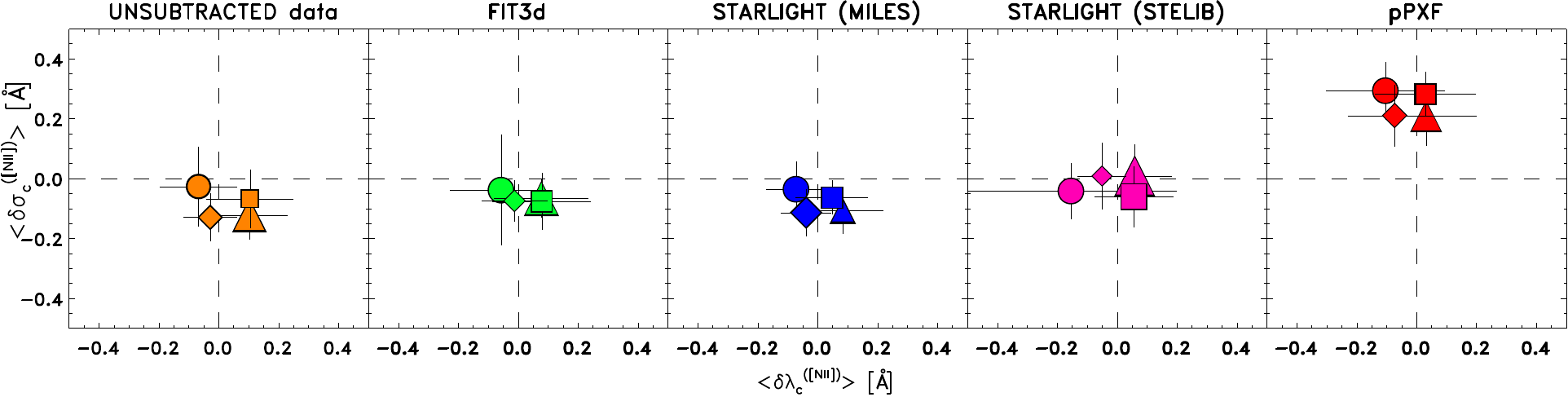}\\
\includegraphics[width=0.53\textwidth]{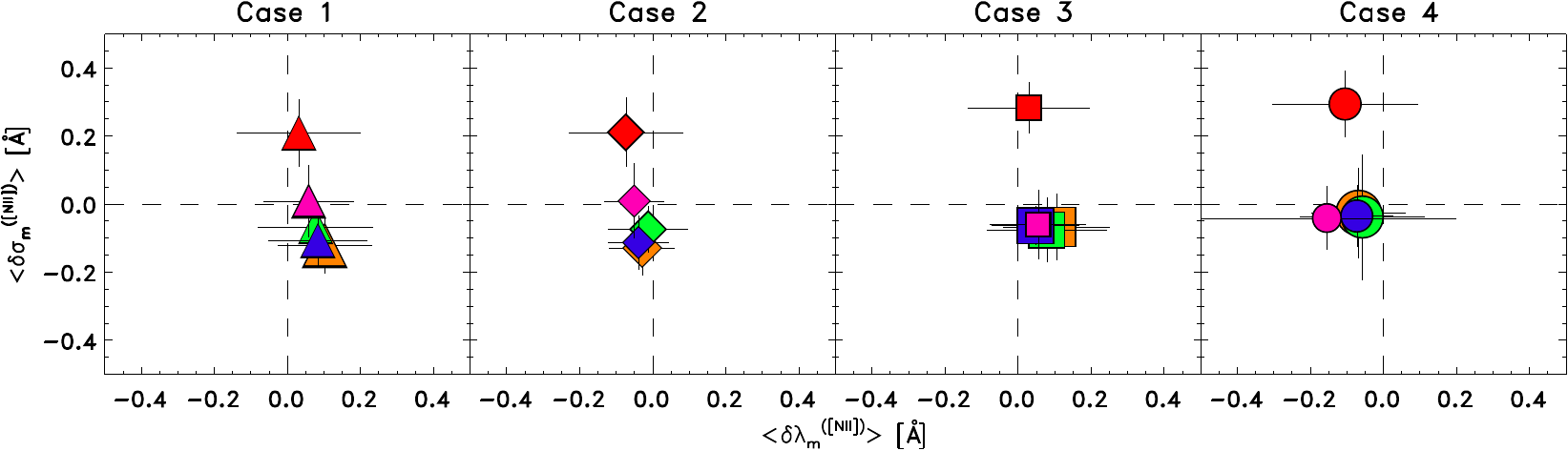}
}
\caption
{
{\it Left:} Mean kinematic results averaged over all individual spectra (see Tab.~\ref{tab_avg_over_spaxels}) for the H$\alpha$ (top) and [NII] (bottom) emission lines.
{\it Right:} Results for different methods/cases separately, as indicated, for the H$\alpha$ (top) and [NII] (bottom).
Symbol shapes and colors denote case and method, respectively. Symbol types are the same than those used in Fig.~\ref{fig_Ha_NII_results}.
}    
\label{fig_avg_over_spaxels}
\end{figure*}

\begin{table*}
\centering
\begin{small}
\begin{tabular}{cl|ccccc} % Column formatting, @{} suppresses leading/trailing space
\hline\hline
& Case &&& Method \\
&				& {\tt Unsubtracted} & {\tt FIT3D} & {\tt STARLIGHT (MILES)} & {\tt STARLIGHT (STELIB)} & {\tt pPXF }\\
\hline\hline
$\langle\delta\lambda_{mc}^{\rm H\alpha}\rangle$ 
& {\tt [1]}   &    0.023 $\pm$  0.043  &    0.000 $\pm$ 0.057   &    0.004 $\pm$ 0.034  &  --0.021 $\pm$ 0.036  &  --0.047 $\pm$ 0.049	  \\
 $\pm$
& {\tt [2]}   &     0.091 $\pm$  0.069  &    0.041 $\pm$ 0.069   &    0.053 $\pm$ 0.029  &  --0.013 $\pm$ 0.039  &  --0.12 $\pm$ 0.13	 \\
  $\Delta\lambda_{mc}^{\rm H\alpha}$
& {\tt [3]}   &   0.024 $\pm$  0.047  &    0.001 $\pm$ 0.055   &   --0.032 $\pm$ 0.069  &  --0.024 $\pm$ 0.0402 &  --0.050 $\pm$ 0.047	 \\
{\tiny [\AA]}
& {\tt [4]}   &   0.101 $\pm$  0.105  &    0.045 $\pm$ 0.072   &    0.054 $\pm$ 0.033  &  --0.018 $\pm$ 0.052  &   --0.28 $\pm$ 0.14	  \\
\hline
$\langle\delta\sigma_{mc}^{\rm H\alpha}\rangle$                                                                                                                 
 & {\tt [1]}    &    --0.092 $\pm$ 0.037	&	 --0.037 $\pm$ 0.038  &  --0.078 $\pm$ 0.027  &    0.038 $\pm$ 0.046 & 0.238 $\pm$ 0.031   \\
 $\pm$
& {\tt [2]}    &  --0.100 $\pm$ 0.035	&	 --0.045 $\pm$ 0.036  &  --0.084 $\pm$ 0.028  &    0.038 $\pm$ 0.045 & 0.240 $\pm$ 0.033   \\
  $\Delta\sigma_{mc}^{\rm H\alpha}$
 & {\tt [3]}    &   --0.28 $\pm$ 0.12	&	 --0.035 $\pm$ 0.018  &  --0.092 $\pm$ 0.037  &    0.134 $\pm$ 0.093 & 0.233 $\pm$ 0.041 \\
 {\tiny [\AA]}
 & {\tt [4]}    & --0.31 $\pm$ 0.13	&	 --0.044 $\pm$ 0.025  &  --0.11  $\pm$ 0.037  &    0.136 $\pm$ 0.097 & 0.240 $\pm$ 0.055  \\
\hline\hline
$\langle\delta\lambda_{mc}^{\rm [NII]}\rangle$               
& {\tt [1]}   &   0.102 $\pm$ 0.058	 	&   0.077 $\pm$ 0.071     &  0.083 $\pm$ 0.061  	&  0.058 $\pm$ 0.055  	&  0.031  $\pm$ 0.076 \\ 
 $\pm$
& {\tt [2]}   &--0.0290 $\pm$ 0.0402   	&  --0.013 $\pm$ 0.049     & --0.039 $\pm$ 0.037  	& --0.051 $\pm$ 0.037  	& --0.0740 $\pm$ 0.0702  \\ 
$\Delta\lambda_{mc}^{\rm [NII]}$
& {\tt [3]}   &   0.104 $\pm$ 0.065     		&   0.078 $\pm$ 0.073     &  0.047 $\pm$ 0.054  	&  0.055 $\pm$ 0.059 	&  0.029  $\pm$ 0.075 \\ 
{\tiny [\AA]}
& {\tt [4]}   &--0.068 $\pm$ 0.058	 	&  --0.058 $\pm$ 0.076     & --0.073 $\pm$ 0.044  	& --0.16 $\pm$ 0.16    	& --0.105  $\pm$ 0.089 \ \\ 
\hline             
$\langle\delta\sigma_{mc}^{\rm [NII]}\rangle$
& {\tt [1]}   &  --0.121 $\pm$ 0.037	&  --0.066 $\pm$ 0.028  &   --0.107 $\pm$ 0.034 &   0.009 $\pm$ 0.047 &  0.209 $\pm$ 0.045   \\
 $\pm$
 & {\tt [2]}   & --0.129 $\pm$ 0.036	&  --0.074 $\pm$ 0.031  &   --0.113 $\pm$ 0.035 &   0.009 $\pm$ 0.049 &  0.211 $\pm$ 0.046 \\
 $\Delta\sigma_{mc}^{\rm [NII]}$
& {\tt [3]}   & --0.067 $\pm$ 0.043	&  --0.076 $\pm$ 0.043  &   --0.063 $\pm$ 0.025 &  --0.060 $\pm$ 0.045 &  0.283 $\pm$ 0.033 \\
{\tiny [\AA]}
& {\tt [4]}   &  --0.026 $\pm$ 0.059	&  --0.038 $\pm$ 0.083  &   --0.035 $\pm$ 0.041 &  --0.041 $\pm$ 0.042 &  0.294 $\pm$ 0.043 \\
\hline\hline
\end{tabular}
\end{small}\vskip2mm
\caption{
Mean and standard deviation of the mean kinematic results averaged over all individual spectra.
}
\label{tab_avg_over_spaxels}
\end{table*}

% -----------------------------------------------------------------------------
\subsection{Trends for different methods and cases}
\label{General_results}

Figure~\ref{fig_avg_over_spaxels} shows the mean and standard deviation

\begin{eqnarray}
\langle \delta\lambda_{mc} \rangle = \frac{ \sum_i \delta\lambda_{i, mc} }{ N_i }
\!\!\!&\!\!\!;\!\!\!&
\Delta\lambda_{mc} = \sqrt{ \frac{ \sum_i (\delta\lambda_{i, mc} - \langle\delta\lambda_{mc}\rangle)^2 }{ N_i } }\\
\langle \delta\sigma_{mc} \rangle = \frac{ \sum_i \delta\sigma_{i, mc} }{ N_i }
\!\!\!&\!\!\!;\!\!\!&
\Delta\sigma_{mc} = \sqrt{ \frac{ \sum_i (\delta\sigma_{i, mc} - \langle\delta\sigma_{mc}\rangle)^2 }{ N_i } }
\end{eqnarray}

averaged over the $N_i=12$ spaxels, for each emission line, stellar continuum subtraction method, and line fitting case.
Numerical values are reported on Tab.~\ref{tab_avg_over_spaxels}.

\begin{figure}[!h]
  \centering
   \includegraphics[width=0.46\textwidth]{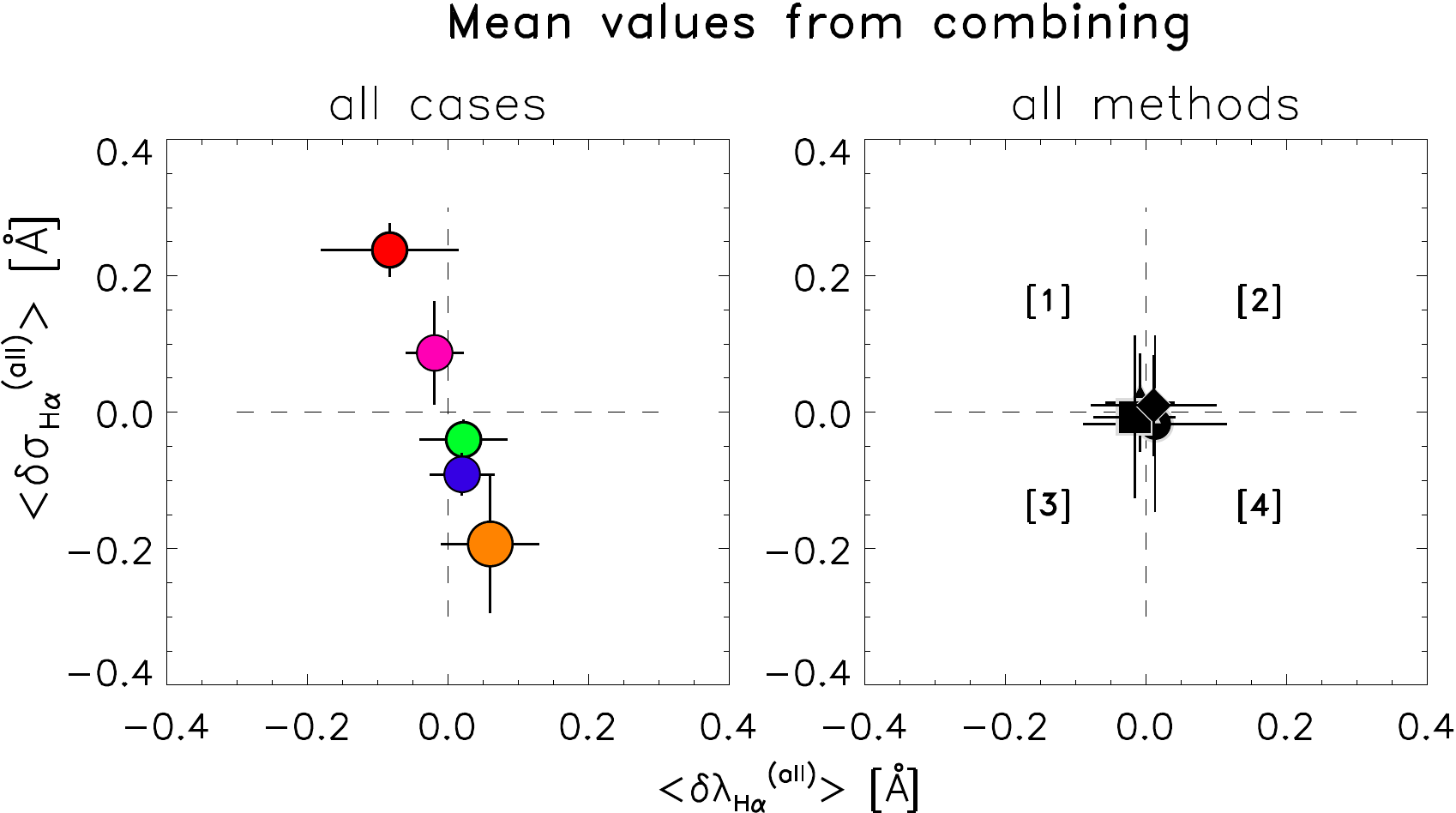}\vskip3mm
\vskip-2mm   \includegraphics[width=0.46\textwidth]{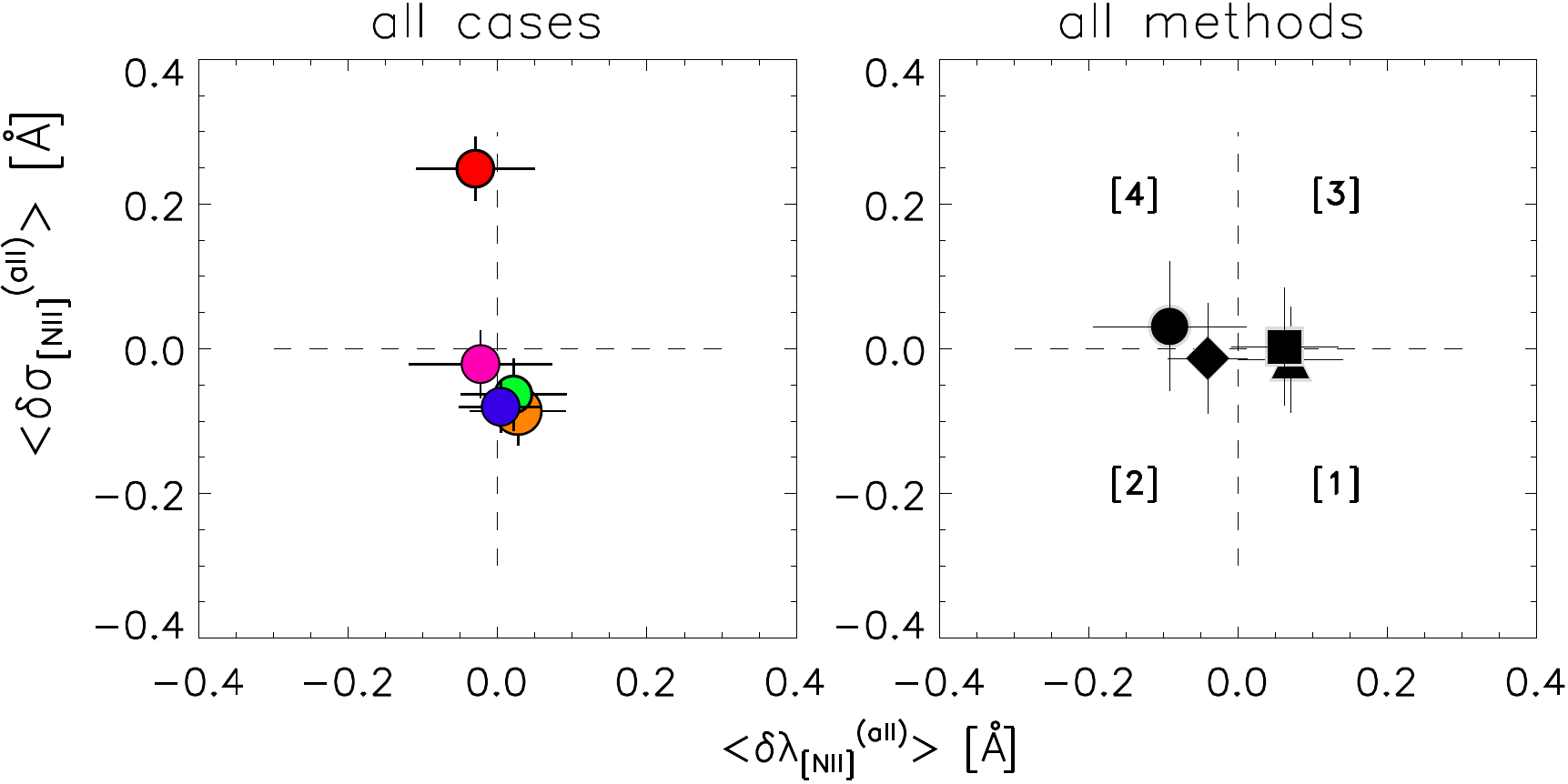}\vskip3mm
\vskip-2mm     \includegraphics[width=0.46\textwidth]{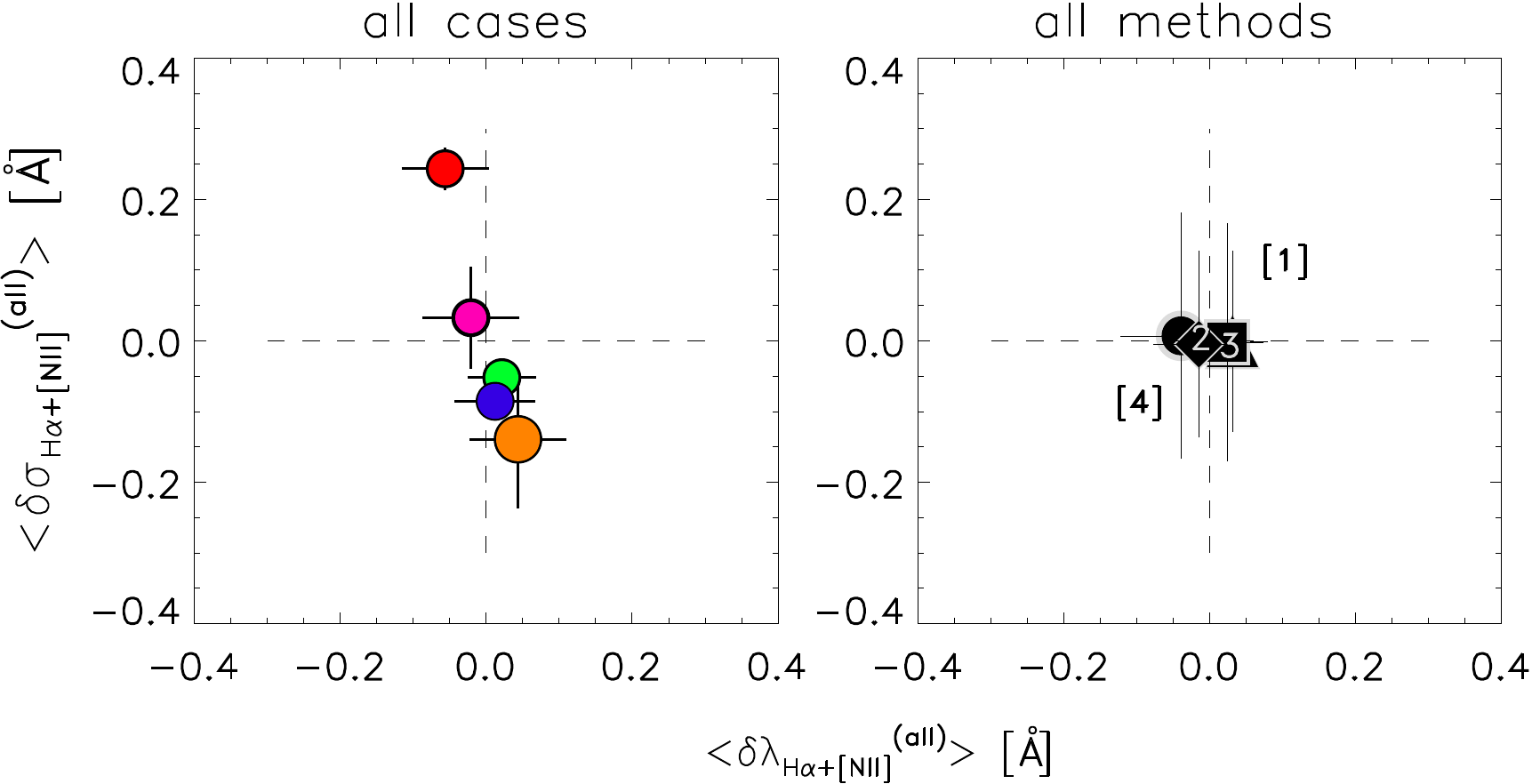}
\caption{\small {\it Top and middle:} Mean H$\alpha$ and [NII] deviations obtained when averaging over all cases (left) and over all methods (right). {\it Bottom}: Mean deviations of the H$\alpha$--[NII] complex for all cases (left) and all methods (right).}    
\label{3_Plots_all}
\end{figure}

In Tabs.~\ref{mean_allcases_samemet} and \ref{mean_allmeth_samemet} the results from further averaging over the $N_m=5$ methods 

\begin{eqnarray}
\langle \delta\lambda_{c} \rangle = \frac{ \sum_m \delta\lambda_{mc} }{ N_m }
\!\!\!&\!\!\!;\!\!\!&
\Delta\lambda_{c} = \sqrt{ \frac{ \sum_{im} (\delta\lambda_{i, mc} - \langle\delta\lambda_{c}\rangle)^2 }{ N_i\,N_m } }\\
\langle \delta\sigma_{c} \rangle = \frac{ \sum_m \delta\sigma_{mc} }{ N_m }
\!\!\!&\!\!\!;\!\!\!&
\Delta\sigma_{c} = \sqrt{ \frac{ \sum_{im} (\delta\sigma_{i, mc} - \langle\delta\sigma_{c}\rangle)^2 }{ N_i\,N_m } }
\end{eqnarray}

for each line fitting case, as well as over the $N_c=4$ cases

\begin{eqnarray}
\langle \delta\lambda_{m} \rangle = \frac{ \sum_c \delta\lambda_{mc} }{ N_c }
\!\!\!&\!\!\!;\!\!\!&
\Delta\lambda_{m} = \sqrt{ \frac{ \sum_{ic} (\delta\lambda_{i, mc} - \langle\delta\lambda_{m}\rangle)^2 }{ N_i\,N_c } }\\
\langle \delta\sigma_{m} \rangle = \frac{ \sum_c \delta\sigma_{mc} }{ N_c }
\!\!\!&\!\!\!;\!\!\!&
\Delta\sigma_{m} = \sqrt{ \frac{ \sum_{ic} (\delta\sigma_{i, mc} - \langle\delta\sigma_{m}\rangle)^2 }{ N_i\,N_c } }
\end{eqnarray}

at fixed stellar subtraction method are considered, respectively. These results are shown in Fig.~\ref{3_Plots_all}.

\begin{table*}
\centering

\begin{tabular}{c|ccccc} 
\hline\hline
 &   \makebox{ {\tt Unsubtracted}}  &   \makebox{  {\tt FIT3D} }  &   \makebox{{\tt STARLIGHT (MILES)} }  &   \makebox{{\tt STARLIGHT (STELIB)}}   &   \makebox{{\tt pPXF }}\\
\hline
$\delta\lambda_m^{\rm H\alpha}$  &    0.060 $\pm$  0.079  		&   0.022 $\pm$ 0.070  &   0.020 $\pm$ 0.052 	&   --0.019 $\pm$ 0.046   & --0.083 $\pm$ 0.109	\\	
$\delta\sigma_m^{\rm H\alpha}$    &   --0.19 $\pm$  0.11  	&  --0.040 $\pm$ 0.033  &  --0.091 $\pm$ 0.034 	&    0.087 $\pm$ 0.085   &  0.238 $\pm$ 0.045	\\	
\hline                                                                                                                                              
 $\delta\lambda_m^{\rm [NII]}$  &    0.027 $\pm$  0.072  		&   0.021 $\pm$ 0.079  &   0.004 $\pm$ 0.063 	&   --0.023 $\pm$ 0.108   & --0.030 $\pm$ 0.089	\\	
 $\delta\sigma_m^{\rm [NII]}$   &   --0.086 $\pm$  0.054  		&  --0.063 $\pm$ 0.056  &  --0.080 $\pm$ 0.041 	&    0.021 $\pm$ 0.052   &  0.249 $\pm$ 0.051	\\	
\hline\hline
\end{tabular}
\vskip2mm
\caption{Mean and root mean square values for each stellar subtraction method.}
\label{mean_allcases_samemet}
\end{table*}

\begin{table*}
\centering
\begin{tabular}{c|cccc} 
\hline\hline
 &   \makebox{ {\tt case 1 }}  &   \makebox{  {\tt case 2} }  &   \makebox{{\tt case 3} }  &   \makebox{{\tt case 4}} \\
 \hline
$\delta\lambda_c^{\rm H\alpha}$   & 		--0.008 $\pm$ 0.045		&  0.011 $\pm$ 0.081	& --0.016 $\pm$ 0.052   &   0.013 $\pm$ 0.091 \\
$\delta\sigma_c^{\rm H\alpha}$     & 		 0.014 $\pm$ 0.065		&  0.010 $\pm$ 0.066	& --0.007 $\pm$ 0.107   &  --0.017 $\pm$ 0.116 \\
\hline
$\delta\lambda_c^{\rm [NII]}$      & 		 0.070 $\pm$ 0.063		&	 --0.041 $\pm$ 0.048	 &  0.062 $\pm$ 0.064  & --0.092 $\pm$ 0.092   \\
$\delta\sigma_c^{\rm [NII]}$       & 		 --0.015 $\pm$ 0.066	&    --0.013 $\pm$ 0.068	 &  0.003 $\pm$ 0.073 &   0.031 $\pm$ 0.081  \\
\hline\hline
\end{tabular}
\vskip2mm
\caption{Mean and root mean square values for each line fitting case.}
\label{mean_allmeth_samemet}
\end{table*}

Our results show clear systematic trends depending on the stellar continuum subtraction algorithm.
The measurements obtained from the {\tt FIT3d} method and those derived after applying the {\tt STARLIGHT (MILES)} method are relatively similar, and both prescriptions tend to yield slightly larger wavelengths and lower widths (with respect to the mean values) for either line, regardless of the line fitting case.
On the other hand, higher line widths are found when the {\tt STARLIGHT (STELIB)} and {\tt pPXF} algorithms are used, with a mild bias towards lower wavelengths.

In particular, for the H$\alpha$ line the wavelength deviations follow a decreasing trend, from higher to lower wavelengths (from 0.06 \AA\ to --0.083 \AA) when moving from the {\tt Unsubtracted} to the {\tt pPXF} methods. On the other hand, the width deviations increase from a mean value of --0.19 \AA\ for the {\tt Unsubtracted} method up to $\sim$ 0.24 \AA\ for the {\tt pPXF} one.
When taking into account the [NII] line, all but {\tt pPXF} method derived similar wavelength and width deviations with respect to the total mean values. However, they still follow a similar trend (i.e., decreasing or increasing if we consider the wavelength or width deviations, respectively)  than that shown for the H$\alpha$ line but covering a smaller range of values (see Tab.~\ref{mean_allcases_samemet}). 
It is important to note, once again, that this does \emph{not} necessarily imply that they are closer to the \emph{unknown} true solution.

Focusing on the results derived from the {\tt STARLIGHT} code with different stellar libraries (i.e., {\tt MILES, STELIB}) and comparing with the measurements obtained with {\tt pPXF} based on the {\tt MILES} library, it is evident that the details of the minimization algorithm are at least as important as the choice of spectral basis. 
Starting from the same data, each program converges to a different solution.
Subtle differences in the residual spectra are apparent in Fig.~\ref{fig_spectra_all_fixed}, and it is not surprising that these differences propagate into the final kinematic parameters.
In our case, the difference when using the same method with different stellar library ({\tt STARLIGHT} using {\tt MILES} or {\tt STELIB}) or the same stellar library in different stellar subtraction method ({\tt STARLIGHT and pPXF} using {\tt MILES}) is at least a factor of $\times$2, deriving a larger gap among the kinematic values ($\lambda$, $\sigma$) when different methods are considered.

Regarding the line fitting approach, we found a certain similarity between the cases that share the same kinematics for the velocity dispersion parameter: i.e., when $\sigma$ is fixed ({\tt cases [1]} and {\tt [2]}) and when $\sigma$ is free to vary ({\tt cases [3]} and {\tt [4]}). We derived the same trends for the wavelength and the velocity dispersion of both H$\alpha$ and [NII] emission lines.
The magnitude of the uncertainties is similar to those stemming from the stellar subtraction method, but the effect is much more random rather than systematic.

In general, the two lines yield consistent measurements, although the differences in the recovered kinematics arising from using different methods are somewhat stronger for the Balmer line.
For [NII], all methods result in fairly similar values, with the only possible exception of the velocity dispersion obtained when {\tt pPXF} is used to subtract the stellar continuum.
Velocity dispersion is actually more sensitive to the adopted prescription than the mean velocity derived from the line centroid, causing the larger spread already observed in Figs~\ref{fig_Ha_NII_results} and~\ref{fig_Dispersion_lam_sig}.

The kinematic results obtained from fitting the H$\alpha$--[NII] complex (see Fig.~\ref{3_Plots_all}, bottom) show a more robust determination of the global kinematic properties with respect to the adopted line fitting approach (right panel), as the strongest (and thus more stable) line tends to dominate the result, and random deviations in each individual line tend to compensate each other on average \citep[see also][]{Belfiore19}. However, the systematic differences arising from the continuum subtraction method (left panel) persist, as they are present in both emission lines.

% -----------------------------------------------------------------------------
\section{Discussion}
\label{discuss_}
% -----------------------------------------------------------------------------

\begin{figure*}
   \centering
  \hskip-9mm\includegraphics[width=0.8\textwidth]{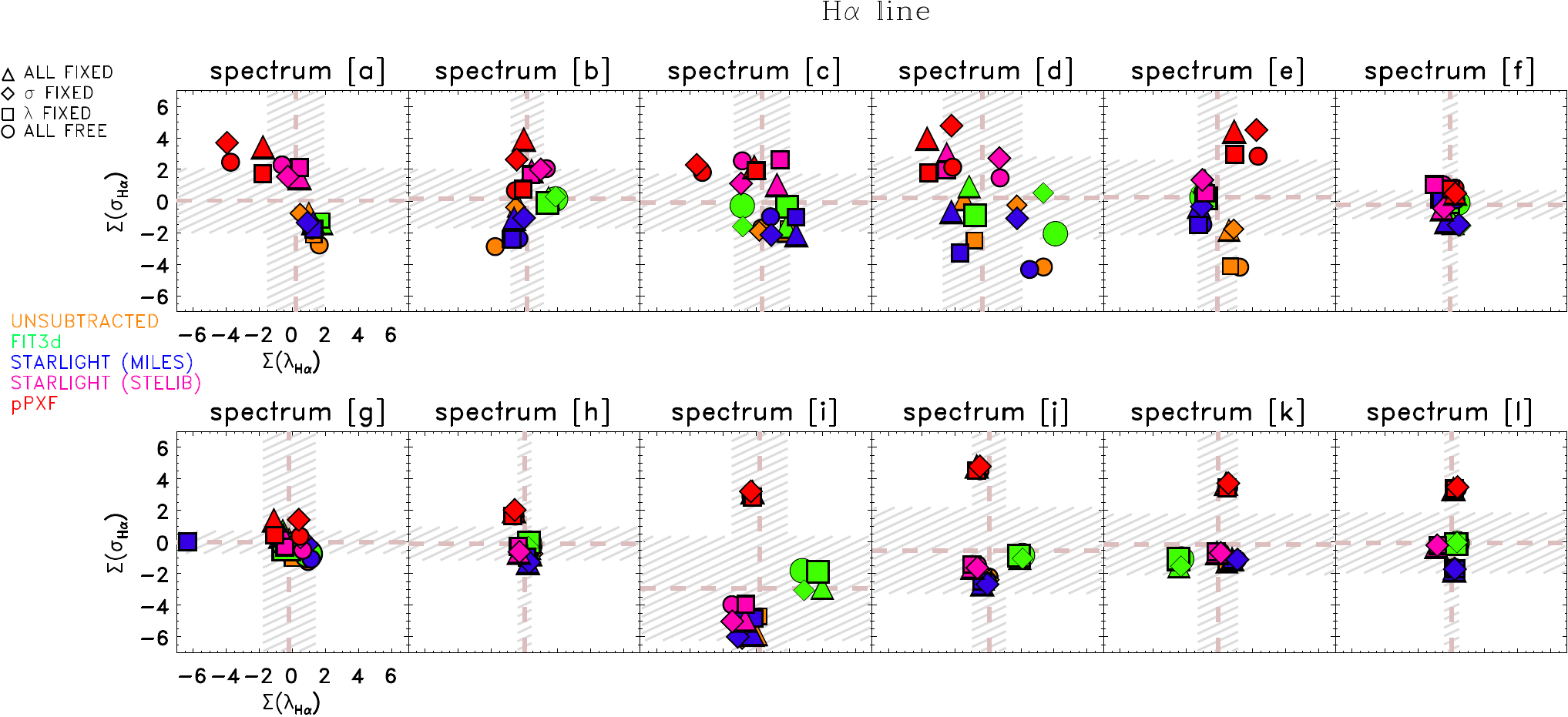}
  \vskip3mm
     \includegraphics[width=0.75\textwidth]{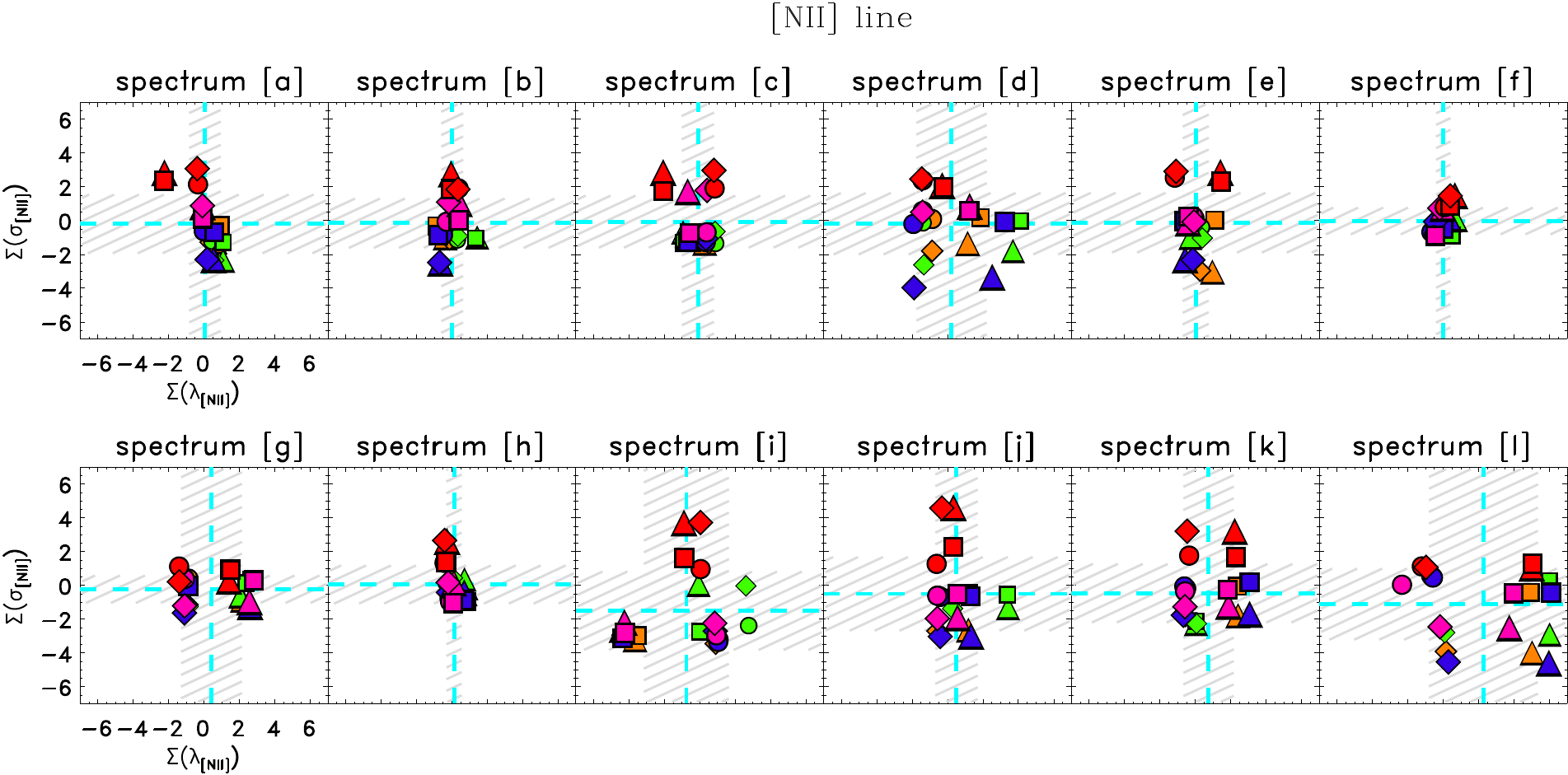}
\caption{\small Kinematic parameter deviations, $\Sigma$($\lambda_{i, mc}$) and $\Sigma$($\sigma_{i, mc}$) computed with respect to their mean values and normalized to the line fitting errors derived using {\tt IDL} for each individual spectrum. The mean deviations as well as their standard deviations are represented using dashed lines (light red for the H$\alpha$ and cyan for the [NII] lines) and the shaded area in grey, respectively.
{\it Top panels:} H$\alpha$ results. {\it Bottom panels:} [NII] results.}
\label{Sigma_Ha_NII_factor}
\end{figure*}

\begin{figure*}
   \centering
  \hskip-1.1cm\includegraphics[width=0.8\textwidth]{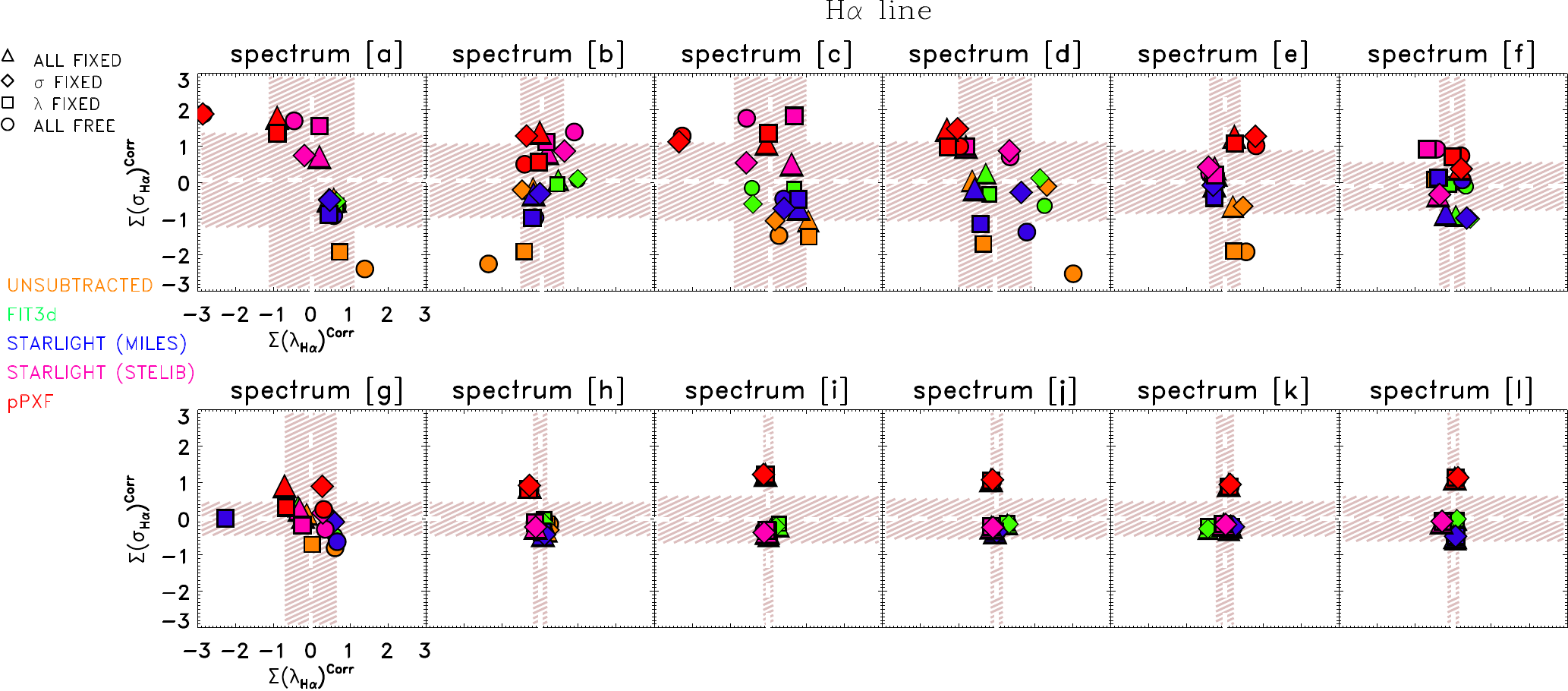}
   \vskip3mm  
    \includegraphics[width=0.74\textwidth]{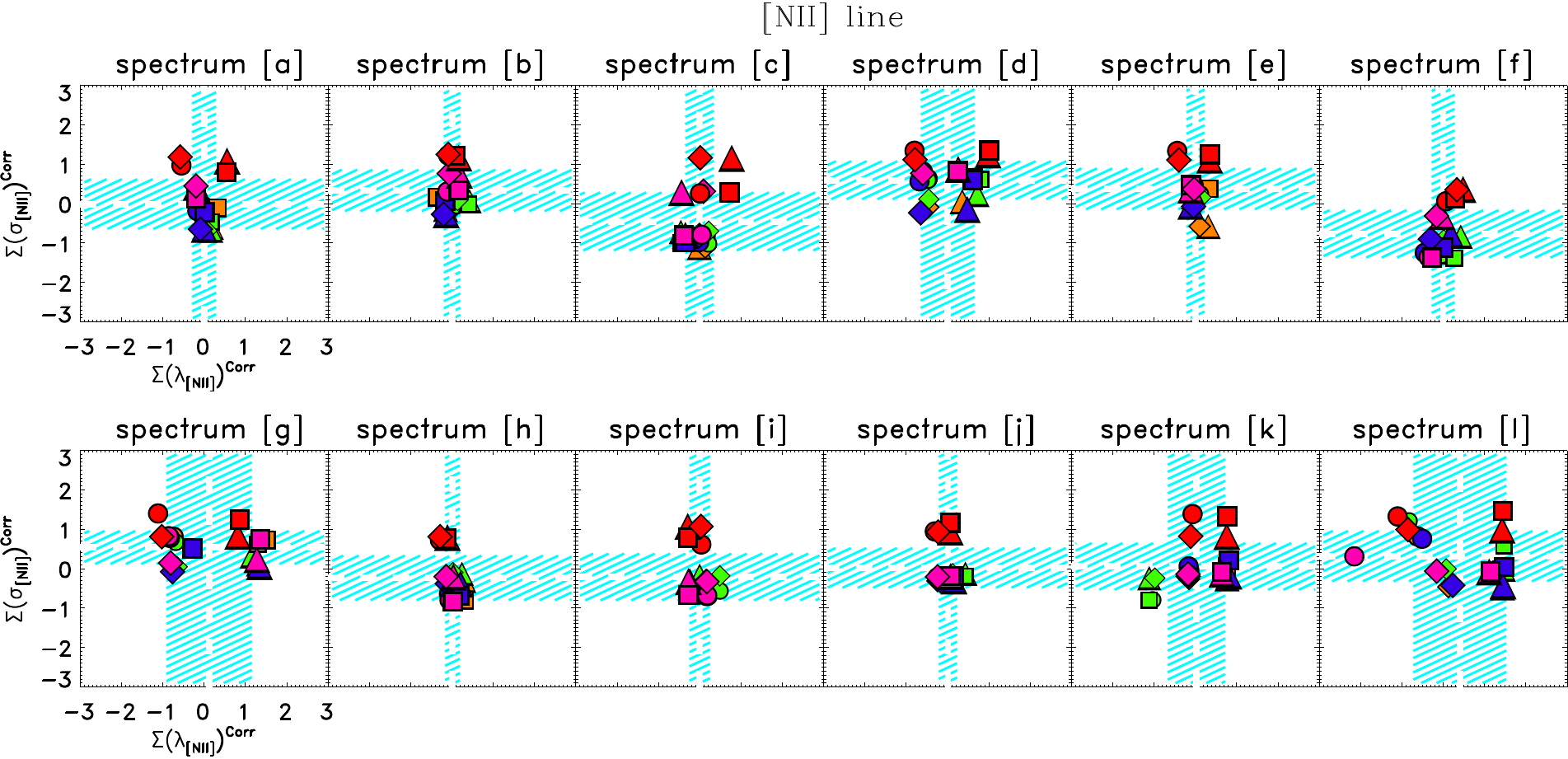}
   \vskip-1mm
\begin{minipage}[h]{18cm}
\caption{\small Corrected deviations $\Sigma$($\lambda_{i, mc}$)$^{i, mc}$ and $\Sigma$($\sigma_{i, mc}$)$^{Corr}$ for the H$\alpha$ ({\it top}) and [NII] ({\it bottom}) lines derived from applying the same mean dispersion factor ($\Delta\overline\lambda$ = $\Delta\overline\sigma\sim$ 0.2 \AA; see also Fig.~\ref{fig_Dispersion_lam_sig}) for the wavelength and the velocity dispersion. In this correction the dispersion generated from using different stellar subtraction methods and line fitting cases is included. The error bars are now of the order $\lesssim$ 1 \AA\ for all spectra and for both parameters (shaded light red and cyan areas for the H$\alpha$ and [NII] lines, respectively). The mean corrected deviations are represented using dashed white lines for both lines.}
\label{Ha_NII_results_corr}
\end{minipage}
\end{figure*}

In order to gauge whether these uncertainties are statistically compatible with the errors returned by the {\tt MPFITEXPR} routine, we have normalized (i.e., standardized) the deviation of each individual measurement from the mean ($\delta\lambda_{i, mc}$, $\delta\sigma_{i, mc}$) using the line fitting errors derived from the output covariance matrix ($\Delta\lambda_{i, mc}$, $\Delta\sigma_{i, mc}$):
\begin{equation}
\hskip5mm\Sigma\lambda_{i, mc} = \frac{\delta\lambda_{i, mc}}{\Delta\lambda_{i, mc}}
~; ~~
\Sigma\sigma_{i, mc} = \frac{\delta\sigma_{i, mc}}{\Delta\sigma_{i, mc}}.
\label{Corrections}
\end{equation}

The normalized deviations give us an estimate of the `number of $\sigma$' that each result departs from the mean value, assuming that the line fitting errors should be representative of the expected standard deviation.
As can be readily seen in Fig.\ref{Sigma_Ha_NII_factor}, the dispersion of these variables is often well above unity, as expected from Gaussian statistics, for both H$\alpha$ and [NII] lines.

In the next step we modified the Eq. \ref{Corrections} including along with the line fitting errors an extra correction term accounting for the contribution coming from the choice of a specific stellar subtraction methods ($\Delta\overline\lambda_i$ and $\Delta\overline\sigma_i$) in the computation of the standard deviation. The new (i.e., corrected) formulas are then modified:

\begin{equation}
\Sigma(\lambda_{i, mc})^{Corr} = \frac{\delta\lambda_{i, mc}}{\Delta\lambda_{i, mc}^{Corr}} \hskip1mm; 
\hskip2mm\Sigma(\sigma_{i, mc})^{Corr} = \frac{\delta\sigma_{i, mc}}{\Delta\sigma_{i, mc}^{Corr}}.
\end{equation}

where

\begin{equation}
\hskip0mm \Delta\lambda_{i, mc}^{Corr} = \sqrt{\Delta\lambda_{i, mc}^2 + \Delta\overline\lambda_i^2};
\hskip1mm \Delta\sigma_{i, mc}^{Corr} = \sqrt{\Delta\sigma_{i, mc}^2 + \Delta\overline\sigma_i^2}.
\end{equation}

For both lines we derive a mean dispersion of $\Delta\bar\lambda\sim$~0.13~\AA\ for the wavelength while $\Delta\bar\sigma\sim$ 0.2 \AA\ for the velocity dispersion. In order to consider a general value for both parameters and for both lines we take into account the highest dispersion value derived for the wavelength and velocity dispersion parameters ($\sim$0.2 \AA), as derived for the velocity dispersion of the H$\alpha$ line.

As can be seen in in Fig.~\ref{Ha_NII_results_corr}, applying the same mean value for $\Delta\bar\lambda$ and $\Delta\bar\sigma$ we finally derived smaller dispersions (of order unity) for both axes and for both lines. Within the (corrected) 1-$\sigma$ error bar, the results of all methods can be considered statistically consistent with each other. 

We then propose the following prescription to compute the correction factor (CF) that needs to be considered to take into account the uncertainty associated to the choice of a specific stellar subtraction method. The formula can be related to the instrumental profile of the instrument, $\Delta_{\tt MUSE}$:

\vskip-3mm
\begin{equation}
\hskip4mm\Delta\overline\lambda = \Delta\overline\sigma = CF \times \frac{{\tt FWHM}_{\hskip1mminstr}^{\tt \hskip2mmMUSE}}{2.354}= CF \times \Delta_{\tt MUSE}
\end{equation}

In the case of the {\tt MUSE} data the (theoretical) instrumental resolution is of the order of 1.1 \AA\ (FWHM=2.6~\AA). Considering the mean uncertainty value of 0.2~\AA, the correction factor (CF) can be derived as:

\begin{equation}
\hskip4mm CF = \frac{\Delta\overline\lambda}{\Delta_{\tt MUSE}} \lesssim \frac{0.2}{1.1} = 0.18 \sim 0.2. 
\end{equation}

In this particular case, the (observed, computed) instrumental resolution $\Delta_{\tt MUSE}$ is of the order of 0.9 \AA, deriving a CF of the order of $\sim$0.22.
To check the validity of this equation the same analysis should be applied to other local sources observed with {\tt MUSE}, as well as other instruments, in order to test whether our simple prescription can be extended to other instrumental resolutions.

% -----------------------------------------------------------------------------
\section{Summary and Conclusions}
% -----------------------------------------------------------------------------

We have compared the kinematic results obtained when applying three different stellar subtraction methods to a sample of twelve spectra in the nearby normal star--forming galaxy NGC2906, observed with {\tt MUSE} instrument on the VLT. The spectra have been selected according to their different ionization phases, located in different parts of the galaxy, from the nuclear regions to the outer spiral arms. 

In order to study and compare the kinematics of the gas component, as that derived from applying the stellar subtraction methods, with that derived from the unsubtracted (gas and stars) one, we have considered the following stellar subtraction methods: {\tt FIT3D, STARLIGHT} and {\tt pPXF}. 
Each of them has been combined with a specific stellar library: {\tt MILES} for the {\tt STARLIGHT} and {\tt pPXF} methods, {\tt STELIB} for the {\tt STARLIGHT} method and a combination of two stellar libraries, {\tt GRANADA \& MILES} for the {\tt FIT3D} one.
From their combination we obtained four different data sets of the gaseous ionized component. 
To better characterize the dispersion when deriving the kinematic parameters we also include different line fitting approaches. In particular, we have considered all possible ways to combine the wavelength and the velocity dispersion parameters allowing them to be free to vary or letting them fixed according to the atomic physics. 
 
The main results of the present study can be summarized as follows:
    
\begin{itemize}

\item[$\bullet$] When comparing the kinematic deviations (for both H$\alpha$ and [NII] lines) derived for the raw unsubtracted data (gas and star; no stellar subtraction is applied) with those derived when applying different stellar subtraction methods for the whole sample of spectra we see that the kinematic results obtained with {\tt FIT3D} are very similar to that derived in the {\tt STARLIGHT (MILES)} case whereas high kinematic deviations are found when using the {\tt STARLIGHT (STELIB)} and {\tt pPXF};
	
\vskip1.5mm
\item[$\bullet$] When averaging over all cases, the velocity dispersion $\sigma$ is the parameter most affected by the choice of the different combinations {\tt method--line fitting}, whose deviations range between $^{+0.3}_{-0.3}$ \AA, while the wavelength $\lambda$ parameter is less affected ($<$$^{+0.1}_{-0.2}$~\AA\ or $^{+5}_{-9}$ km s$^{-1}$). 

\vskip2mm
\item[$\bullet$] When averaging over all methods, the kinematic deviations obtained for the H$\alpha$ and [NII] lines ($\langle\delta\lambda_m\rangle$, $\langle\delta\sigma_m\rangle$) show {\it point--}symmetric distribution for the {\tt cases~[1]--[4]} and the {\tt cases~[2]--[3]}. 

\vskip2mm
\item[$\bullet$] When analyzing the kinematic results for each individual spectrum we derive larger dispersion (error bars) in the kinematic parameters when the line taken into account show a low S/N ratio: this happened for both H$\alpha$ and [NII] lines, deriving a typical value of $\lesssim$ 0.2 \AA. They represent the systemic error as due to the selection of a specific stellar subtraction method.

\end{itemize}

The implication of these results allow us to draw the following conclusions:

\begin{enumerate}

\item In order to properly derive the kinematics of the ionized gas component the subtraction of the stellar continuum is not crucial, although it is when the real flux of the line has to be recovered; however, the determination of the H$\alpha$ width is quite affected from the choice of the stellar subtraction method. This factor has to be considered in presence of outflow; 

\vskip1mm
\item 
Any method (including none) can be used to measure the gas kinematics as long as an additional term ($\Delta\overline\lambda$ = $\Delta\overline\sigma$ $\sim$ 0.2~$\Delta_{\tt MUSE}$) is added to the error budget. 
Further analysis with other local sources observed with {\tt MUSE}, as well as other instruments, should be considered to establish the validity of this equation;

\vskip1mm
\item
The kinematic results obtained from combining the line fitting results of the H$\alpha$--[NII] complex show a robust determination of the global kinematic properties, as the strongest (and thus more stable) line tends to dominate the result and the most extreme deviations, which could be derived when individual lines are considered, are thus diluted. This is supported by the fact that the [NII] line is not affected by stellar absorption.

\end{enumerate}

\begin{acknowledgements}
We thank the anonymous referee for useful comments and suggestions which helped us to improve the quality and presentation of the manuscript. 
This paper is based on observations carried out at the European Southern Observatory, Paranal (Chile), Program 60.A-9100(B) with MUSE at Very Large Telescope (VLT).
This research made use of the NASA/IPAC Extragalactic Database (NED), which is operated by the Jet Propulsion Laboratory, California Institute of Technology, under contract with the National Aeronautics and Space Administration. 
EB acknowledges support from the Spanish Ministerio de Econom\'ia y Competitividad through the grant AYA2016-79724-C4-1-P as well as the exchange programme `Study of Emission--Line Galaxies with Integral--Field Spectroscopy' (SELGIFS, FP7--PEOPLE--2013--IRSES--612701), funded by the EU through the IRSES scheme. YA is also supported by the Ram\'on y Cajal programme (RyC--2011--09461; Mineco, Spain). 
HIM wants to acknowledge the Mexican CONACyT grant (Ciencia B\'asica) 285080 and PAPIIT--DGAPA--IA104118 (UNAM) project for funding the develop of the Pipe3D code. SFS thanks Conacyt CB--285080, FC--2016--01--1916 and PAPIIT--DGAPA--IN100519 projects. \end{acknowledgements}

% -----------------------------------------------------------------------------
\bibliographystyle{aa} 
%\bibliography{biblio}    
% -----------------------------------------------------------------------------

\appendix  
% -----------------------------------------------------------------------------
\section{Complementary line fitting results}
\label{sec_appendix}
% -----------------------------------------------------------------------------

\begin{figure*}
   \centering
\vspace{5mm} 
   \includegraphics[width=0.19\textwidth, height=0.13\textwidth]{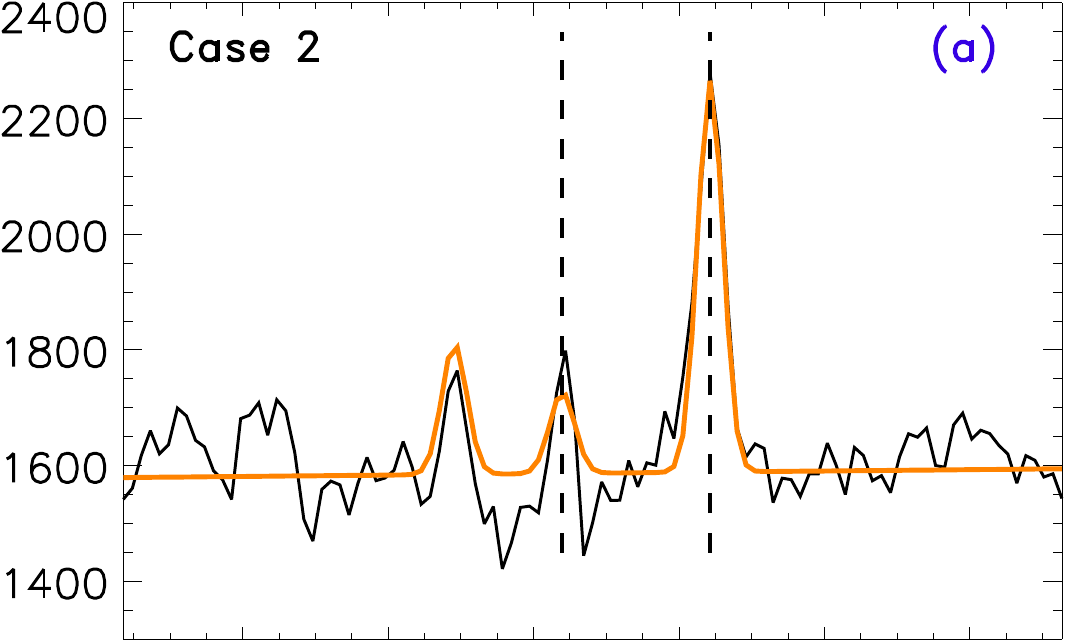}   
   \includegraphics[width=0.19\textwidth, height=0.13\textwidth]{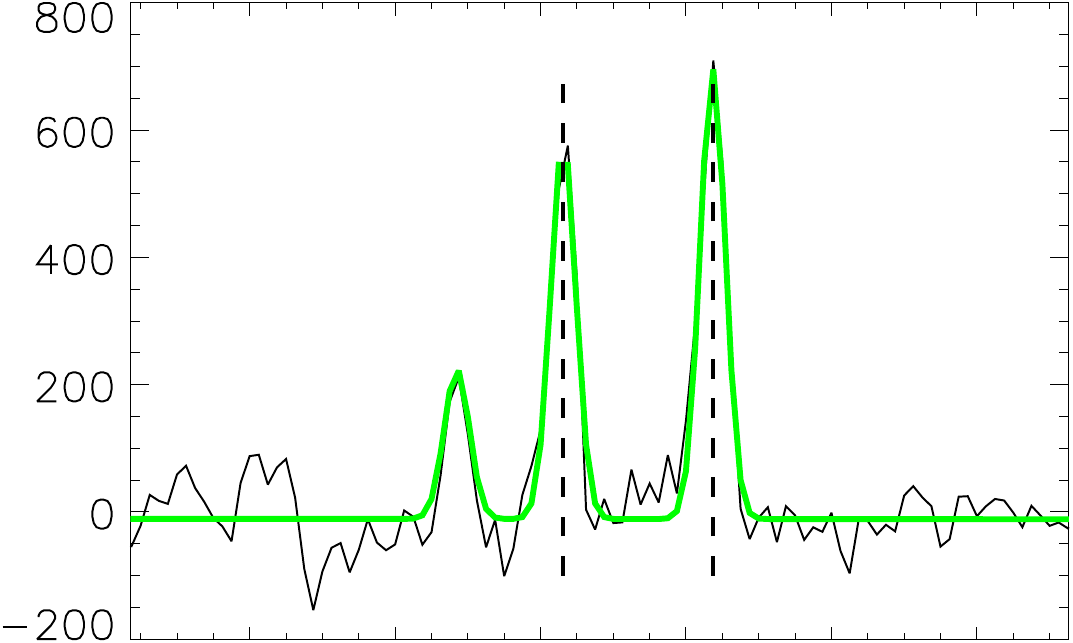}   
   \includegraphics[width=0.19\textwidth, height=0.13\textwidth]{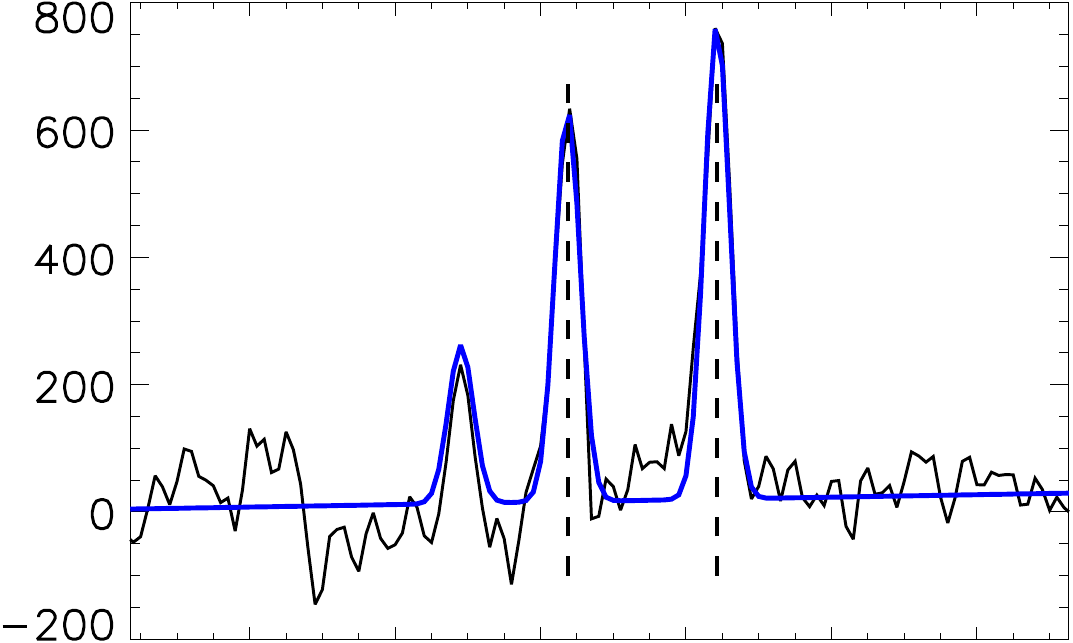}   
   \includegraphics[width=0.19\textwidth, height=0.13\textwidth]{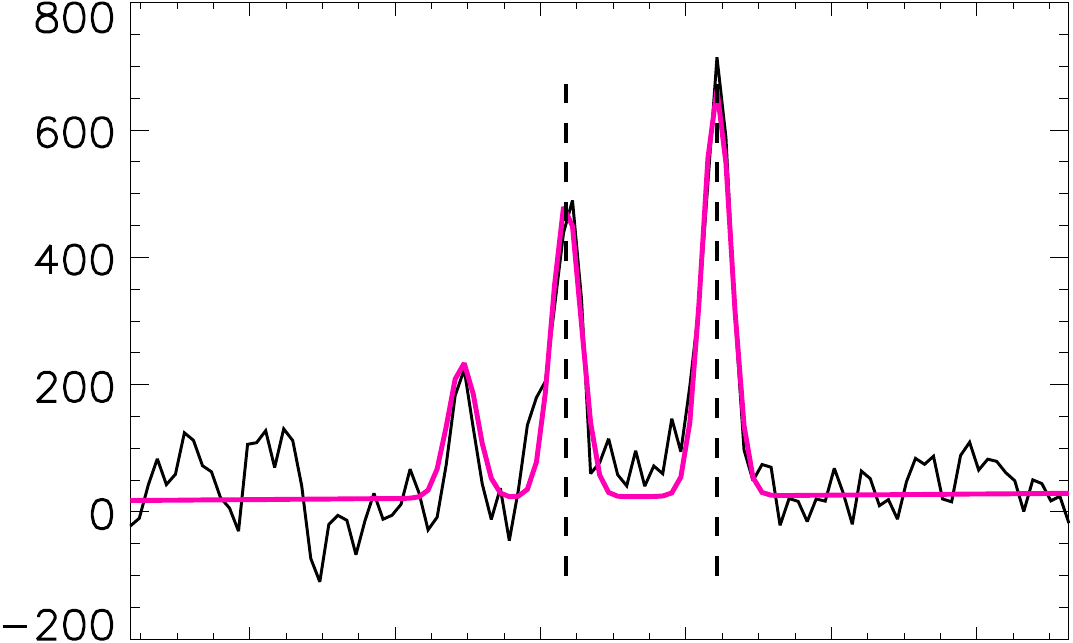}   
   \includegraphics[width=0.19\textwidth, height=0.13\textwidth]{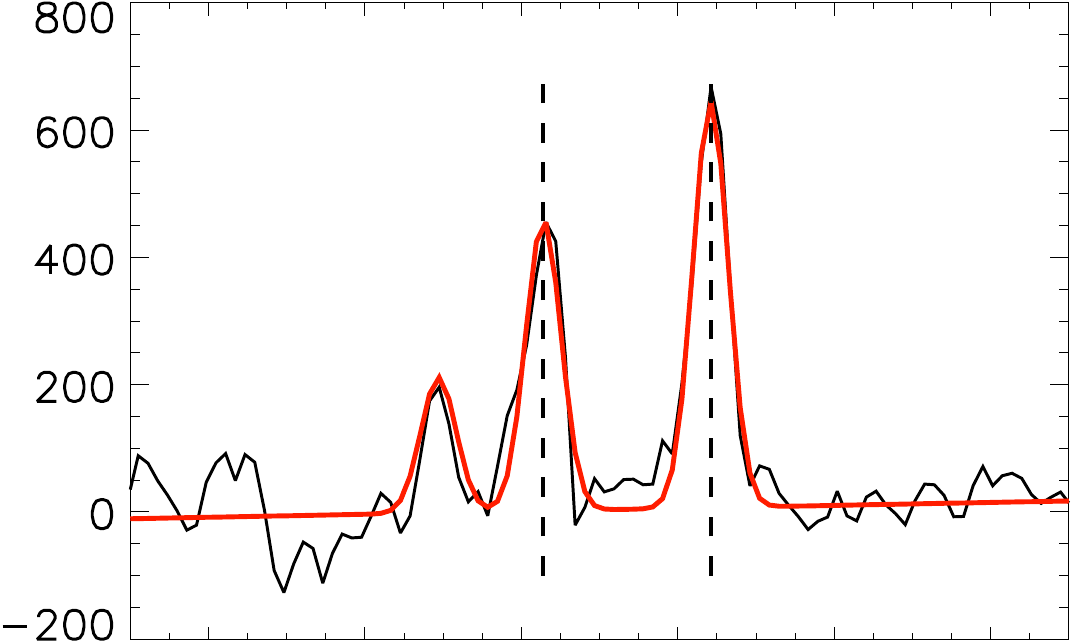}   
    
   \includegraphics[width=0.19\textwidth, height=0.13\textwidth]{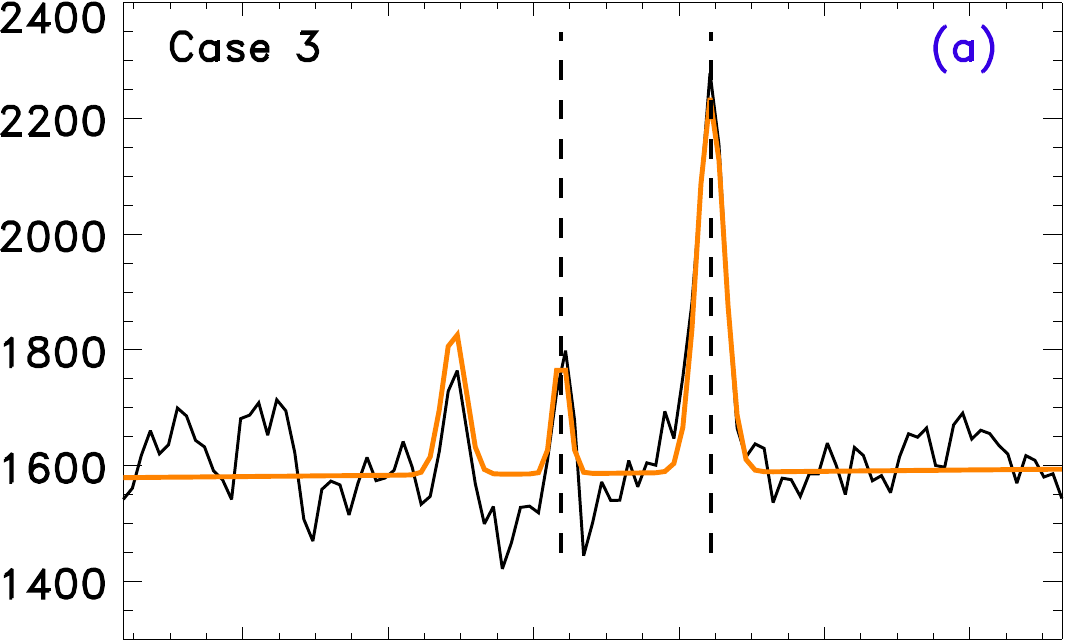}
   \includegraphics[width=0.19\textwidth, height=0.13\textwidth]{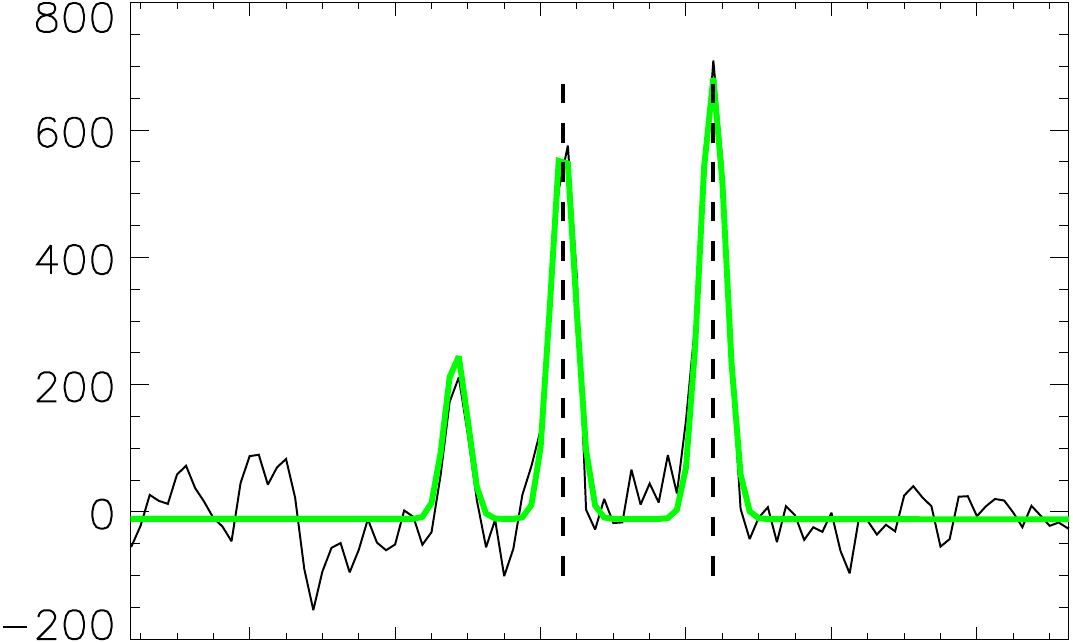}   
   \includegraphics[width=0.19\textwidth, height=0.13\textwidth]{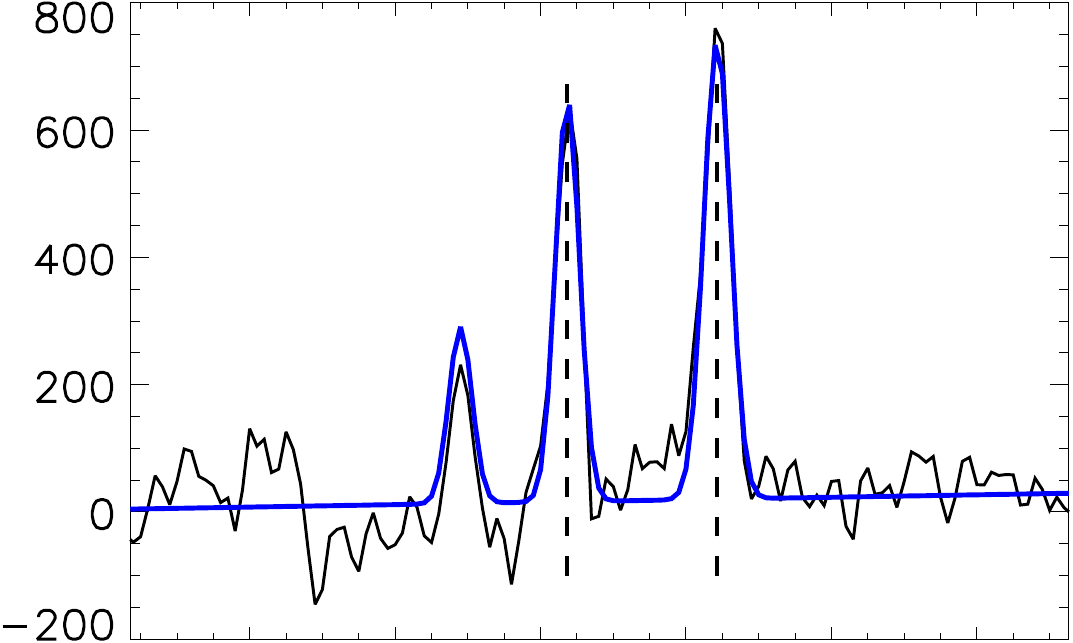}
   \includegraphics[width=0.19\textwidth, height=0.13\textwidth]{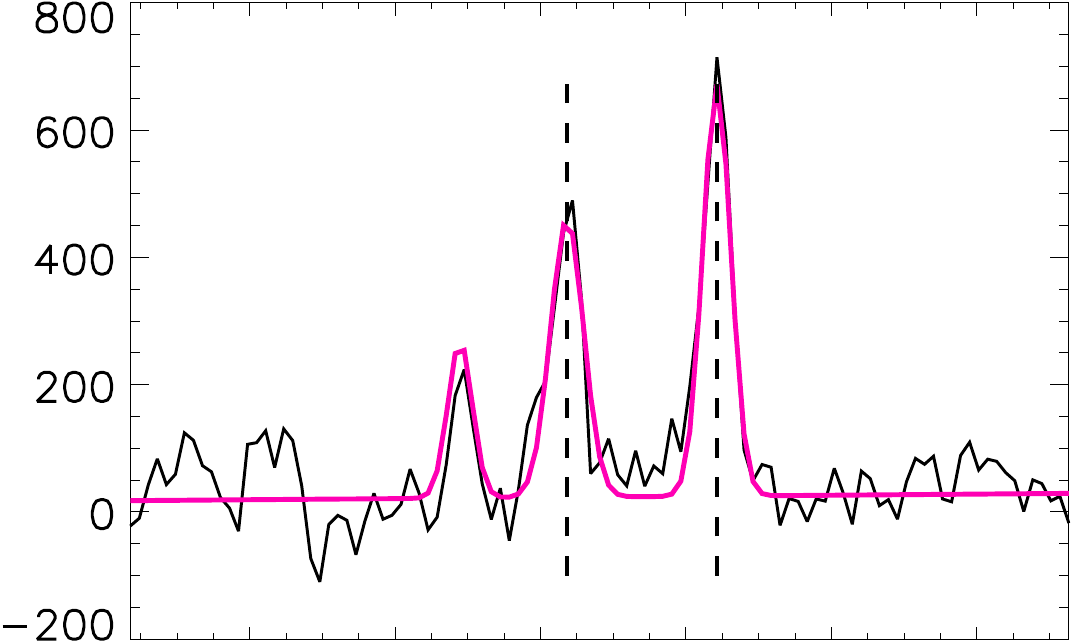}   
   \includegraphics[width=0.19\textwidth, height=0.13\textwidth]{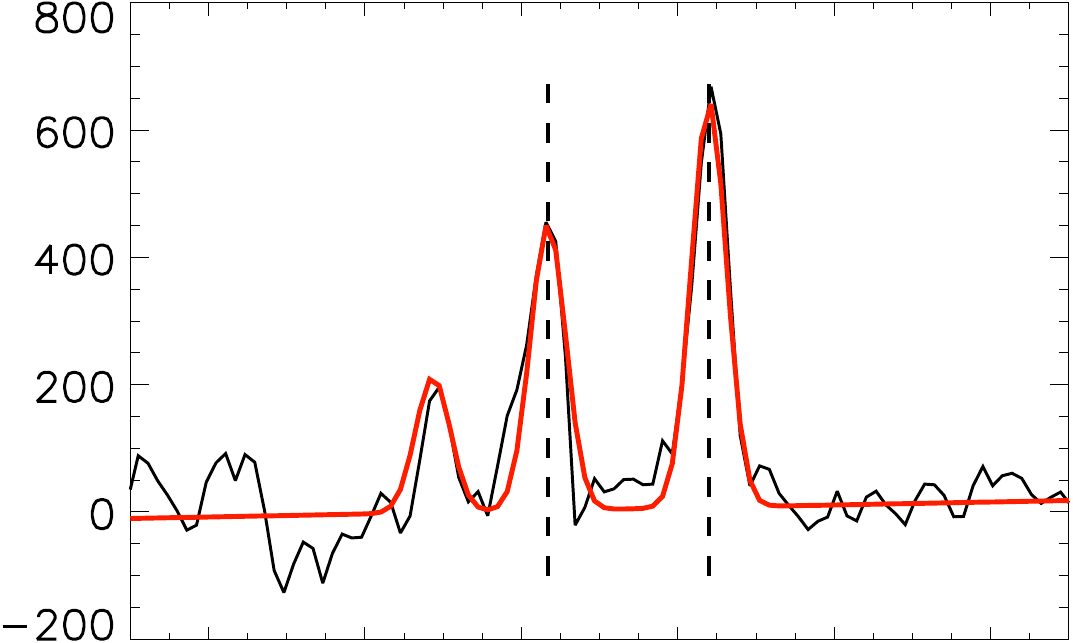}

  \includegraphics[width=0.19\textwidth, height=0.13\textwidth]{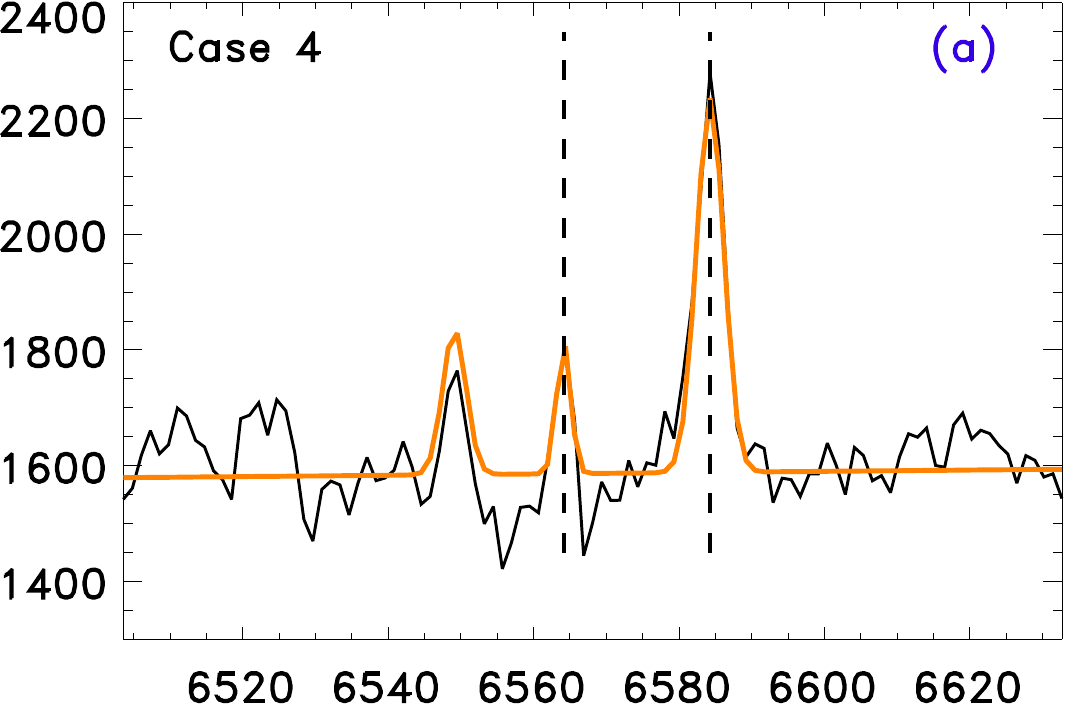}
  \includegraphics[width=0.19\textwidth, height=0.13\textwidth]{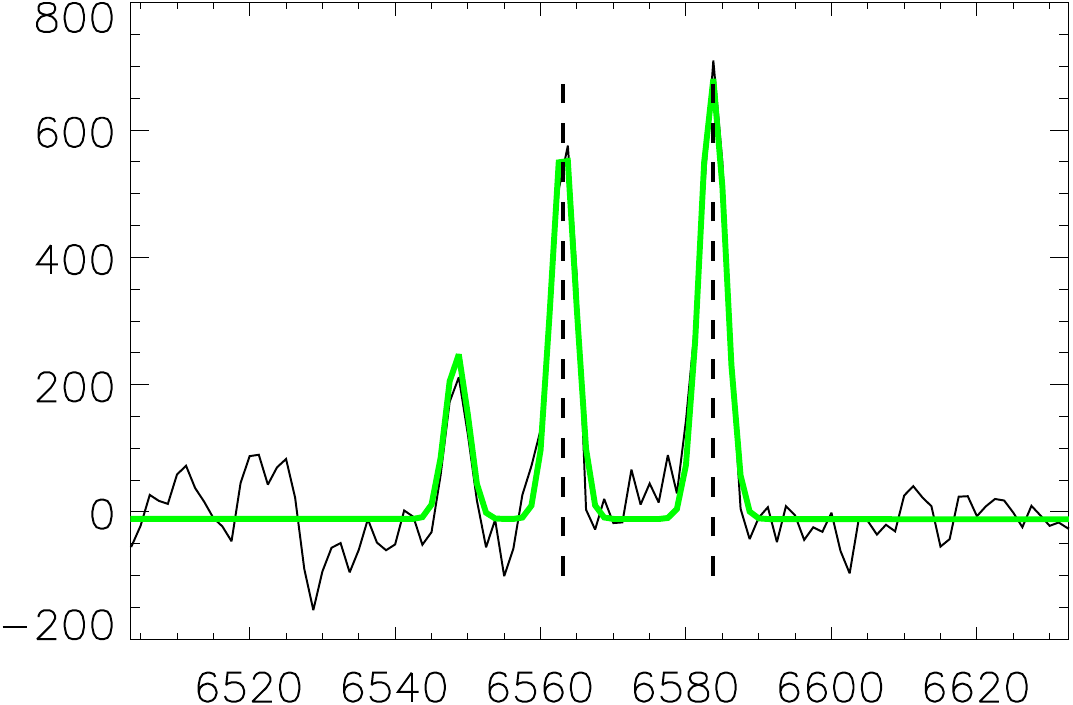}   
  \includegraphics[width=0.19\textwidth, height=0.13\textwidth]{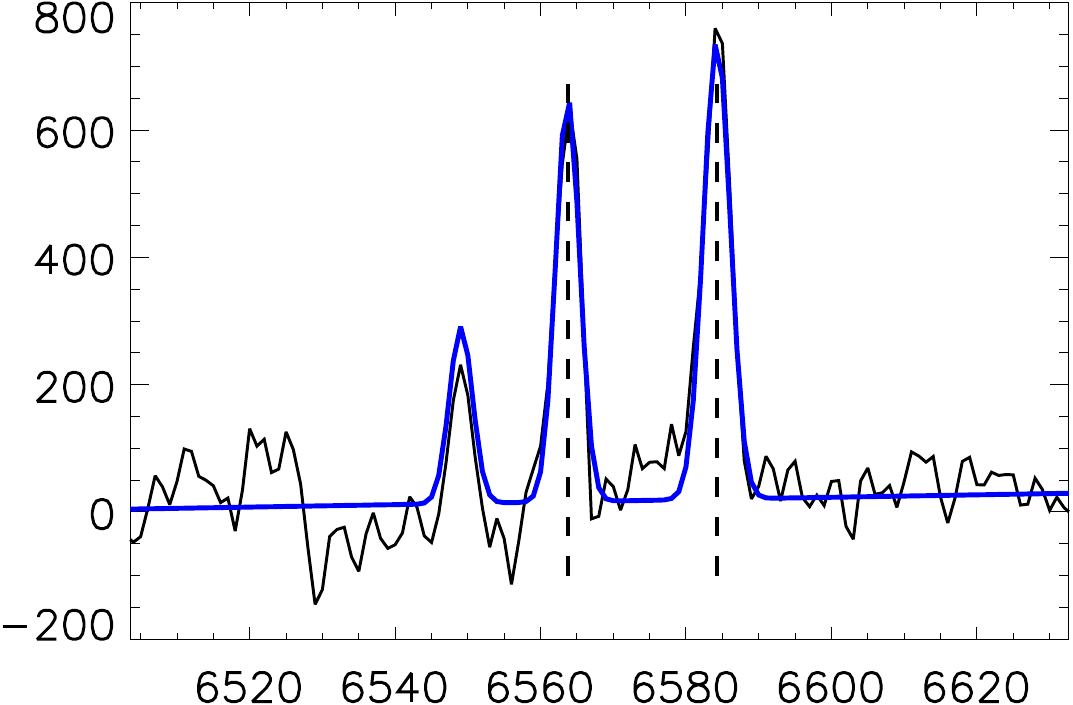}
  \includegraphics[width=0.19\textwidth, height=0.13\textwidth]{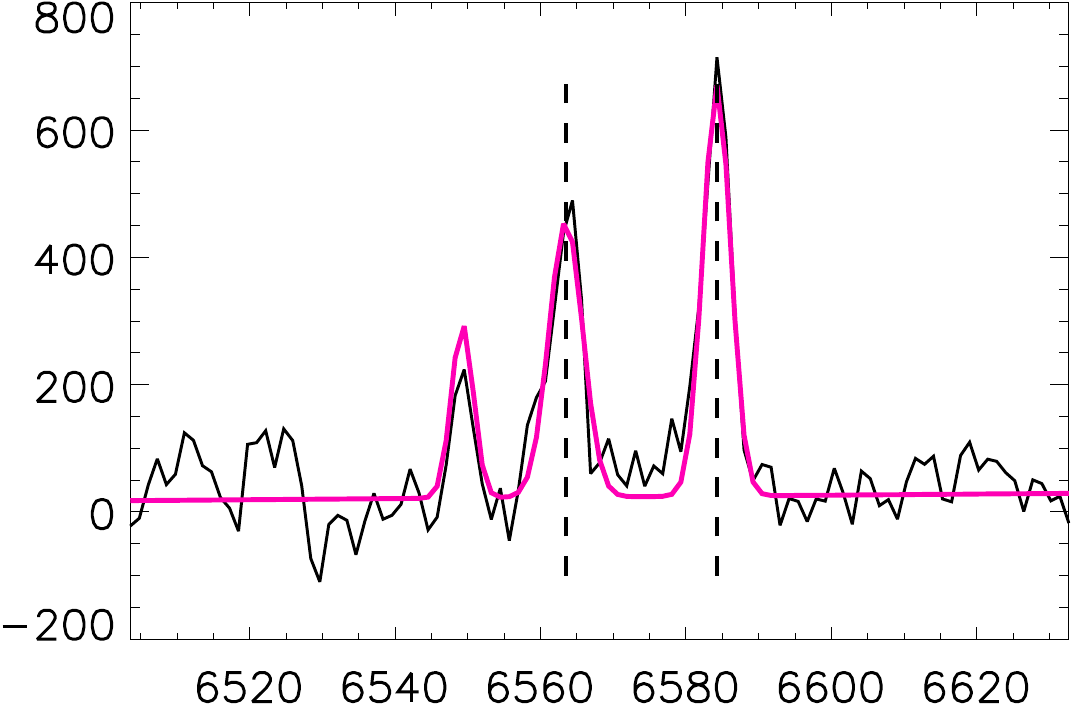}   
  \includegraphics[width=0.19\textwidth, height=0.13\textwidth]{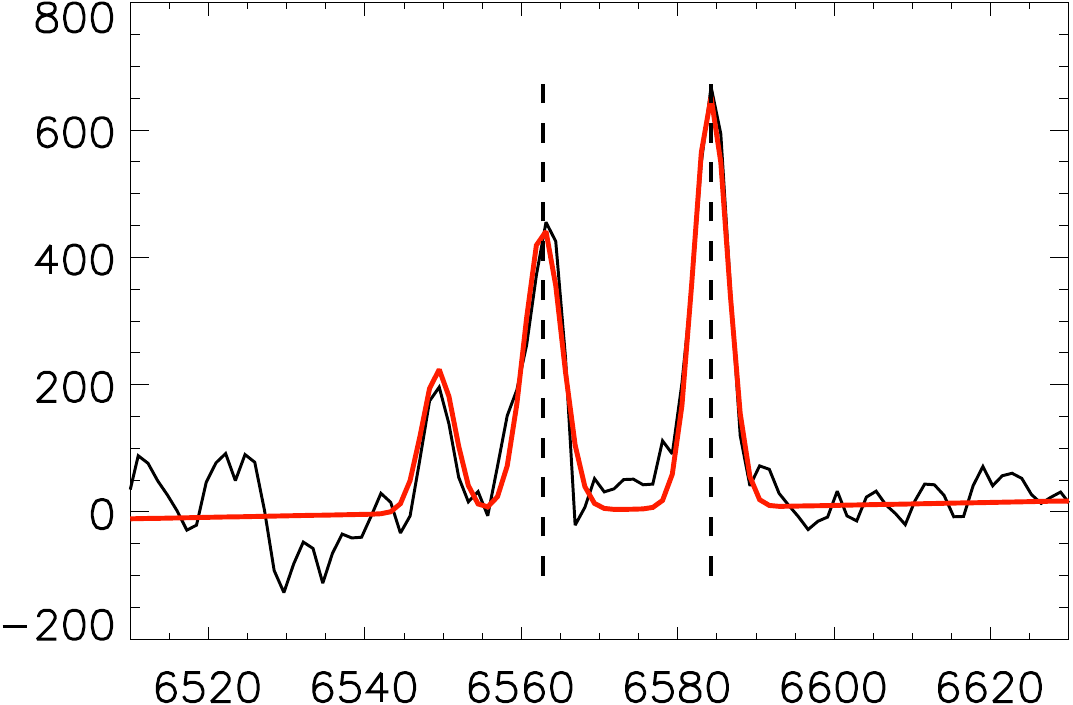}   
\vskip3mm\caption{(General comments about the panels as in Fig.~\ref{fig_spectra_all_fixed}) Line fitting results for the spectrum {\tt `a'} obtained following the assumptions of {\tt case 2} [$\lambda$ {\tt free}, $\sigma$ {\tt fixed}] ({\it top panels}), {\tt case 3} [$\lambda$ {\tt fixed}, $\sigma$ {\tt free}] ({\it middle panels}), and {\tt case 4} [$\lambda$ and $\sigma$ {\tt free}] ({\it bottom panels}). For each case and method the centroid of the H$\alpha$ and [NII] lines is highlighted using the dashed lines. The line fitting results are shown using the colored solid line: the `unsubtracted' data are in orange, {\tt FIT3D} data in green, {\tt STARLIGHT (MILES)} data in dark blue, {\tt STARLIGHT (STELIB)} ones in magenta and {\tt pPXF} ones in red. 
}
 \label{app_cases}
\end{figure*}

In this Appendix we present complementary kinematic results which have not been shown in the main text. In Fig.~\ref{app_cases} we present the line fitting results obtained when considering the {\tt case 2,3 and 4} applied to the H$\alpha$--[NII] complex to the spectrum {\tt `a'} as an example. This kind of analysis has been applied to the whole sample of spectra to derive the kinematic results discussed in the text. 

\begin{figure*}
\vskip4.8mm
   \centering
%\vspace{-5mm} 
   \includegraphics[width=0.87\textwidth, height=0.22\textwidth]{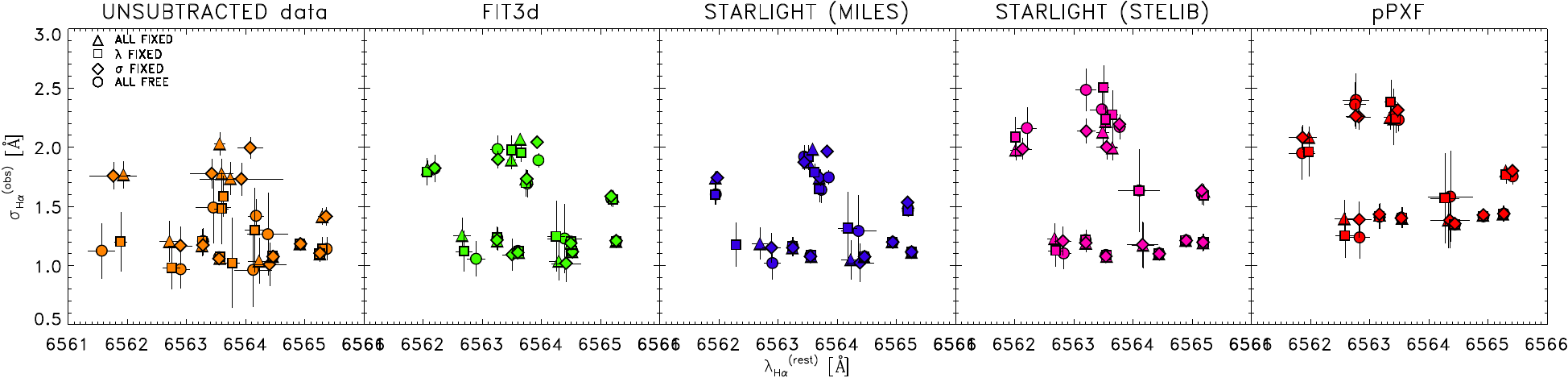}   
\vskip1cm
   \includegraphics[width=0.87\textwidth, height=0.22\textwidth]{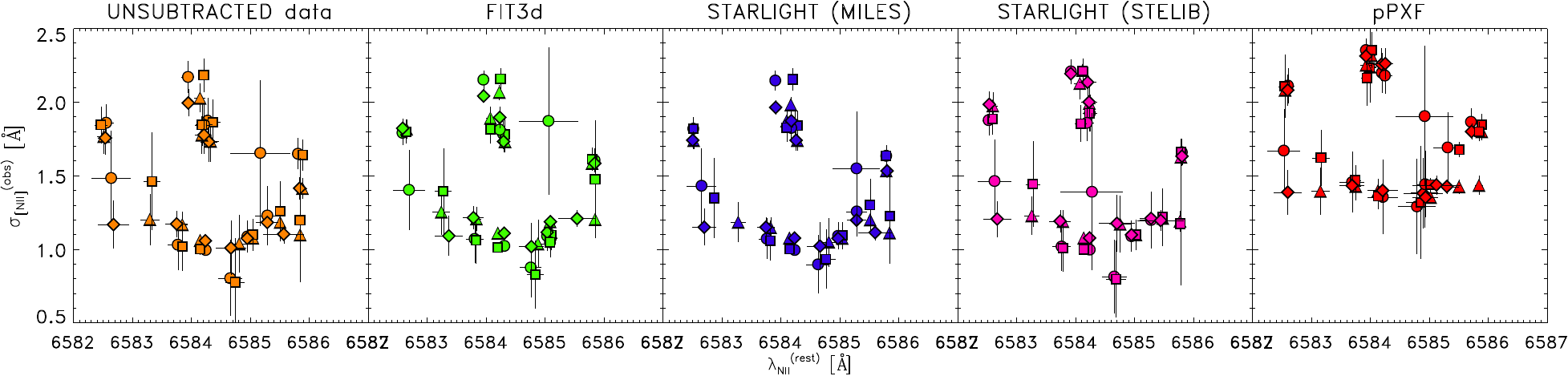}   
      \caption{Individual kinematic values for the H$\alpha$ ({\it top panel}) and [NII] ({\it bottom panel}) lines as derived from combining the four cases with the different methods. The error bar for each value is derived from the line fitting method (see text for details). The symbol types are the same than those used in the text.  The color code is the same than in previous figures: the `unsubtracted data are in orange, {\tt FIT3D} data in green, {\tt STARLIGHT (MILES)} data in dark blue, {\tt STARLIGHT (STELIB)} ones in magenta and {\tt pPXF} ones in red.      
}
 \label{spectra_sigma_all}
\end{figure*}

In Fig.~\ref{spectra_sigma_all} we show the kinematic results ($\lambda_{rest}$, $\sigma_{obs}$) obtained for all spectra when applying the four cases for the five methods. 
When no stellar subtraction method ({\tt Unsubtracted}) is applied to the H$\alpha$ line, we derive the smallest value of the velocity dispersions (i.e., $\sigma\sim$1.3 \AA) followed by that obtained when applying the {\tt STARLIGHT (MILES)} method. When considering the {\tt pPXF} routine we get the highest velocity dispersion (i.e., $\sigma\sim$1.7 \AA). The [NII] velocity dispersion values cover a similar range (1.4 --~1.8 \AA) and show an analogous trend than that derived for the Balmer line.

We also present in Tabs.~\ref{individual_total}, \ref{mean_allcases_samemet_orig} and \ref{mean_allmeth_samemet_orig} the main results obtained when combining different line fitting approaches with different stellar subtraction methods.
In particular, in Tab.~\ref{individual_total} the respective mean values for the wavelength and width of the two emission lines are shown for all different cases and methods. In Tabs.~\ref{mean_allcases_samemet_orig} and \ref{mean_allmeth_samemet_orig} we complement the information given in the previous table considering the mean values of the four cases for each method and vice versa (i.e., mean values of the five methods for each case).

%********************************************************************************************
% SIGMA FIXED

\begin{table*}
\centering
\begin{small}
\caption{Mean value (and standard deviations) for the H$\alpha$ and [NII] lines among the different methods and different cases.}
\label{individual_total}
\begin{tabular}{@{} l|ccccc @{}} % Column formatting, @{} suppresses leading/trailing space
\hline\hline
{\backslashbox{{\tt Case}\\\strut \hskip5mm{\tt Param}\\\ \hskip8mm{\tt Line}}{\strut \\\ \hskip-5mm{\tt Method} }} 
 &   \makebox{ {\tt  {\bf Unsubtracted} }}  &   \makebox{  {\tt FIT3D} }  &   \makebox{{\tt STARLIGHT (MILES)} }  &   \makebox{{\tt STARLIGHT (STELIB)}}   &   \makebox{{\tt pPXF }}\\
 \hline
 \hskip13mm(1) & (2)& (3)  & (4) & (5)  &(6)  \\
\hline\hline
{\tt [1]} $\lambda_{H\alpha}$ &    6563.87 $\pm$ 1.02 	&    6563.85 $\pm$ 0.96  &   6563.85 $\pm$ 1.00	&	   6563.83 $\pm$ 0.98	&	    6563.81 $\pm$ 1.05  \\
{\tt [2]} $\lambda_{H\alpha}$  &    6563.94 $\pm$ 1.04	&    6563.89 $\pm$ 0.93  &   6563.91 $\pm$ 0.98	&	   6563.84 $\pm$ 0.96	&	    6563.74 $\pm$ 1.13  \\
{\tt [3]} $\lambda_{H\alpha}$  &   6563.88 $\pm$ 1.01	&    6563.85 $\pm$ 0.95  &   6563.82 $\pm$ 1.05	&      6563.83 $\pm$ 0.98		&           6563.81 $\pm$ 1.05  \\
{\tt [4]} $\lambda_{H\alpha}$  &  6563.95 $\pm$ 1.08	&    6563.89 $\pm$ 0.92  &   6563.91 $\pm$ 0.99	&      6563.83 $\pm$ 0.95		&	    6563.73 $\pm$ 1.13   \\
\hline
{\tt [1]} $\sigma_{H\alpha}$  &       1.38 $\pm$ 0.35	&       1.43 $\pm$ 0.36	&      1.39 $\pm$ 0.35	&	     1.51 $\pm$ 0.45	&     1.71 $\pm$ 0.40         \\
{\tt [2]} $\sigma_{H\alpha}$   &       1.37 $\pm$ 0.35	&       1.43 $\pm$ 0.36	&      1.39 $\pm$ 0.35	&        1.51 $\pm$ 0.47	&     1.71 $\pm$ 0.40         \\
{\tt [3]} $\sigma_{H\alpha}$    &       1.19 $\pm$ 0.18	&       1.44 $\pm$ 0.34	&      1.38 $\pm$ 0.29	&     	 1.61 $\pm$ 0.53	&     1.71 $\pm$ 0.40         \\
{\tt [4]} $\sigma_{H\alpha}$    &       1.17 $\pm$ 0.16	&       1.43 $\pm$ 0.34	&	   1.36 $\pm$ 0.30	&	     1.61 $\pm$ 0.54	&     1.71 $\pm$ 0.42         \\
\hline\hline
{\tt [1]} $\lambda_{[NII]}$  &      6584.46 $\pm$ 1.01	&	   6584.44 $\pm$ 0.96		&	    6584.44 $\pm$ 1.01	&	    6584.42 $\pm$ 0.99	&     6584.39 $\pm$ 1.05    \\
{\tt [2]} $\lambda_{[NII]}$  &      6584.33 $\pm$ 1.03	&	   6584.34 $\pm$ 1.01		&	    6584.32 $\pm$ 1.03	&	    6584.31 $\pm$ 1.02	&	  6584.28 $\pm$ 0.99    \\
{\tt [3]} $\lambda_{[NII]}$  &      6584.46 $\pm$ 1.01	&	   6584.44 $\pm$ 0.96		&	    6584.41 $\pm$ 1.06	&	    6584.41 $\pm$ 0.98	&	  6584.39 $\pm$ 1.05    \\
{\tt [4]} $\lambda_{[NII]}$  &      6584.29 $\pm$ 0.99	&	   6584.30 $\pm$ 0.96		&	    6584.28 $\pm$ 1.00	&	    6584.20 $\pm$ 0.96	&     6584.25 $\pm$ 0.98    \\
\hline
{\tt [1]} $\lambda_{[NII]}$  &        1.37 $\pm$ 0.35	&	      1.43 $\pm$ 0.36	&	     1.39 $\pm$ 0.35	&	        1.51 $\pm$ 0.47	&	       1.71 $\pm$ 0.39    \\
{\tt [2]} $\lambda_{[NII]}$  &        1.37 $\pm$ 0.35	&	      1.43 $\pm$ 0.36	&	     1.39 $\pm$ 0.35	&	        1.51 $\pm$ 0.47	&	       1.71 $\pm$ 0.40    \\
{\tt [3]} $\lambda_{[NII]}$  &        1.43 $\pm$ 0.44	&	      1.42 $\pm$ 0.42	&	     1.44 $\pm$ 0.40	&	        1.44 $\pm$ 0.45	&	       1.78 $\pm$ 0.36    \\
{\tt [4]} $\lambda_{[NII]}$  &        1.48 $\pm$ 0.44	&	      1.46 $\pm$ 0.43	&	     1.47 $\pm$ 0.40	&	        1.46 $\pm$ 0.44	&	       1.79 $\pm$ 0.36	  \\
\hline\hline
\end{tabular}
\begin{minipage}[h]{18cm}\vskip3mm
{\bf Notes.} Col (1): Line fitting case ({\tt [1],~[2],~[3],~[4]}), parameter ($\lambda$, $\sigma$) and emission line (H$\alpha$, [NII]) considered. Cols (2--6): Kinematic (mean) results obtained when analyzing, respectively: the {\tt Unsubtracted} data and those derived from applying the following stellar subtraction methods: {\tt FIT3D}, {\tt STARLIGHT (MILES)}, {\tt STARLIGHT (STELIB)} and {\tt pPXF}.
\end{minipage}
\end{small}
\end{table*}

\begin{table*}
\vskip0.8cm
\centering
\begin{small}
\caption{Mean values (and standard deviations) for the H$\alpha$ and [NII] lines when considering the four cases of line fitting for each method.}
\label{mean_allcases_samemet_orig}
\begin{tabular}{@{} l|ccccc@{}} 
\hline\hline
{\backslashbox{ {\tt Param}\\\ \hskip2mm{\tt Line}}{\strut \\\ \hskip-5mm{\tt Method} }} 
 &   \makebox{ {\tt Unsubtracted }}  &   \makebox{  {\tt FIT3D} }  &   \makebox{{\tt STARLIGHT (MILES)} }  &   \makebox{{\tt STARLIGHT (STELIB)}}   &   \makebox{{\tt pPXF }}\\
\hline
  \hskip11mm	(1) & (2)& (3)  & (4) & (5)  &(6)  \\
\hline\hline
$\lambda_{H\alpha}$&  6563.91 $\pm$ 0.50	&	    6563.87 $\pm$ 0.46	&	   6563.87 $\pm$ 0.49	&	  6563.83 $\pm$ 0.47	&	  6563.77 $\pm$ 0.53   \\
 $\sigma_{H\alpha}$  &  1.28 $\pm$ 0.14	&	     1.43 $\pm$ 0.17 	&	     1.38 $\pm$ 0.16	&	    1.56 $\pm$ 0.23		&	     1.71 $\pm$ 0.20    	\\
\hline
  $\lambda_{[NII]}$  &  6584.39 $\pm$ 0.49	&	   6584.38 $\pm$ 0.47	&	   6584.36 $\pm$ 0.50	&	  6584.33 $\pm$ 0.48	&	  6584.33 $\pm$ 0.50 	\\
  $\sigma_{[NII]}$ &    1.43 $\pm$ 0.19	&	     1.46  $\pm$ 0.19 	&	     1.44 $\pm$ 0.18	&	    1.51 $\pm$ 0.22		&	     1.76 $\pm$ 0.19 		 \\
\hline\hline
\end{tabular}
\begin{minipage}[h]{18cm}\vskip3mm
{\bf Notes.} Col (1): Kinematic parameter ($\lambda$, $\sigma$) derived for each emission line (H$\alpha$, [NII]). Cols (2--6): Mean kinematic values when considering: Col (2): The {\tt Unsubtracted} data. 
Col (3): the {\tt FIT3D} data. 
Col (4): the {\tt STARLIGHT (MILES)} data. 
Col (5): the {\tt STARLIGHT (STELIB)} data. 
Col (6): the {\tt pPXF} data.
\end{minipage}
\end{small}

\centering
\vskip1cm
\begin{small}
\caption{Mean values (and standard deviations) for H$\alpha$ and [NII] lines when considering the five methods for each line fitting case.}
\label{mean_allmeth_samemet_orig}
\begin{tabular}{@{} c|cccc@{}} 
\hline\hline 
{\backslashbox{ {\tt Param}\\\ \hskip2mm{\tt Line}}{\strut \\\ \hskip0mm{\tt Case} }} 
 &   \makebox{ {\tt  [1] }}  &   \makebox{  {\tt [2]} }  &   \makebox{{\tt [3]} }  &   \makebox{{\tt [4]}} \\
 \hline
 	(1) & (2)& (3)  & (4) & (5)    \\
\hline\hline
$\lambda_{H\alpha}$ &   6563.84 $\pm$ 0.44	&	6563.86 $\pm$ 0.44	&	 6563.83 $\pm$ 0.44		&	 6563.86 $\pm$ 0.44      \\
 $\sigma_{H\alpha}$   &     1.49 $\pm$ 0.17	&	    1.48 $\pm$ 0.17	&	  1.46 $\pm$ 0.18		&	 1.46 $\pm$ 0.18     \\
\hline
  $\lambda_{[NII]}$     &     6584.43 $\pm$ 0.44	   &	  6584.32 $\pm$0.44	&	  6584.42 $\pm$ 0.44	&	  6584.27 $\pm$ 0.43    \\
  $\sigma_{[NII]}$       &   1.49 $\pm$ 0.17		   & 	1.48 $\pm$ 0.17	&	   	1.50 $\pm$ 0.19	&   	1.53 $\pm$ 0.19     	\\
\hline\hline
\end{tabular}
\begin{minipage}[h]{18cm}
\vskip4mm
{\bf Notes.} Col (1): Kinematic parameter ($\lambda$, $\sigma$) derived for each emission line (H$\alpha$, [NII]). Cols (2--5): Mean kinematic values when considering, respectively, the four line fitting cases: {\tt [1], [2], [3] and [4]}.
\end{minipage}
\end{small}
\vskip5mm
\end{table*}

\section{Chi--square results}
\label{Chi2}

In this section we present the results obtained for each spaxel when computing the chi--square ($\chi^2$) values in order to test which stellar continuum subtraction code more faithfully explains the raw observed data (within the errors).

The goodness of the fits has been evaluated by means of the reduced $\chi^2$ ($\chi^2_{red}$), where the number of degrees of freedom are the number of data points used in the fit, following the equation: 

\begin{equation} 
\hskip10mm   \chi^2_{red} = \sum_{i=1}^{N_\lambda} \left( \frac{O_i - M_i}{\sigma_i}\right)^2 / N_\lambda\\
\end{equation}

where $N_\lambda$ is the number of observed data points, $O_i$ and $M_i$ are the raw observed flux data and the total flux model values for the $i$--th point, and $\sigma_i$ is the corresponding observed error. 

We constrain the derivation of the $\chi^2_{red}$ values in the range defined by the lines of our interest, which involves the H$\alpha$--[NII] complex (rest frame wavelength range from 6540 \AA\ to 6600 \AA).

In particular, the total model emission in the wavelength range of our interest ($M_i$) is given by the combination of the stellar continuum model ($M_{i,\star}$) and the pure gas model ($M_{i, gas}$) emissions. The pure gas emission ($M_{i,gas}$) has been derived as the average of the models obtained when applying the four line fitting cases ($N_c$ = 4, see text for details). Thus, each model can be described as follows:

\begin{equation} 
  \hskip2mm M_i = M_{i,\star}  + \sum_{k=1}^{N_c} M_{k}^{i, gas} / N_c = M_{i,\star}  + M_{i, gas}. \\
\label{ff_total}
\end{equation}

\begin{figure*}[b]
   \centering
\vspace{2cm} 
   \includegraphics[width=0.32\textwidth, height=0.42\textwidth]{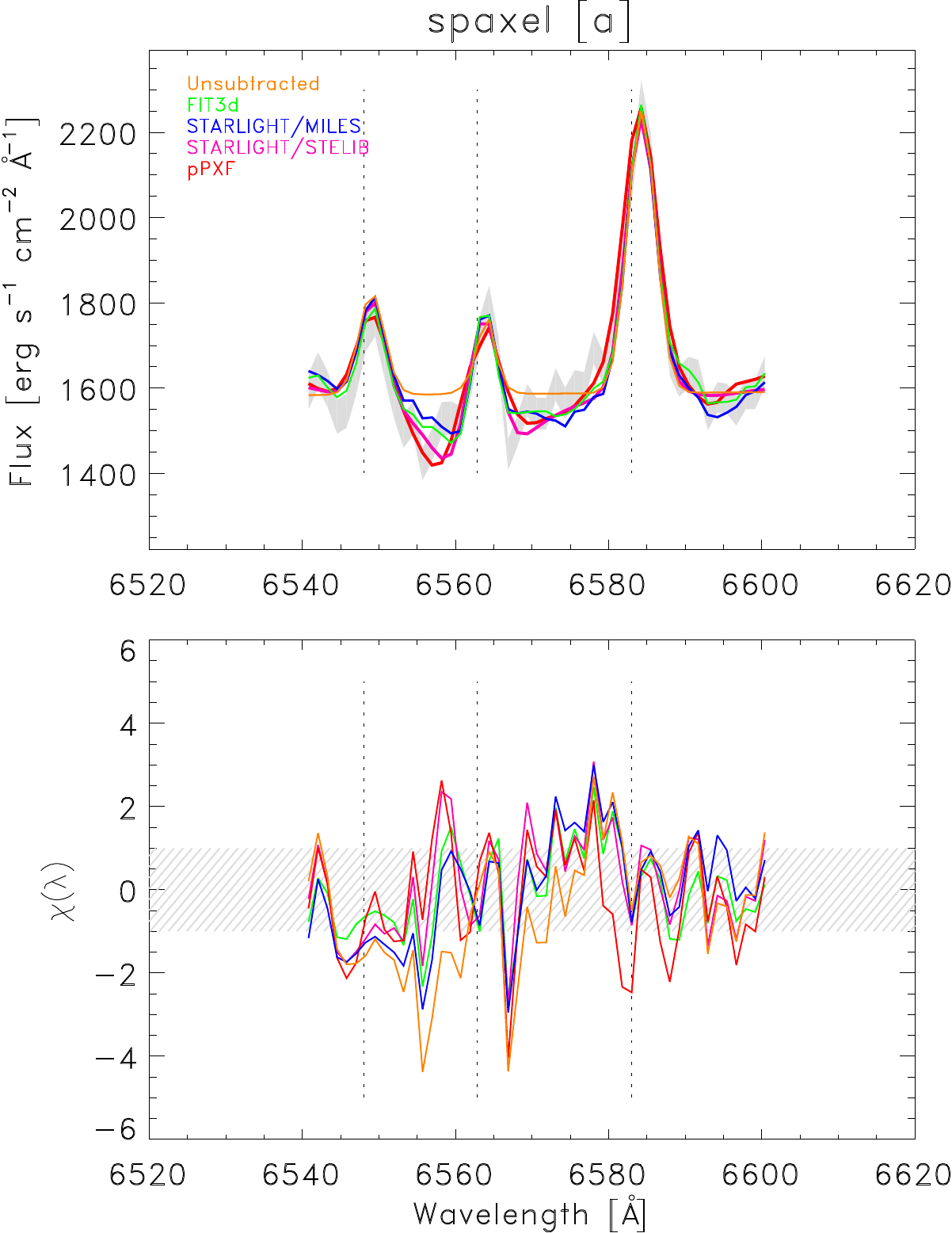}   
   \includegraphics[width=0.32\textwidth, height=0.42\textwidth]{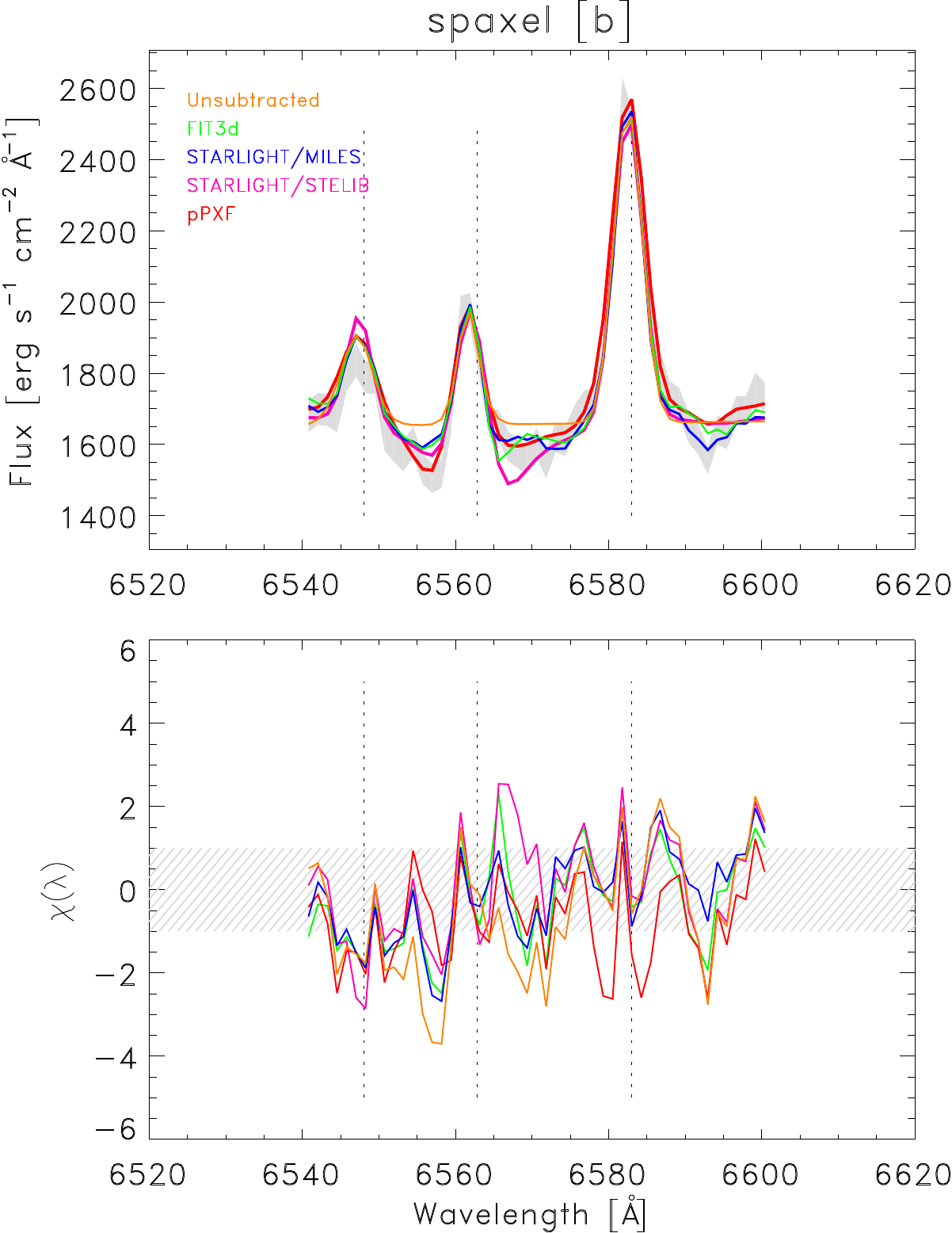}   
   \includegraphics[width=0.32\textwidth, height=0.42\textwidth]{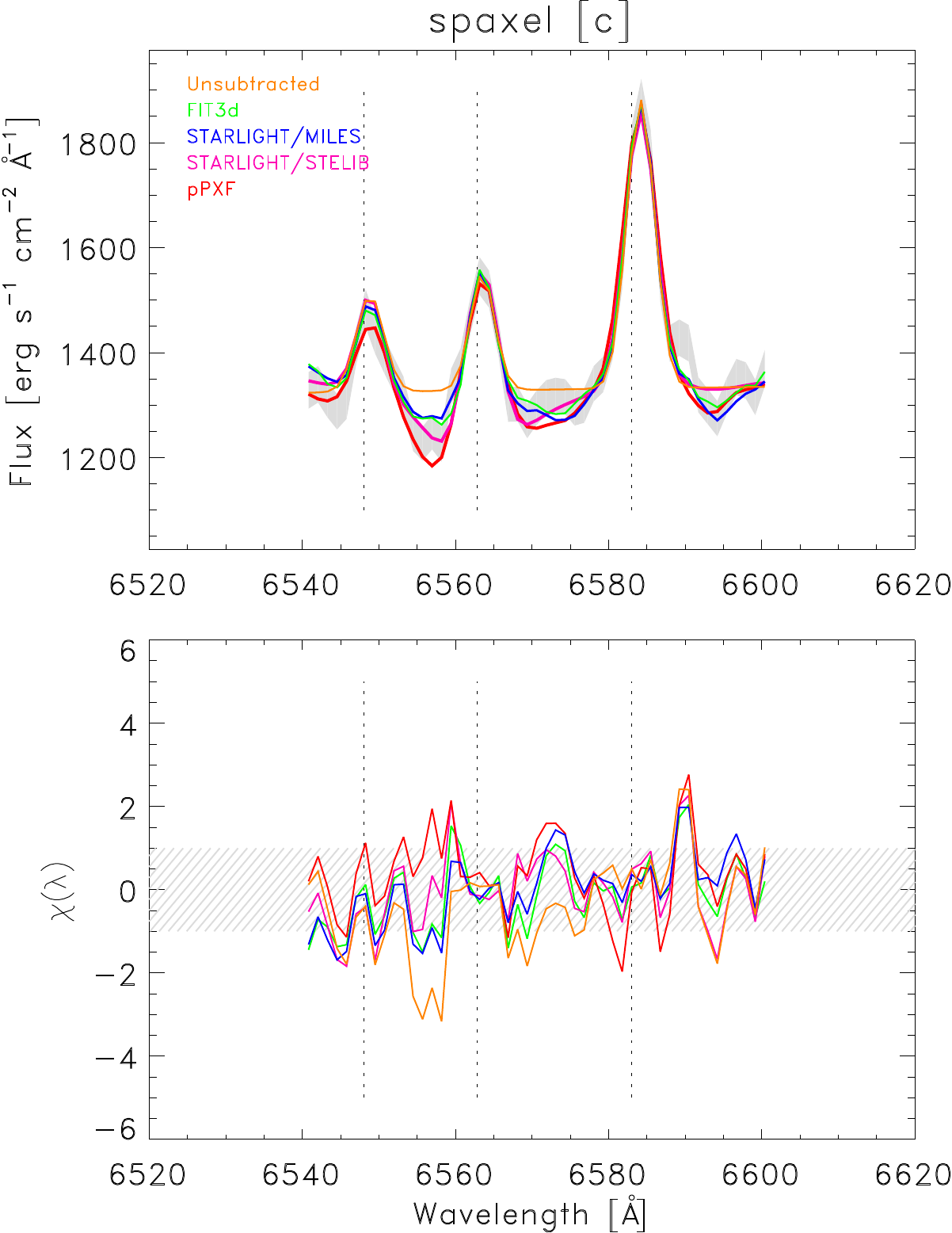}   

\vspace{2cm} 

   \includegraphics[width=0.32\textwidth, height=0.42\textwidth]{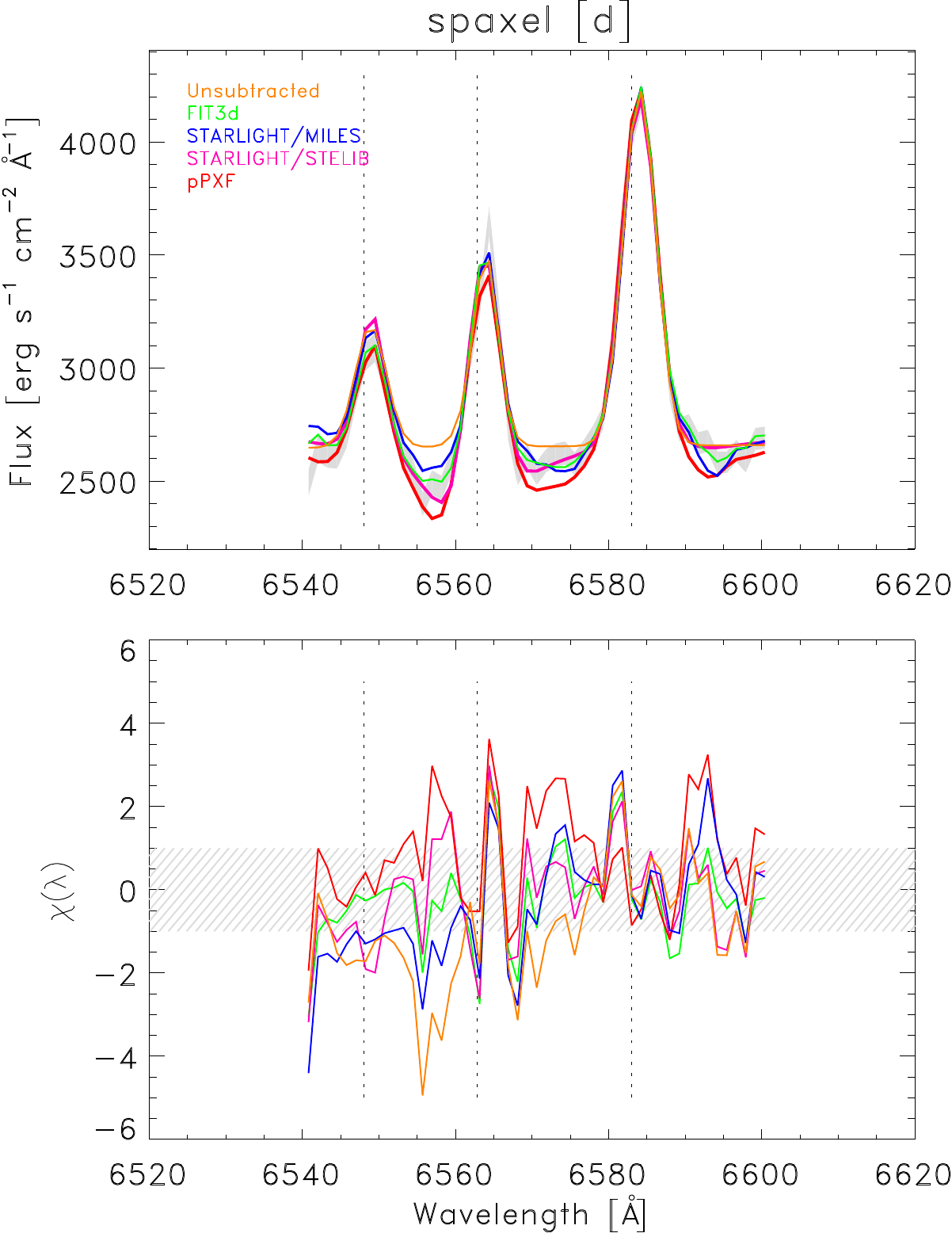}   
   \includegraphics[width=0.32\textwidth, height=0.42\textwidth]{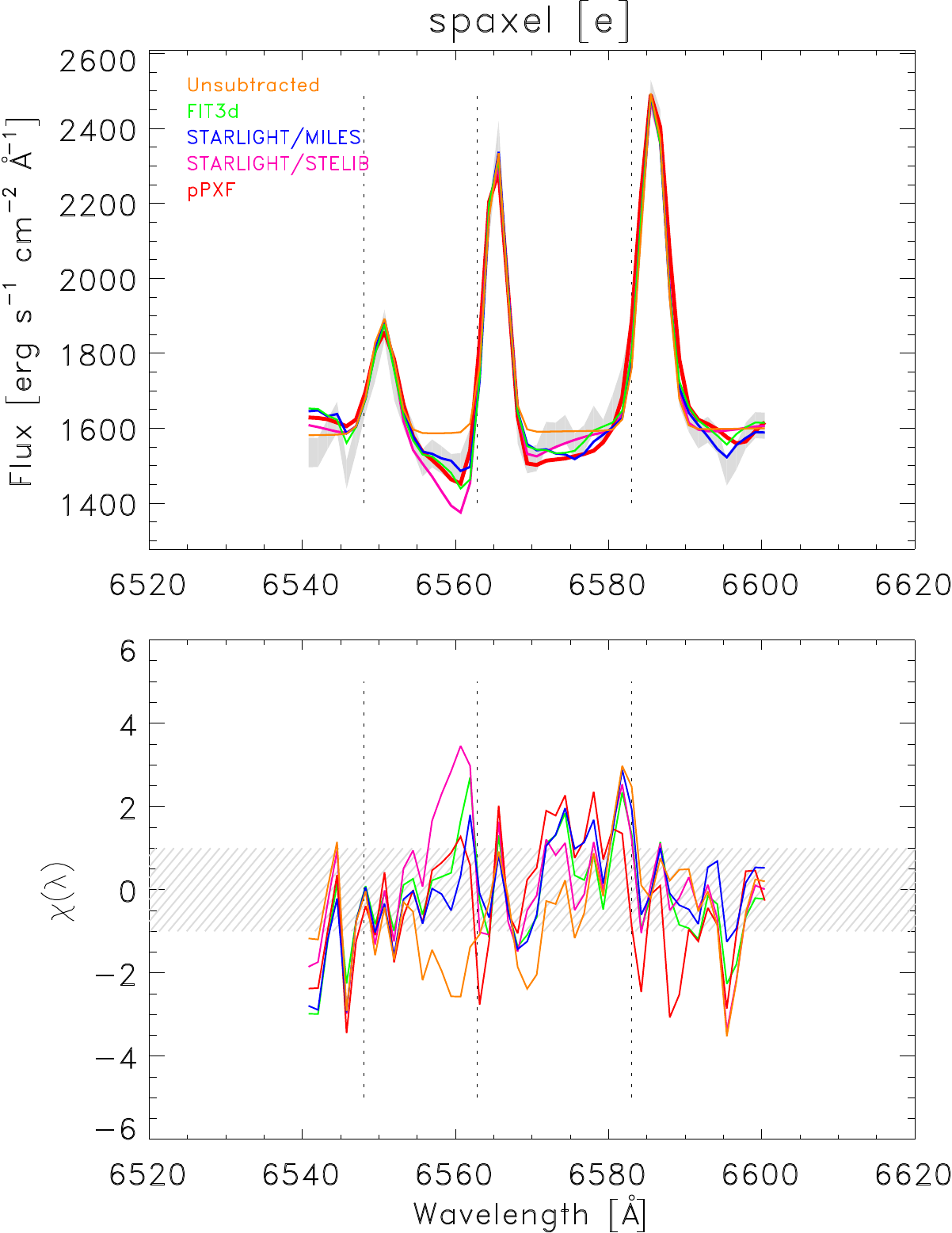}   
   \includegraphics[width=0.32\textwidth, height=0.42\textwidth]{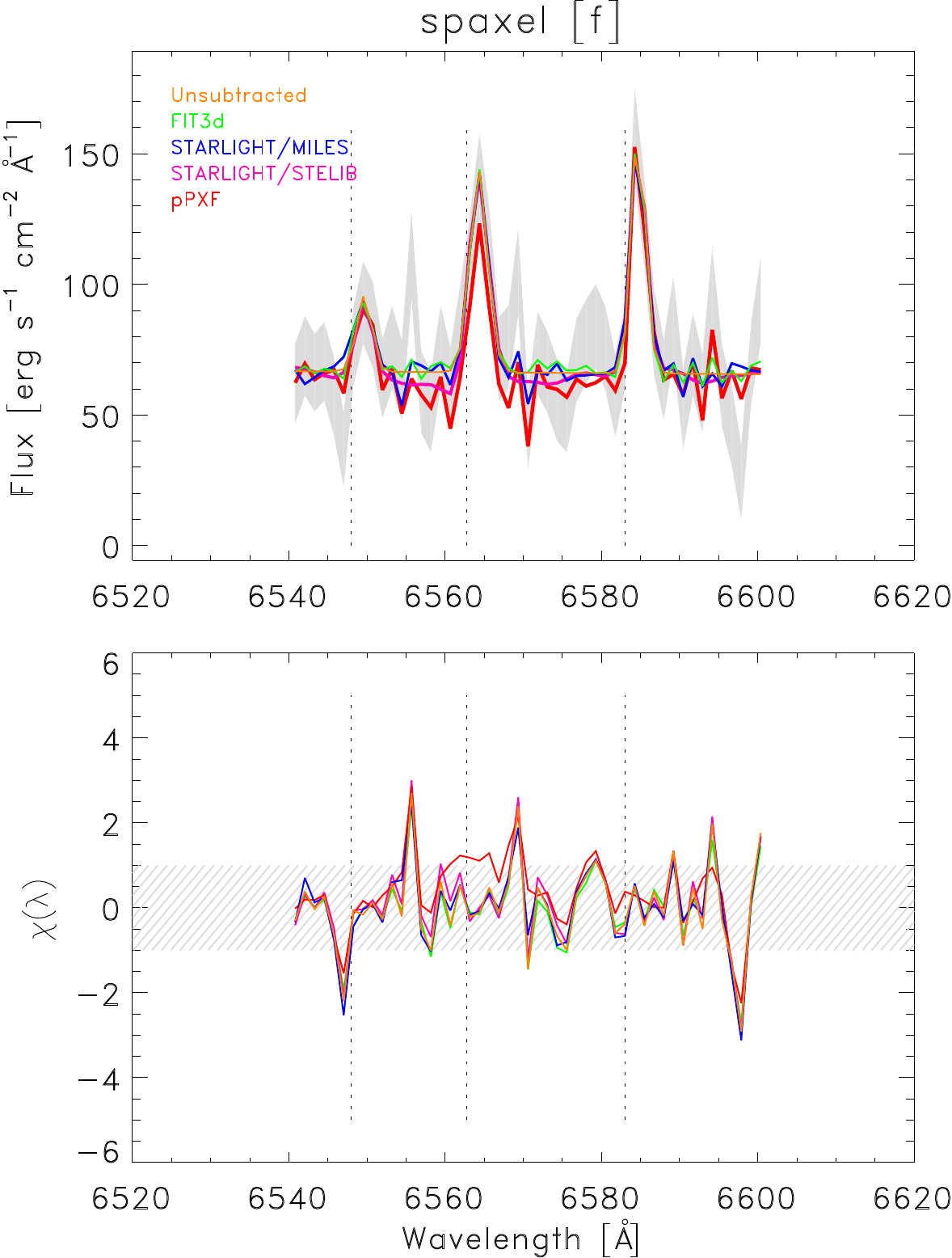}   
     \caption{{\it Top:} Comparison between the raw observed data (within the errors; shaded grey area) and the results derived when using the five models. The color code is the same as in previous figures ({\tt Unsubtracted} model is in orange, {\tt FIT3d} is in green, {\tt STARLIGHT/MILES} in dark blue, {\tt STARLIGHT/STELIB} in magenta and {\tt pPXF} is in red). The vertical dot lines identify the rest--frame wavelength of the H$\alpha$--[NII] complex. {\it Bottom:} Chi ($\chi$) distribution obtained for each method in the rest frame wavelength range 6540--6600 \AA\ (see text in App.~\ref{Chi2}). The vertical dot lines identify the rest--frame wavelength of the H$\alpha$--[NII] complex. The horizontal area highlighted by the dashed grey lines represents the $\pm$1 $\chi$ range values.}
\label{chi2_plots}
\end{figure*}

\begin{figure*}
   \centering
   \includegraphics[width=0.32\textwidth, height=0.42\textwidth]{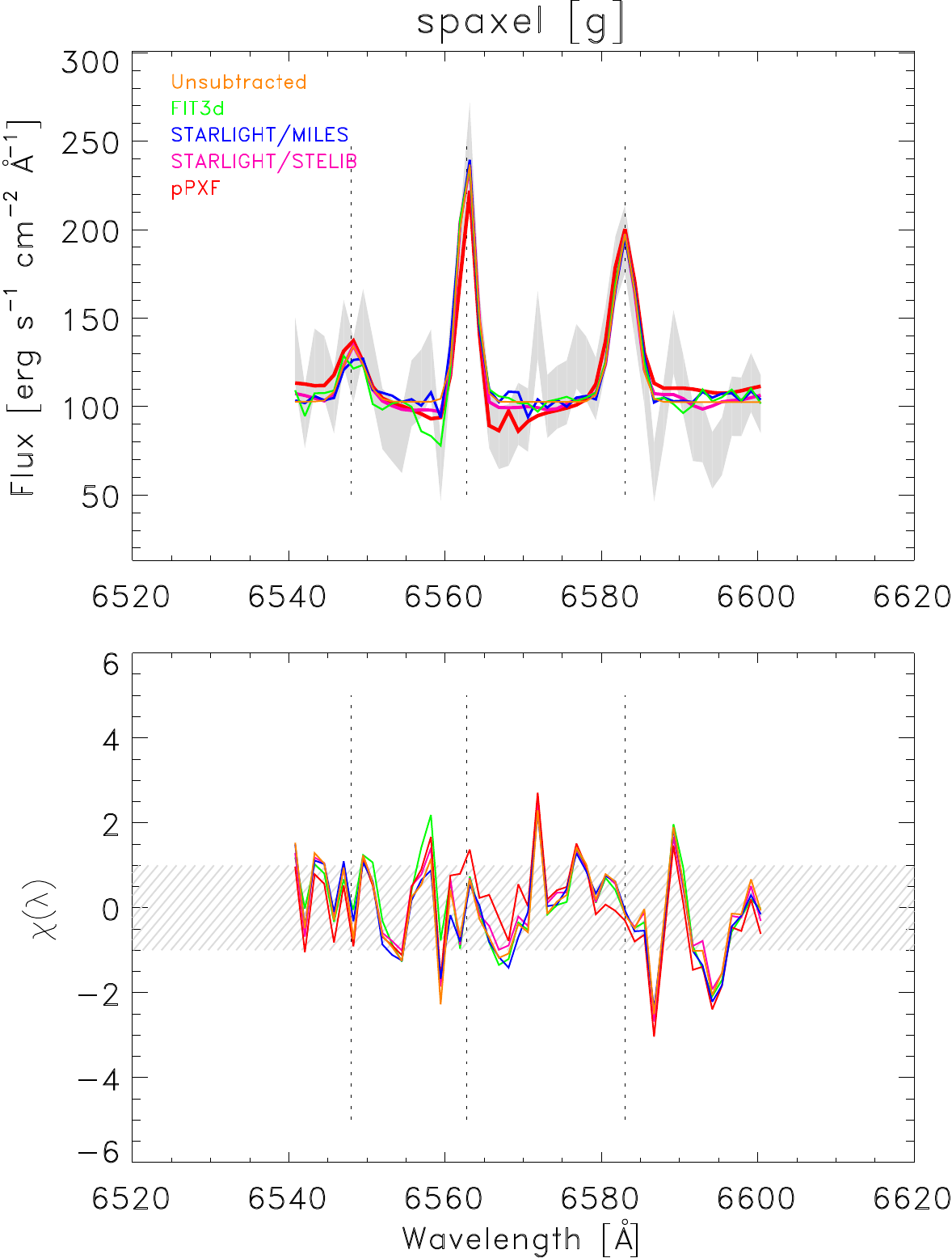}   
   \includegraphics[width=0.32\textwidth, height=0.42\textwidth]{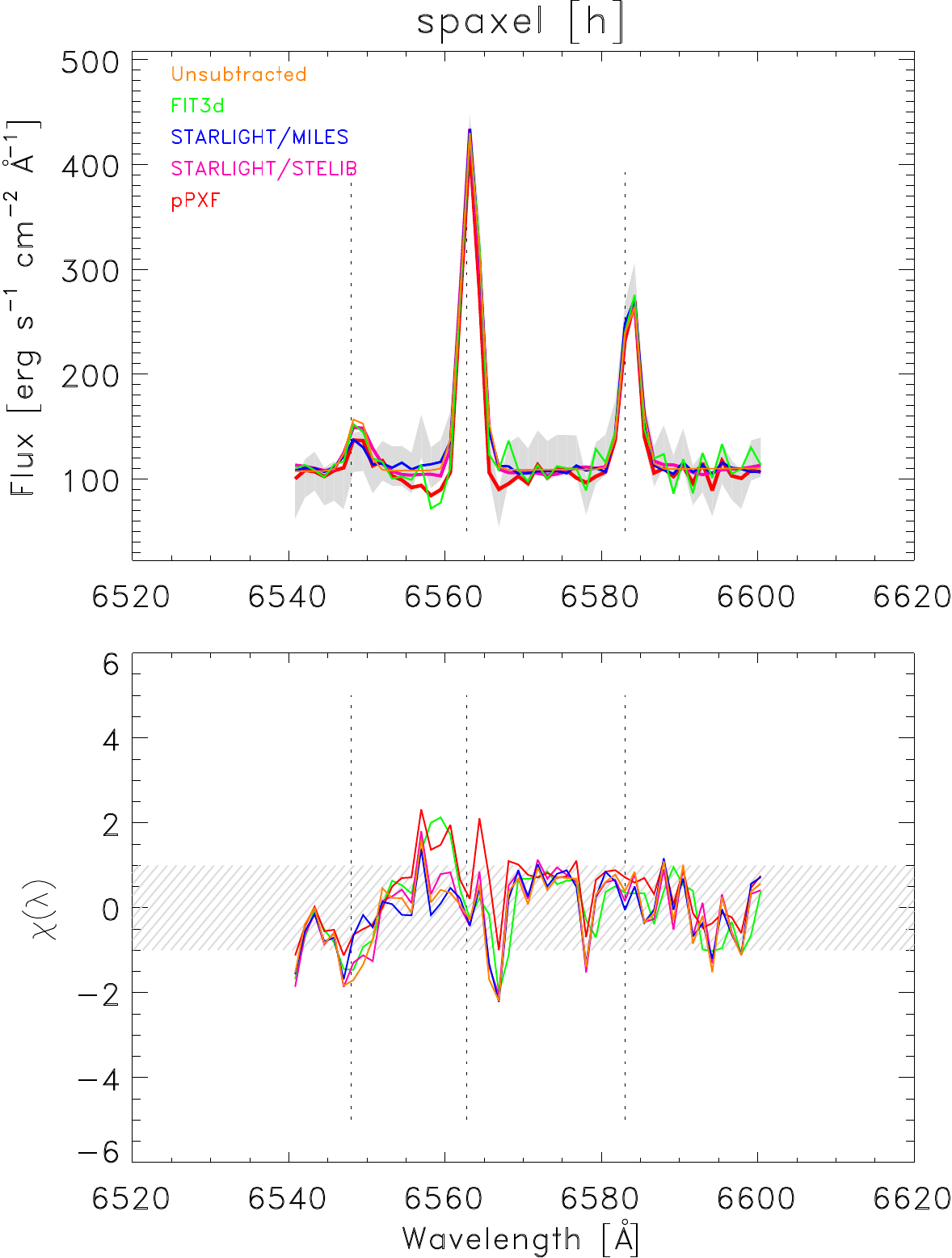}   
   \includegraphics[width=0.32\textwidth, height=0.42\textwidth]{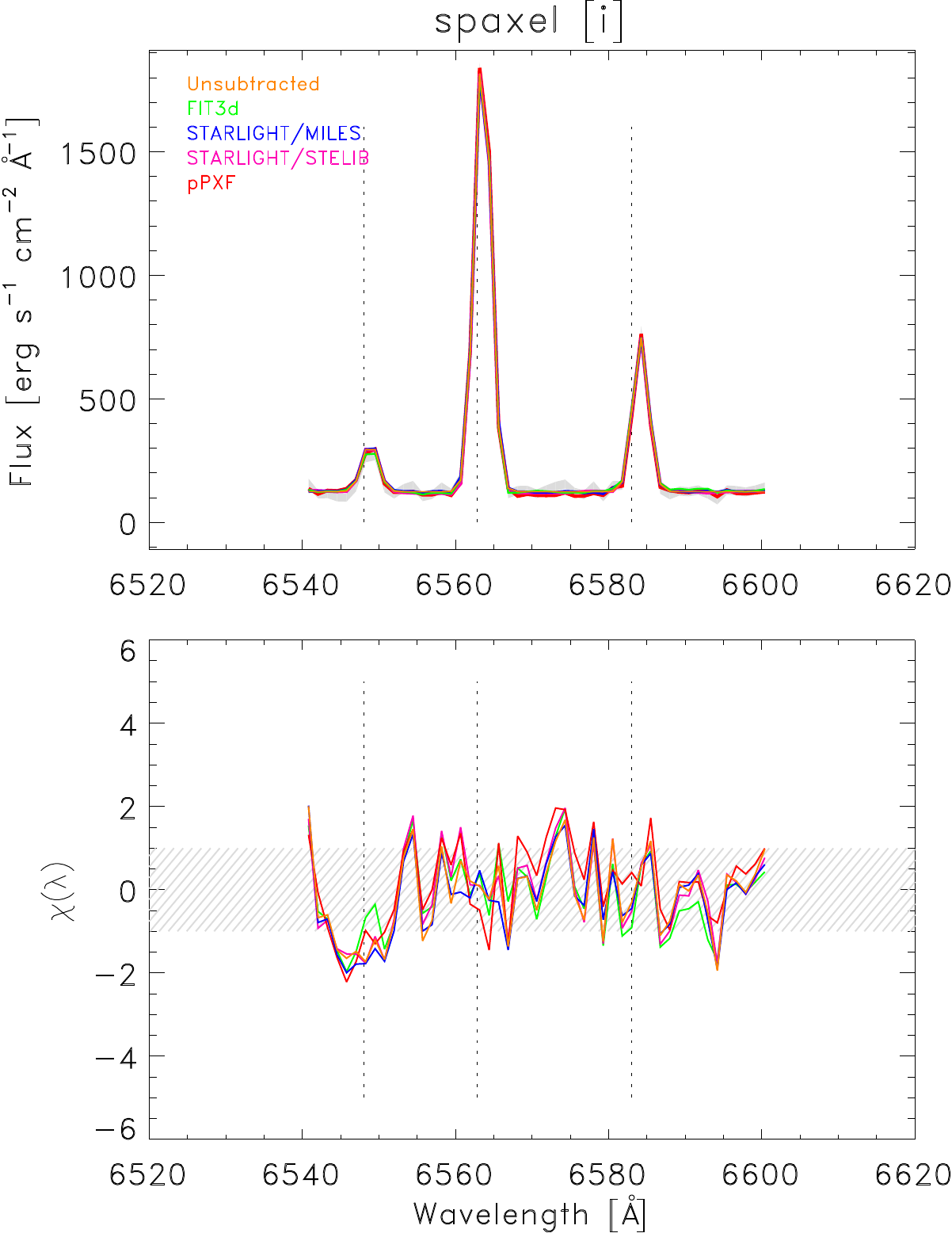}   

\vspace{5mm} 

   \includegraphics[width=0.32\textwidth, height=0.42\textwidth]{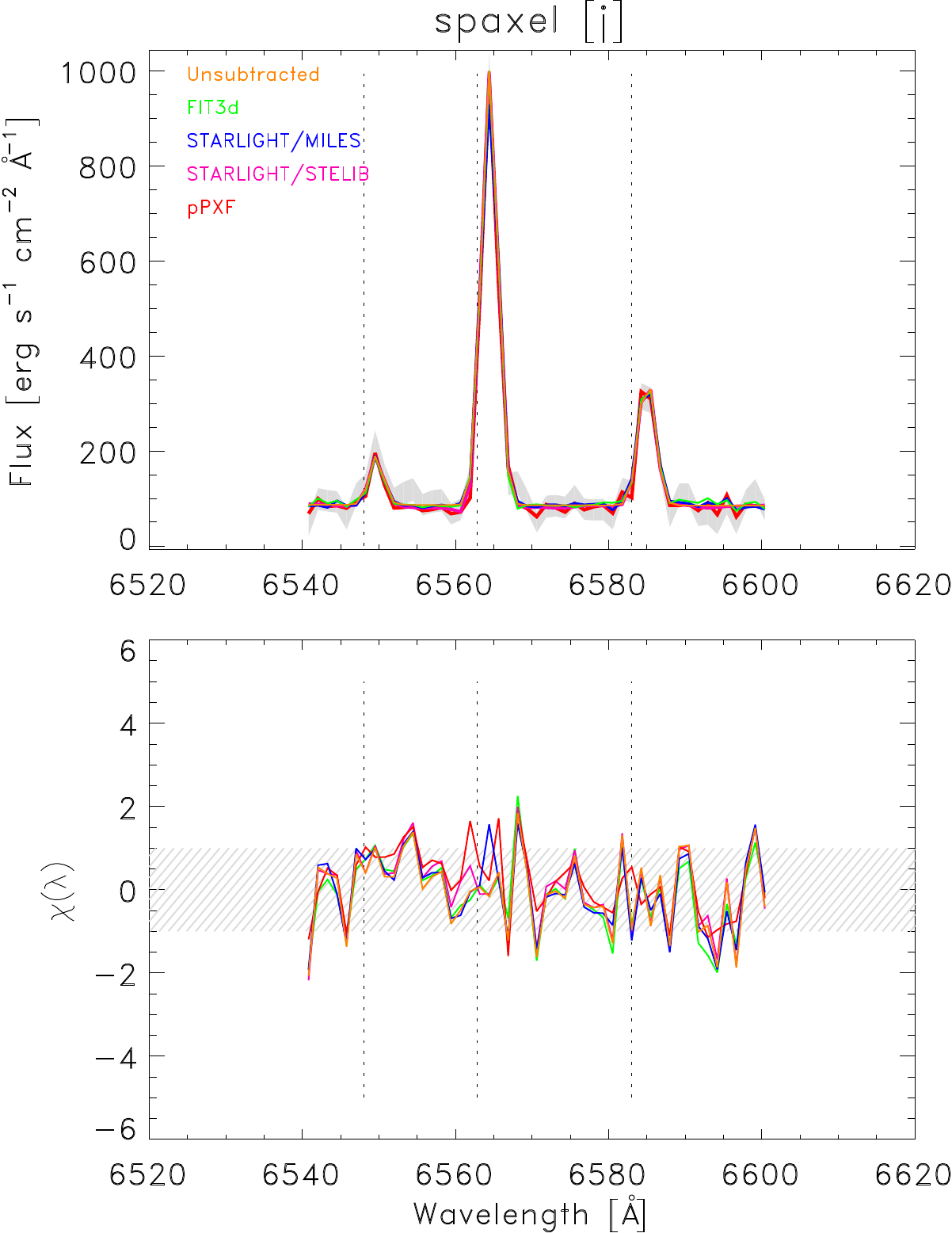}   
   \includegraphics[width=0.32\textwidth, height=0.42\textwidth]{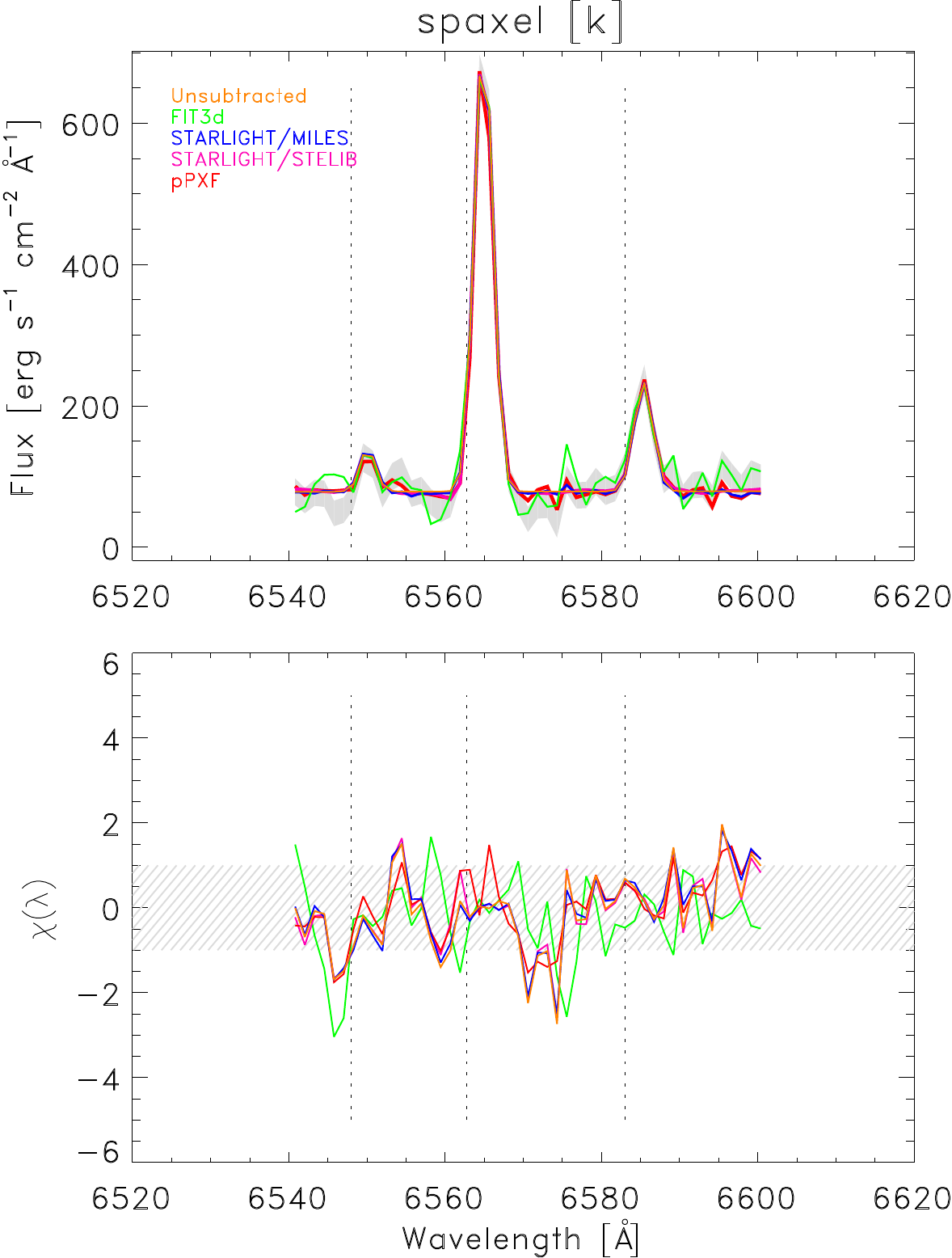}   
   \includegraphics[width=0.32\textwidth, height=0.42\textwidth]{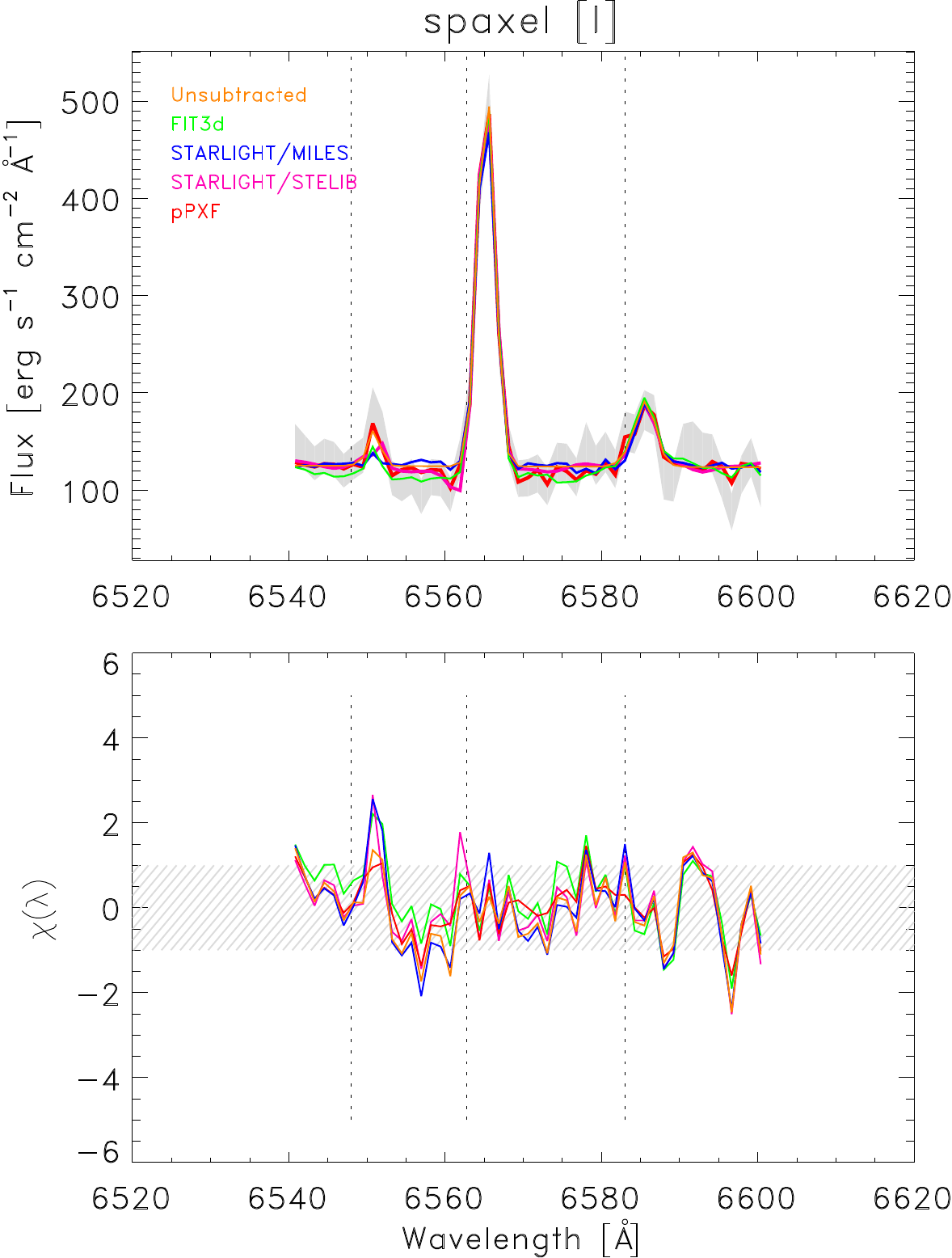}   
 \caption{Same figure caption as in Fig.~\ref{chi2_plots}.}
 \label{chi2_plots_2}
\end{figure*}

In Figs.~\ref{chi2_plots} and \ref{chi2_plots_2} we compare the data $O_i$ with the different models $M_i$, derived as described in Eq.~\ref{ff_total}, for each spaxel. The $\chi$ values as a function of the wavelength ($\lambda$) are also shown for each model.

\begin{figure*}
\vskip5mm
   \centering
  \includegraphics[width=0.56\textwidth]{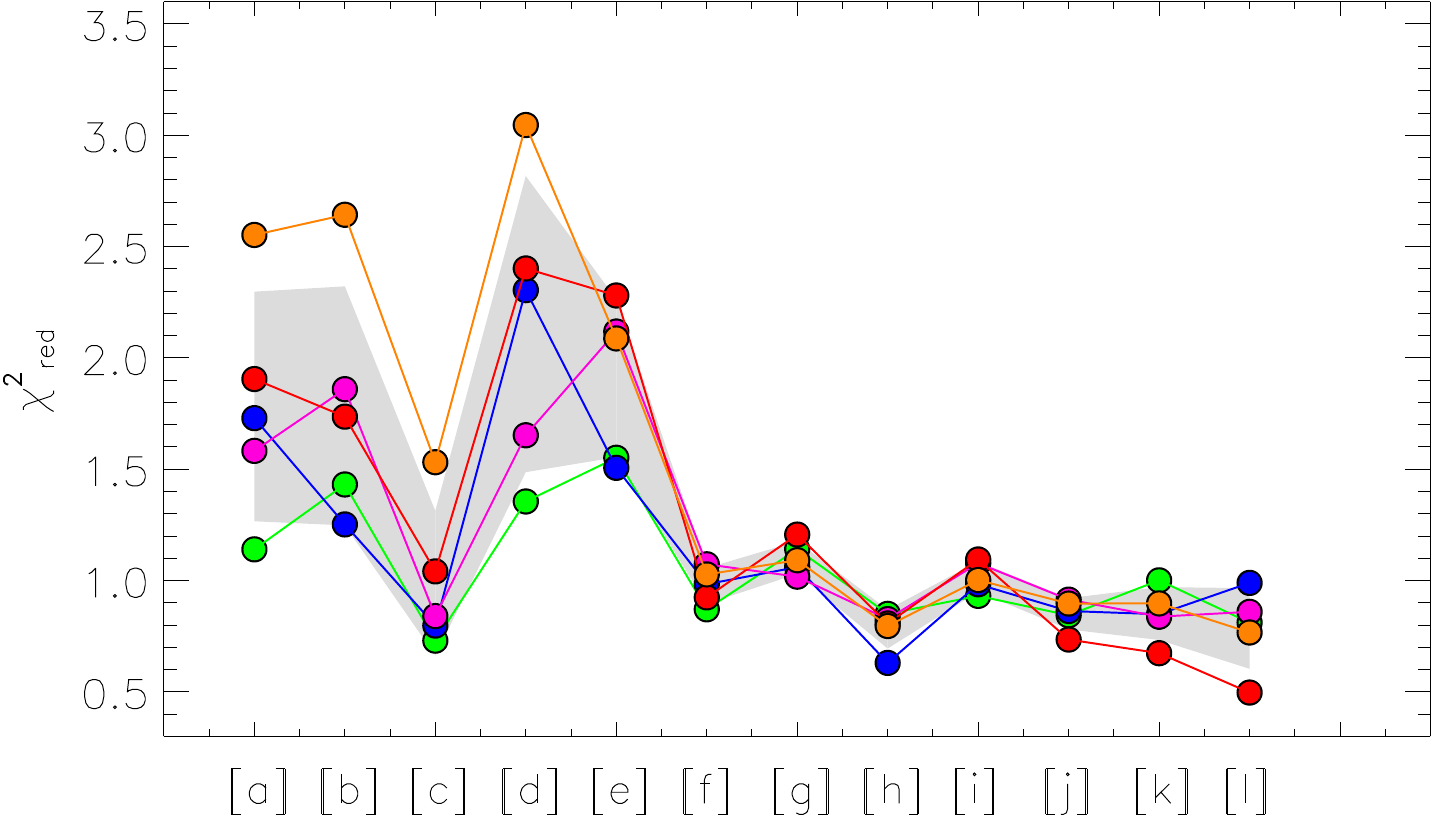}   
     \caption{Reduced chi--square distribution for each individual spaxel as derived when applying the five stellar subtraction methods (color--coded as in Fig.~\ref{chi2_plots}). The grey area represents the range of values covered by the mean chi--square within the respective standard deviation derived for each $i$-th spaxel ($\overline\chi^2_i \pm \Delta\overline\chi^2_i$).}
 \label{chi2_values_fig}
\end{figure*}

The reduced chi--square $\chi^2_{red}$ values for all spaxels are reported in Tab.~\ref{chi2_values_tab} and they are also shown in Fig.~\ref{chi2_values_fig}. The best solution (lowest $\chi^2_{red}$) as well as the secondary best $\chi^2_{red}$ results are highlighted in Tab.~\ref{chi2_values_tab}.

Our results suggest that for the majority of the spectra, especially those characterized by a log([NII]/H$\alpha$) ratio in between --0.49 and $\sim$0.0 (low H$\alpha$/[NII] ratio; spectra from {\tt a} to {\tt f}), the best solution is achieved when using the {\tt FIT3d} method. 
The {\tt pPXF} method seems to give good fits only when spectra with higher H$\alpha$/[NII] ratio are considered (i.e., {\tt j, k, l}).
In between these two ranges (i.e., spectra {\tt g, h, i}) {\tt FIT3d and STARLIGHT (MILES and STELIB)} methods derive the best fits.

However, the {\tt Unsubtracted} model does not give good solutions in any case, deriving for the innermost spectra (i.e., from [a] to [d] with low H$\alpha$/[NII] ratio) the most discrepant results (a factor of $\times\sim$2) among all methods considered in this analysis. When considering spectra with higher H$\alpha$/[NII] ratio, the {\tt Unsubtracted} model can also give reasonable secondary fits in a few cases, although for the majority of the spectra the best candidates for a secondary best $\chi^2_{red}$ are {\tt FIT3D and STARLIGHT (MILES and STELIB)}.

\begin{table*}
\centering
\begin{small}
\caption{Reduced chi--square $\chi^2_{red}$ values for the different spaxels and models.}
\label{chi2_values_tab}
%\hskip-8mm 
\begin{tabular}{c|ccccc|cc} 
\hline\hline  
{\tt  Spectrum} &  $\chi^2_{red}$ & $\chi^2_{red}$	&	$\chi^2_{red}$  &$\chi^2_{red}$  & $\chi^2_{red}$	&	Best fit 		& Secondary   \\
{\tt name} & ({\tt \small Unsubtracted}) & ({\small \tt FIT3d})  &  ({\tt \small STARLIGHT/MILES})  &  ({\tt \small STARLIGHT/STELIB}) & ({\tt \small pPXF}) 	& model&	best model	\\ 
  (1) & (2) & (3) & (4) & (5) & (6) & (7) &(8)\\ 
\hline
{\tt a}		&	 2.55		& {\bf 1.14}	&	1.73	&	1.58		&	 1.91	&		{\tt \small \textcolor{green}{FIT3d}} 	& {\tt \small \textcolor{magenta}{STELIB}}	\\
{\tt b}		&	 2.64		& 1.43	&	{\bf 1.25}	&	1.86		&	 1.74	&		{\tt \small \textcolor{blue}{STARLIGHT/MILES}}	& {\tt \small \textcolor{green}{FIT3d}} 	\\
{\tt c}		&	 1.53		& {\bf 0.73}	&	0.80	&	0.84		&	 1.04	&		{\tt \small \textcolor{green}{FIT3d}} 	& {\tt \small \textcolor{blue}{MILES}}	\\
{\tt d}		&	 3.05		& {\bf 1.36}	&	2.30	&	1.65		&	 2.40	&		{\tt \small \textcolor{green}{FIT3d}} 	& {\tt \small \textcolor{magenta}{STELIB}}	\\
{\tt e}		&	 2.09		& 1.55	&	{\bf 1.51}	&	2.11		&	 2.28	&		{\tt \small \textcolor{blue}{STARLIGHT/MILES}}		& {\tt \small \textcolor{green}{FIT3d}}  \\
{\tt f}		&	 1.03		& {\bf 0.87}	&	0.98	&	1.07		&	 0.92	&		{\tt \small \textcolor{green}{FIT3d}} 	&{\tt \small \textcolor{red}{pPXF}}			\\
{\tt g}		&	 1.09		& 1.14	&	1.06	&	{\bf 1.02}		&	 1.21	&		{\tt \small \textcolor{magenta}{STARLIGHT/STELIB}}	& {\tt \small \textcolor{blue}{MILES}}	\\
{\tt h}		&	 0.79		& 0.85	&	{\bf 0.63}	&	0.83		&	 0.81	&		{\tt \small \textcolor{blue}{STARLIGHT/MILES}} 	&	{\tt \small \textcolor{orange}{Unsubtracted}} \\
{\tt i}		&	 1.00		& {\bf 0.93}	&	0.98	&	1.08		&	 1.09	&		{\tt \small \textcolor{green}{FIT3d}} & {\tt \small \textcolor{blue}{MILES}}	\\
{\tt j}		&	 0.90		& 0.84	&	0.86	&	0.91		&	 {\bf 0.74}	&		{\tt \small \textcolor{red}{pPXF}} 	& {\tt \small \textcolor{green}{FIT3d}} 	\\
{\tt k}		&	 0.90		& 1.00 	&	0.85	&	0.84		&	 {\bf 0.67}	&	 	{\tt \small \textcolor{red}{pPXF}} 	& {\tt \small \textcolor{magenta}{STELIB}}	\\
{\tt l}		&	 0.77		& 0.81	&	0.99	&	0.86		&	 {\bf 0.50}	&		{\tt \small \textcolor{red}{pPXF}} 	&	{\tt \small \textcolor{orange}{Unsubtracted}} \\
\hline\hline
\end{tabular}
\begin{minipage}[h]{18cm}\vskip3mm
{\bf Notes.} Col (1): Name of the spectrum (see text). 
Cols (2 -- 6): Reduced chi--square values obtained when comparing the data with the five models ({\tt Unsubtracted}, {\tt FIT3d}, {\tt STARLIGHT/MILES}, {\tt STARLIGHT/STELIB} and {\tt pPXF}). The lowest $\chi^2_{red}$ value is pointed out.
Col (7): Best fit model with the lowest $\chi^2_{red}$ value. 
Col (8): Secondary best model solution.
\end{minipage}
\end{small}
\end{table*}

%----------------------------------

% -----------------------------------------------------------------------------
\end{document}